\documentclass[twocolumn, twocolappendix]{aastex631}

\shorttitle{A new look into the atmospheric composition of WASP-39\,b}
\shortauthors{Ma et al.}

\graphicspath{{./}{figures/}}

\usepackage{savesym}
\savesymbol{tablenum}
\usepackage{siunitx}
\usepackage{chemfig}
\usepackage{amssymb} %for the \shortminus
\usepackage{amsmath}
\usepackage{latexsym}
\usepackage{graphicx}
\usepackage{longtable}
\usepackage{epsf}
\usepackage{color, colortbl}

\definecolor{Gray}{gray}{0.9}
\definecolor{Dark}{gray}{0.7}
\definecolor{LightCyan}{rgb}{0.88,1,1}
\definecolor{LightYellow}{rgb}{1,1,0.88}
\definecolor{LightPink}{rgb}{1,0.88,1}
\definecolor{LightRed}{rgb}{1,0.88,0.88}
\definecolor{LightPurple}{rgb}{0.88,0.88,1}

\begin{document}

\title{A new look into the atmospheric composition of WASP-39\,b}

\correspondingauthor{Sushuang Ma}
\email{sushuang.ma@kcl.ac.uk}
\footnote{This is the Accepted Manuscript version of an article accepted for publication in \apj. IOP Publishing Ltd is not responsible for any errors or omissions in this version of the manuscript or any version derived from it. This Accepted Manuscript is published under a CC BY licence. The Version of Record is available online at \url{http://doi.org/10.3847/1538-4357/ae5620}.}

\author[0000-0001-9010-0539]{Sushuang Ma}
\affiliation{Department of Physics, Faculty of Natural, Mathematical \& Engineering Sciences, \\King's College London, Strand, WC2R 2LS London, UK}

\author[0000-0002-1437-4228]{Arianna Saba}
\affiliation{Department of Physics, Faculty of Natural, Mathematical \& Engineering Sciences, \\King's College London, Strand, WC2R 2LS London, UK}

\author[0000-0003-2241-5330]{Ahmed Faris Al-Refaie}
\affiliation{Department of Physics and Astronomy, University College London, \\Gower Street, WC1E 6BT London, UK}

\author[0000-0001-6058-6654]{Giovanna Tinetti}
\affiliation{Department of Physics, Faculty of Natural, Mathematical \& Engineering Sciences, \\King's College London, Strand, WC2R 2LS London, UK}

\author[0000-0001-9286-9501]{Sergei N. Yurchenko}
\affiliation{Department of Physics and Astronomy, University College London, \\Gower Street, WC1E 6BT London, UK}

\author[0000-0002-4994-5238]{Jonathan Tennyson}
\affiliation{Department of Physics and Astronomy, University College London, \\Gower Street, WC1E 6BT London, UK}

\author[0000-0001-7480-0324]{Cesare Cecchi-Pestellini}
\affiliation{INAF—Osservatorio Astronomico di Palermo, P.za Parlamento 1, 90134 Palermo, Italy}

\received{2025 April 10}
\revised{2026 March 10}
\accepted{2026 March 19}
\published{2026 May 7}
\submitjournal{\apj}

\begin{abstract}
Being one of the first exoplanets observed by the James Webb Space Telescope, WASP-39\,b has become an iconic target and many transit spectra recorded with different instruments (NIRISS, NIRCAM, NIRSpec G395H, NIRSpec PRISM and MIRI) are currently available, allowing in-depth studies of its atmosphere. We present here a novel approach to interpret WASP-39\,b's transit spectroscopic data, consisting of a multi-step process where ab initio equilibrium chemistry models and blind retrievals are used iteratively to find physically robust, optimal solutions. Following this approach, we have identified a new scenario to explain WASP-39\,b's atmospheric composition, in which silicon-based chemistry plays a major role. In this scenario, SiO may explain the spectral absorption at 4.1~$\mu$m, currently interpreted as being due to SO$_2$. SiO and the other gas species identified by the retrieval models, i.e. H$_2$O, CO$_2$, Na and K, are consistent with an atmosphere in chemical equilibrium with a temperature-pressure profile constrained by H$_2$O and CO$_2$ absorption bands. In addition, silicate clouds and hazes can produce the spectral features observed by MIRI in the spectral window 5--12~$\mu$m. While we advocate the need for more data, possibly at higher spectral resolution, to confirm our results for WASP-39\,b's atmospheric composition, we highlight a refined atmospheric retrieval strategy with pre-selection and post-reconstruction to guide the next generation of transit spectroscopy.
\end{abstract}

%% Keywords should appear after the \end{abstract} command. 
%% The AAS Journals now uses Unified Astronomy Thesaurus concepts:
%% https://astrothesaurus.org
%% You will be asked to selected these concepts during the submission process
%% but this old "keyword" functionality is maintained in case authors want
%% to include these concepts in their preprints.

%% From the front matter, we move on to the body of the paper.
%% Sections are demarcated by \section and \subsection, respectively.
%% Observe the use of the LaTeX \label
%% command after the \subsection to give a symbolic KEY to the
%% subsection for cross-referencing in a~\ref command.
%% You can use LaTeX's~\ref and \label commands to keep track of
%% cross-references to sections, equations, tables, and figures.
%% That way, if you change the order of any elements, LaTeX will
%% automatically renumber them.
%%
%% We recommend that authors also use the natbib \citep
%% and \citet commands to identify citations. The citations are
%% tied to the reference list via symbolic KEYs. The KEY corresponds
%% to the KEY in the \bibitem in the reference list below. 

%% Introduction

\section{Introduction} \label{sec:intro}

\subsection{First WASP-39 b observations}
% History of W observation
WASP-39\,b has been among the first exoplanets observed by the James Webb Space Telescope (JWST) shortly after its commissioning phase in 2022 \citep{2023PASP..135d8001R}. 
The planet was discovered by the ground-based transit survey SuperWASP \citep{Faedi2011WASP39b}. 
Their analysis indicated WASP-39\,b to be a highly-inflated Saturn-like exoplanet, with a mass of $0.28 \pm 0.03 \; \mathrm{M_J}$ and a radius of $1.27 \pm 0.04 \; \mathrm{R_J}$. The study also estimated an equilibrium temperature of ${\sim}1100$~K.
The parent star, WASP-39\,A, was studied by \citet{Faedi2011WASP39b, Mancini2018WASP39b, Bonomo2017WASP39b, Maciejewski2016WASP39b}, among others:
it is a G8 type star with a mass of $0.93 \pm 0.03 \, \mathrm{M_\odot}$, a radius of $0.895 \pm 0.023 \, \mathrm{R_\odot}$, an effective temperature of $5400 \pm 150 \, \mathrm{K}$ and [Fe/H] of $- 0.12 \pm 0.1$. 
The stellar age was estimated to be 8.5--9 Gyrs \citep{Mancini2018WASP39b,Bonomo2017WASP39b}.

%% Hubble results
Prior to JWST, observations of WASP-39\,b using the Wide Field Camera 3 (WFC3) onboard the Hubble Space Telescope indicated the presence of water vapour in its atmosphere \citep{wakeford2018cloudretrieval, Tsiaras2018hot-jupiter, Pinhas2018WASP39b, Pinhas2019hot-jupiterH2O, FisherHeng2018WFC3retrieval}. Additionally, sodium and potassium were suggested by observations using the Hubble Space Telescope Imaging Spectrograph \citep[HST STIS,][]{Sing2016WASP39b, Fischer2016WASP39b}, the ground-based VLT \citep[0.411--0.810 $\mu\text{m}$,][]{Nikolov2016WASP39b} and LRG-BEASTS survey \citep[0.4--0.9 $\mu\text{m}$,][]{Kirk2019WASP39b}. \citet{wakeford2018cloudretrieval} and \citet{Kirk2019WASP39b} reported highly supersolar metallicities to interpret these datasets. \citet{ThorngrenFortney2019WASP39b} estimated the maximum atmospheric metallicity of WASP-39\,b to be $54.5 \times$ solar.

\citet{Saba2024STIS} analysed archive data obtained with Hubble STIS and
WFC3 instruments of WASP-39 at various epochs. The metrics used as stellar activity indicators leaned towards WASP-39 being an inactive star. These results are in agreement with
previous studies \citep{Sing2016WASP39b, Mancini2018WASP39b}, with the exception of \citet{Pinhas2018WASP39b},
who suggested stellar heterogeneities.

\subsection{JWST observations of WASP-39 b} \label{sec:data}
% JWST observation of W
Transit spectra of WASP-39\,b using different JWST's instrument modes were published by \citet[][NIRCam, 2.4--4.0 $\mu\text{m}$]{Ahrer2023WASP39b}, \citet[][NIRISS, 0.6--2.8 $\mu\text{m}$]{Feinstein2023WASP39b}, \citet[][NIRSpec G395H, 2.7--5.2 $\mu\text{m}$]{Alderson2023WASP39b}, \citet[][NIRSpec PRISM, 0.5--5.5 $\mu\text{m}$]{Rustamkulov2023WASP39b}, and \citet[][NIRSpec G395H $\&$ PRISM]{Sarkar2024WASP39b}. 
 A single transit of WASP-39\,b was observed by NIRISS in SOSS mode on 26 July 2022; by NIRCAM in both long-wavelength (LW) and short-wavelength (SW) channels on 22$-$23 July 2022; by NIRSpec with its G395H grating using the Bright Object Time Series mode (BOTS) on 30--31 July 2022 and in the PRISM mode on 10 July 2022. Additionally, it was observed with MIRI LRS, from 5 to 12 $\mu\text{m}$, on 14 February 2023. By combining all available JWST's observations of WASP-39\,b, we obtain a spectral coverage from 0.6 to 12 $\mu\text{m}$.

The transit data obtained with each of these instruments have undergone extensive scrutiny by the exoplanet community \citep{2023Natur.614..649J, 2023ApJ...949L..15G, Tsai2023WASP39b, Esparza-Borges2023WASP39b, 2023ApJ...956L..19L, 2023ApJ...959L..30T, 2024A&A...685A..64K, Arfaux2024Cloud, Roy-Perez2025phAerosol_R}. A variety of data analysis pipelines, saturation handling techniques \citep{Sarkar2024WASP39b}, noise mitigation methods \citep{Holmberg2023_NIRISS} and spectral extraction algorithms \citep{2022PASP..134i4502D} have been developed. Beyond providing open access to the data, one of the goals of the ERS program was to invite the community to perform independent data analysis and compare results. NIRISS data were reduced with six different pipelines: \texttt{nirHiss} \citep{Feinstein2023WASP39b,Bell2022Eureka}, \texttt{exoTEDRF} \citep{Radica2024exoTEDR,Radica2023exoTEDRF}, \texttt{Transitspectroscopy} \citep{espinoza_2022_6960924}, \texttt{Iraclis} \citep{2016ApJ...832..202T}, \texttt{NAMELESS}, \texttt{FIREFLy} \citep{2022ApJ...928L...7R}. NIRCAM data was processed using \texttt{Eureka!} \citep{Bell2022Eureka}, \texttt{tshirt}, \texttt{HANSOLO} \citep{2016A&A...587A..67L, 2017A&A...606A..18L} and \texttt{chromatic-fitting}. NIRSpec G395H data was reduced by six separate teams using \texttt{ExoTiC-JEDI} \citep{lili_alderson_2022_7185855}, \texttt{Aesop}, ExoTEP \citep{2019ApJ...887L..14B}, \texttt{Tiberius} \citep{Kirk2019WASP39b} and \texttt{Transitspectroscopy}. For NIRSpec PRISM data, \texttt{Eureka!}, \texttt{Tiberius}, \texttt{FIREFLy}, and \texttt{tshirt} were utilised. MIRI LRS data was processed with \texttt{Eureka!}, \texttt{Tiberius}, and \texttt{SPARTA} \citep{2023Natur.620...67K}.

The data obtained with different JWST instrument modes have not always suggested a consistent story, possibly due to uncertainties in instrumental performance and/or reduction processes \citep[e.g.,][]{Holmberg2023_NIRISS, davey2024binning, Lueber2024WASP39b}. For instance, depending on the instrument mode used, WASP-39\,b's C/O ratio has been reported to be sub-stellar \citep{Ahrer2023WASP39b}, sub-solar \citep{Feinstein2023WASP39b}, sub-solar to solar \citep{Alderson2023WASP39b} or super-solar \citep{Rustamkulov2023WASP39b}. 
It is known that different data pipelines can potentially produce inconsistent results \citep{Mugnai2024population}: this inconsistency can pose challenges when combining datasets processed by different pipelines. Offsets in the spectral data may arise due to different data analysis approaches, variations in intrinsic instrumental calibrations, and different choices of system parameters \citep[see e.g.,][]{yip2021wasp96b}. Currently, no single pipeline has been used to uniformly reduce WASP-39\,b's data from all JWST instruments and their corresponding observing modes. We use here the spectral data published in \cite{Carter2024JWSTWASP39b} and \cite{Powell2024WASP39bMIRI} for the reasons described below. 

Although the datasets for WASP-39\,b were processed using different data reduction pipelines, \cite{Carter2024JWSTWASP39b} selected the nominal reductions for each available dataset. They used \texttt{Eureka!} to extract a white light curve from each JWST dataset, resulting in a total of seven light curves: NIRISS order 1 and order 2, NIRSpec G395H NRS 1 and NRS 2, NIRSpec PRISM, NIRCam SW photometric channel, and LW spectroscopic channel. A joint white light curve fitting with \texttt{juliet} \citep{2019MNRAS.490.2262E} was performed, incorporating one TESS light curve, six NGTS light curves, and radial velocity measurements from CORALIE and SOPHIE. This combined fitting yielded highly precise constraints on the system parameters, achieving a minimum precision of 0.5\,\%. \texttt{Eureka!} was also used to extract chromatic light curves at native resolution for each JWST dataset. The spectral light curve fitting was performed by fixing the orbital parameters to those derived from the joint fitting. This procedure ensured a reduction in the observable offsets among the initial data release presented in \cite{Lueber2024WASP39b}. \cite{Carter2024JWSTWASP39b} applied an additional manual offset of --177 ppm to the NIRSpec PRISM data. Moreover, their final combined spectrum excluded the saturated region of the NIRSpec PRISM data between 0.62--2.1~$\mu\text{m}$.

MIRI observations were scheduled approximately seven months after the observations were recorded with NIRISS, NIRSpec, and NIRCam. MIRI transit data were observed in the low-resolution spectroscopy (LRS) slitless prism mode at an average resolution of approximately 100. \cite{Powell2024WASP39bMIRI} reports that the white light-curve fitting of MIRI data performed with \texttt{Eureka!} adopted the same orbital parameters as those used by \cite{Carter2024JWSTWASP39b}. Although the MIRI data were reduced with three different pipelines, we include here only the reduction obtained with \texttt{Eureka!} to minimise offsets. This choice ensures further consistency across all instruments. Meanwhile, \citet{Flagg2024WASP39b} have suggested that MIRI data beyond 10\,$\mu$m might be contaminated by circumstellar debris disk features that need to be considered in data analysis.

\subsection{Scientific interpretation of WASP-39 b transit spectra recorded with JWST} \label{sec:intro_so2}

Various molecules were reported in the literature to be present in the atmosphere of WASP-39\,b, including H$_2$O, CO, CO$_2$, K, H$_2$S and CH$_4$ \citep{Lueber2024WASP39b, constantinou2024viraexoplanetatmosphericretrieval, Ahrer2023WASP39b, Alderson2023WASP39b, Rustamkulov2023WASP39b, Feinstein2023WASP39b, Esparza-Borges2023WASP39b}. There are also tentative detections of SO$_2$ from \citet{Tsai2023WASP39b, Alderson2023WASP39b, Rustamkulov2023WASP39b, constantinou2024viraexoplanetatmosphericretrieval} and \citet{Powell2024WASP39bMIRI}. \citet{Lueber2024WASP39b} re-examined the spectral feature around $4.1 \, \mu\text{m}$ attributed to SO$_2$ absorption and suggested that its interpretation appears to be model-dependent: models with clouds favour the detection of SO$_2$ in the NIRSpec PRISM spectrum. They also discussed the use of grey clouds and non-grey ones, as implemented in \texttt{BeAR} \citep{Kitzmann2020helios}, in the simulations. 

%% elemental ratio
\citet{Lueber2024WASP39b} considered CH$_4$, H$_2$O, CO, CO$_2$ and SO$_2$ as C- and O-bearing molecules in their retrieval simulations of NIRCAM and NIRSpec PRISM ERS spectra. Their retrieved values of C/O using random forest are consistent with the stellar value of 0.46 $\pm$ 0.09 from \citet{Polanski2022CO}. Their NIRSpec PRISM free retrieval results are closer to the solar value of 0.55 from \citet{Asplund2009solarabundance} \citep[see Fig. 13 in][]{Lueber2024WASP39b}. Their nested-sampling retrievals suggest C/H and O/H values to be consistent with solar.

%%% Metallicity
The spread in the reported metallicities in the literature suggests that factors beyond instrumental modes may play a role in these discrepancies. For instance, when using different data analysis techniques, \citet{Alderson2023WASP39b}, \citet{Feinstein2023WASP39b} and \citet{Rustamkulov2023WASP39b} reported super-solar metallicities, \citet{Ahrer2023WASP39b} constrained it to be solar to super-solar, while \citet{Lueber2024WASP39b} found sub- to solar metallicity in most of the free retrievals and super-solar metallicity in the random-forest retrieval. Another identified source of discrepancy is the spectroscopic line lists used to interpret the data. For example, \citet{Niraula2023WASP39b} tested different opacity sources and found different abundances of the same molecules.

\subsection{Retrieval techniques}
% Retrieval and TauREx
%% The difference with forward methods
Spectral retrieval methods have often been used in recent years to interpret exoplanetary atmospheric spectra. These algorithms use statistical techniques and high-performance computers to sample a broad parameter space and to search for optimised solutions through a considerable number of iterations. Robust convergence, especially in high-dimensional retrievals, requires a large number of iterations and steps, making the computing power requirements more onerous. A notable example of such sampling algorithms is Multinest \citep{Feroz2019multinest, Feroz2009Multinest, Feroz2008multinest}, a Bayesian inference tool that calculates the evidence with an associated error estimate while generating posterior samples from distributions that may exhibit multiple modes and complex degeneracies in high dimensions.

Constraining the atmospheric temperature--pressure ($T$--$p$) profile in the case of exoplanets has been a challenge prior to the launch of JWST \citep[see e.g.,][]{changeat2022five,  Novais2025paramdegene, Bourrier2020wasp121b, Kitzmann2020helios}. These studies indicated that the quality of the observations was not good enough to constrain $T$--$p$ profiles in retrievals.
By contrast, \citet{Rocchetto2016Tp} and \citet{Schleich2024Tp} have demonstrated the necessity of considering non-isothermal retrievals to interpret transit spectra correctly, especially when thermal inversions are present. \citet{Changeat2025titan} performed retrievals using the $N$-point profiles on both Titan occultation data from Cassini and JWST and HST transit data of hot Jupiters, including WASP-39\,b. 
Three-dimensional effects, particularly the inhomogeneity of $T$--$p$ profiles across the limbs and poles, are likely to have an impact as well in the retrieval of vertical thermal profiles \citep{Skinner2025WASP39b, Tada2025WASP39b, Chen2025WASP39b, Fisher2024jwst}. However, it is not easy to isolate these effects from other contributions. Phase-curve observations are, in principle, more informative to constrain said inhomogeneities.

Our paper focuses on improving the atmospheric retrieval strategy, using WASP-39 b as an example. Our approach has led to a different interpretation of the available JWST data, opening up a new perspective in our understanding of the chemical composition of WASP-39\,b, and other similar planets. 

%% Methodology

\section{Methodology} \label{sec:method}
\subsection{Possible offsets among spectra recorded with different instruments} \label{sec:specoffset}
\begin{figure*}[ht!]
\plotone{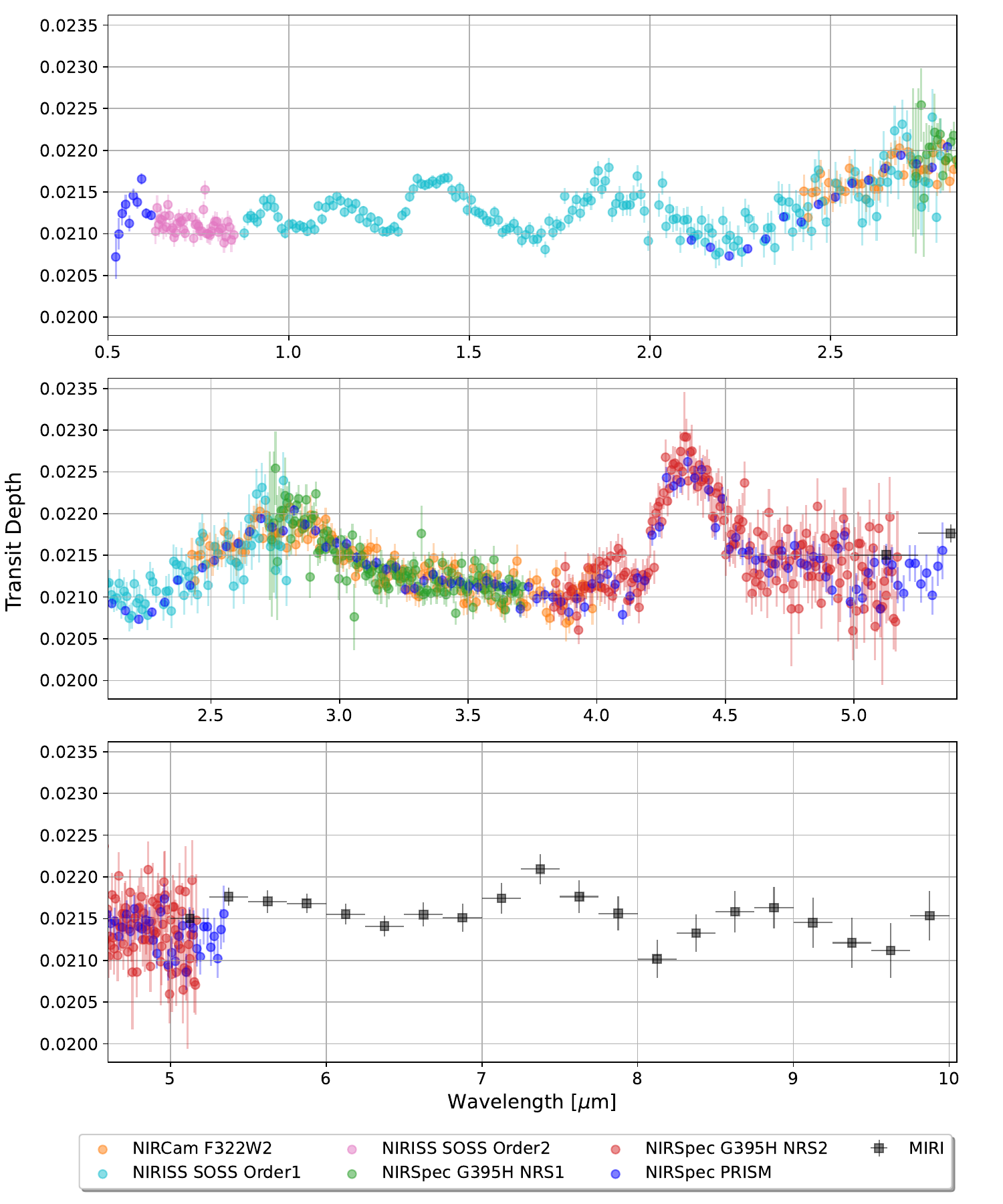}
\caption{JWST NIRISS, NIRCam, NIRSpec and MIRI observations of WASP-39\,b with offsets. The original NIRISS, NIRCam and NIRSpec data are from \citet{Carter2024JWSTWASP39b} and the MIRI data are from \citet{Powell2024WASP39bMIRI}. We tuned the offsets according to data in resolutions available; from resolving power of 100 to native resolutions, and here we plotted the native resolution of NIRSpec PRISM data and 1/5 native resolution for the rest of the instruments and modes. We added offsets to data from each instrument using the NIRSpec data as a baseline, as detailed in the main text. 
\label{fig:observation}}
\end{figure*}

We used the observational data described in Section~\ref{sec:data} and detailed in Table~\ref{tab:datasets}. For our retrieval experiments, we used a resolution (${\sim}100$) for each instrument to maintain a uniform resolution across instruments and modes. However, we have tested our results at higher spectral resolution in the spectral ranges where higher resolution data are available. 
We followed \citet{Carter2024JWSTWASP39b} and applied a manual offset of --177 ppm to the NIRSpec PRISM data, while adding 68 ppm offset to the NIRISS (we assumed orders 1 and 2 have the same offsets) data and 132 ppm to NIRCam according to the NIRSpec data. To align with the shorter wavelengths, we add a 420 ppm offset to MIRI data. The final data we use in further atmospheric analysis is shown in Fig.~\ref{fig:observation}

\begin{deluxetable*}{llll}
\tablecaption{A list of the datasets used in this study, the reduction algorithms they were processed with, the native spectral resolution and the offsets applied. \label{tab:datasets}}
% \tablewidth{0pt}
\tabletypesize{\scriptsize}
\tablehead{\colhead{Instrument Mode} & \colhead{Reduction Algorithms} &  \colhead{Native Res.} & \colhead{Offset [ppm]}
}
% \decimalcolnumbers
\startdata
NIRISS SOSS & \texttt{exoTEDRF} & 350 - 1,390 & 68 \\
NIRSpec G395H & \texttt{ExoTiC-JEDI} & 1,340 - 2,630 & 0 \\
NIRSpec PRISM & \texttt{FIREFLy} & 20 - 290 & -177 \\
NIRCam LW & \texttt{Eureka!} & 850 - 1,360 & 132 \\
MIRI LRS & \texttt{Eureka!} & 100 & 420 \\
\enddata
\end{deluxetable*}

\subsection{Transit and atmospheric models}
\subsubsection{TauREx}
\texttt{TauREx\,3} \citep{alrefaie2021taurex31, Al-Refaie2022taurex} is a state-of-the-art modelling framework to simulate exoplanetary atmospheres and to interpret exoplanet atmospheric data collected with different techniques through inverse models based on Bayesian statistics. Several plugins can be activated to simulate, e.g., atmospheric chemistry \citep{Al-Refaie2022taurex}, cloud microphysics \citep{Ma2023YunMa}, stellar activity \citep{Thompson2024stellar} and phase-curves interpretation \citep{Changeat2024}. Parameters that are included and tested in retrievals with \texttt{TauREx\,3} include but are not limited to instrumental, atmospheric thermal, chemical, and cloud profiles, as well as planetary and stellar parameters. The radiative transfer calculations performed by \texttt{TauREx\,3} include molecular/atomic/ionic opacities, Rayleigh and Mie scattering, and collision-induced absorptions (CIAs). 

\subsubsection{GGChem}
\texttt{GGChem} \citep{Woitke_2018} is an equilibrium chemistry model that estimates the chemical abundances by minimising the system's Gibbs' Free Energy, including scenarios with equilibrium condensation and charged particles. The elemental abundances adopted as initial conditions can be customised. A comparison and cross-validation of all the chemical equilibrium models implemented in \texttt{TauREx\,3} can be found in \citet{Al-Refaie2022taurex}. These include \texttt{GGChem}, \texttt{FastChem} \citep{Stock2018Fastchem} and \texttt{ACE} \citep{Agundez2020ACE, Agundez2012HD209}.

\subsubsection{YunMa}
\texttt{YunMa} \citep{Ma2023YunMa} includes cloud microphysics models to estimate the vertical size distributions of cloud particles expected in an atmosphere according to its thermodynamical and chemical conditions, following, e.g., the approach described in \citet{ackerman2001precipitating} (A–M hereafter). \texttt{YunMa} can also adopt simplified assumptions, such as homogeneous particle size (HPS) distributions, when there is limited knowledge of the atmospheric conditions and/or efficiency needed for retrieval studies. The current \texttt{YunMa} estimates cloud opacities using the BH-Mie approach \citep{bohren2008absorption} of the cloud particles' Mie scattering and absorption, from refractive indices available in the literature. When integrated into \texttt{TauREx\,3}, \texttt{YunMa} enables the generation of, or retrieval from, transmission and emission spectra of cloudy atmospheres.

\subsection{Hybrid approach adopted here}
\label{sec:method_hybrid}
\begin{figure*}[ht!]
\centering
\includegraphics[width=1.\textwidth]{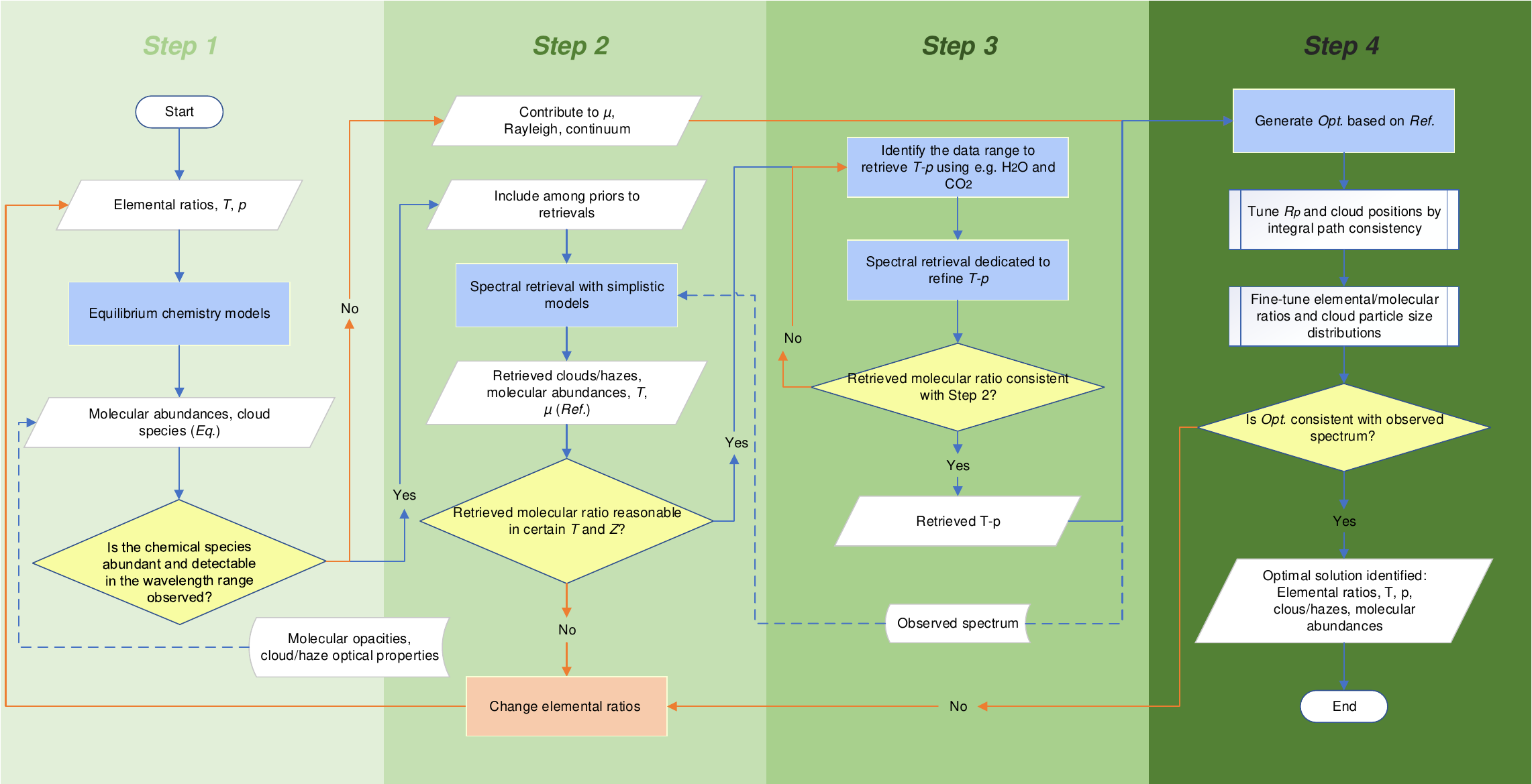}
\caption{Sketch of the approach to analyse exoplanet transit data adopted in this paper.}
\label{fig:approach}
\end{figure*}

To interpret the JWST's observations of WASP-39\,b, we adopt here a novel approach that combines free retrievals and ab initio models to explore more efficiently the parameter space of possible, physically viable solutions compatible with the data, as depicted in Fig.~\ref{fig:approach}. 
We first use equilibrium chemistry models, assuming certain metallicity values, elemental ratios and thermodynamical conditions, to estimate the most plausible gaseous and condensed species -- \textit{Eq.} -- expected to be present in the atmosphere (Step~1). Then, we perform a spectral retrieval analysis (with \texttt{TauREx\,3} and \texttt{YunMa}) with a sub-sample of the chemical species identified in Step~1 as priors, selecting those that have a detectable spectral signature in the wavelength range probed by the observations. We then compare the results from retrievals -- \textit{Ref.} -- to the ones from \textit{Eq.}: if they do not match, we change the assumptions of elemental ratios and/or thermodynamical conditions and repeat the process until convergence is reached (Step~2). To identify optimal solutions -- \textit{Opt.} -- to explain the observed spectra, we first consider the $T$--$p$ profile, which can be refined through retrieval simulations focused on selected spectral bands that contain information about the atmospheric thermal structure (Step~3). After checking the consistency between results from Steps 2 \& 3, finally, we use the optimal atmospheric thermal profile from Step~3 to better constrain the chemical environment identified in Step~2, including condensates from equilibrium chemical models (Step~4).

% to the transmission spectra generated with \textit{Eq.} as input (Step~3). 

We note that the approach adopted in Steps 3 and 4 cannot guarantee the uniqueness of the optimal solutions found.
We explain the key steps of our approach in the following sections.

\subsubsection{Pre-selection of chemical and cloud species as priors to retrievals (Step~1)} \label{sec:method_chem_select}
%%Abundance selection
To identify the most abundant molecular species within the expected temperature and metallicity ranges, we start by assuming the atmosphere to be in chemical equilibrium. We used the \texttt{GGChem} model integrated into \texttt{TauREx\,3} to identify the chemical species and their expected abundances. 
Elements used in the simulations include
H, He, Li, C, N, O, F, Na, Mg, Al, Si, P, S, Cl, K, Ca, Ti, V, Cr, Mn, Fe, Ni, Zr and W
\citep{Woitke_2018} with the elemental ratios of the outer solar convection zone, mainly derived from
spectroscopy of the solar photosphere, as provided by \citet{Lodders2021solar_photo_abund}.
For the pre-selection of the chemical species, we have run the simulations with metallicities ranging from 0.01 to 100\,$\times$\,solar.

%% Cloud selection
In Step~1, instead of simulating condensations with \texttt{GGChem} in the broad parameter space considered, we compared the condensation curves of relevant cloud species at different metallicities occurring in the temperature--pressure regime of WASP-39\,b \citep[see Fig. 10 in][]{Woitke_2018} to exclude cloud species that are unlikely to form in the atmosphere of WASP-39\,b.
MgSiO$_3$ and SiO$_2$ are the most common condensable species forming in the atmosphere of WASP-39\,b in our estimations.

\subsubsection{Free retrieval simulations (Step~2)} \label{sec:retrievalup}
As input to the free retrieval simulations, we adopted the values of planetary and stellar parameters reported by \citet{Faedi2011WASP39b}; we started with assuming an isothermal $T$--$p$ profile. The chemical priors identified in Step~1 were organised into two sets, as detailed in Table~\ref{tab:priors}: we define the active/inactive gases as the molecules in the gas phase being radiatively active/inactive in the relevant wavelength range. 

We narrowed the list of chemical species from Step~1 to a sub-sample selected according to their contributions to the atmospheric opacity, either because of their predicted abundance at the equilibrium or because of the strengths of their molecular/atomic transitions in the wavelength range observed by JWST. Opacity data for 80 species were sourced from the ExoMol database\footnote{https://www.exomol.com/} \citep{chubb2021exomol,jt939}. The versions of cross-sections used in this work and the references align with \citet{Chubb2024Database_Whitepaper_ph}. Among the key opacities used in our simulations and plots, we should mention: 
SiO \citep{Yurchenko2022SiO}, CO$_2$ \citep{Yurchenko2020CO2}, H$_2$O \citep{Polyansky2018pokazatel}, Na \citep{Allard2019Na}, K \citep{Allard2016K}, NaH \citep{Rivlin2015NaH}, CO \citep{Polanski2022CO}, CH$_4$\citep{Yurchenko2024CH4}, 
PN \citep{Semenov2025PN}, PO \citep{Prajapat2017POPS}, TiO \citep{McKemmish2019TiO}, 
VO \citep{McKemmish2016VO}, 
SiS \citep{Upadhyay2018SiS}, PS \citep{Prajapat2017POPS}, NaOH \citep{Owen2021NaOHKOH}, H$_2$S \citep{Azzam2016H2S}, NH$_3$ \citep{Coles2019NH3},
PH$_3$ \citep{Sousa-Silva2014PH3}, SO$_2$ \citep{Underwood2016SO2} and HS \citep{Gorman2019HS}.

The ExoMol database offers cross-sections at a resolving power of 1.5\,$\times 10^4$ and $k$-tables for the wavelength range of 0.3--50~$\mu\text{m}$, which can be directly used by \texttt{TauREx\,3} and many other retrieval models.
In addition, we also included collisional-induced absorption (CIA) and Rayleigh scattering. For CIA, we took opacity data for H$_2$--H$_2$ and H$_2$--He pairs from the HITRAN database \citep{Karman2019hitran}\footnote{https://hitran.org/}. The Rayleigh scattering estimates for H$_2$, He, H$_2$O, and CO$_2$ were taken from \citet{Cox2015allen}. 

To simulate the impact of clouds/hazes in the retrievals, we adopted the HPS model to provide a first-order representation of the upper cloud layers, which empirically have smaller optical densities than the deeper layers. We added the contribution of grey clouds to mimic the optically thick cloud layers, as caused by large cloud particle size and/or number density in more complex models. With the addition of grey clouds, the retrieval analysis can focus on exploring the contribution of the smaller particles. 
Refractive indices of different types of MgSiO$_3$ clouds and temperature conditions at which these can form are available in the literature \citep[e.g.,][]{Jager2003RIMgSiO3, Jaeger1994RIMgSiO3, Dorschner1995RIMgSiO3, Jaeger1998RIMgSiO3, Scott1996RIMgSiO3, Fabian2000RIMgSiO3, Zeidler2015Database}. For our analysis, we chose refractive indices of amorphous silicates from \citet{Jager2003RIMgSiO3} to estimate their scattering contributions. We chose not to include the absorption estimation for MgSiO$_3$ clouds due to uncertainties linked to the Lorentz-oscillator fit method for amorphous particles in the infrared, especially beyond 8~$\mu$m, as discussed in \citet{Jager2003RIMgSiO3}. Additionally, our decision was influenced by the fact that scattering contributions are more significant at wavelengths shorter than 8~$\mu$m (refer to the original data paper and our simulations in Section~\ref{subsec:yunma} for more details).
The refractive indices of crystalline SiO$_2$ are from \citet{Zeidler2013SiO2} collected by the HITRAN database. 
A summary of databases with these refractive indices is provided in \citet{Chubb2024Database_Whitepaper_ph}. 

In addition, we tested the impact of including CO among our priors. We also added the inactive gases N$_2$ and Li as proxies of undetectable species potentially contributing to the atmospheric mean molecular weights ($\mu$) and Rayleigh scattering slopes.

\begin{deluxetable*}{lllll}
\tablecaption{Parameter definitions and priors of spectral retrieval experiments in Step~2 \& 3 (if shown in posteriors). \label{tab:priors}}
% \tablewidth{0pt}
\tabletypesize{\scriptsize}
\tablehead{\colhead{Parameter} & \colhead{Definition} &  \colhead{Unit} & \colhead{Mode} & \colhead{Priors}
}
% \decimalcolnumbers
\startdata
$R_{\mathrm{p}}$ & Distance from planet center to atmosphere base & \nomJovianEqRadius & factor & 0.80 -- 1.80
\\ 
$T$ & Atmospheric isothermal temperature (Step~2) & K & linear & 500 -- 2000
\\
$T_\mathrm{surf}$ & Atmospheric temperature at 10~bar (Step~3) & K & linear & 500 -- 3000
\\
$T_\mathrm{top}$ & Atmospheric temperature at $10^{-8}$~bar (Step~3) & K & linear & 500 -- 3000
\\
$T_\mathrm{1}$ & Atmospheric temperature at $10^{-2}$~bar (Step~3) & K & linear & 500 -- 3000
\\
$T_\mathrm{2}$ & Atmospheric temperature at $10^{-4}$~bar (Step~3) & K & linear & 500 -- 3000
\\
$T_\mathrm{3}$ & Atmospheric temperature at $10^{-6}$~bar (Step~3) & K & linear & 500 -- 3000
\\
\hline
$X_{\mathrm{H_2O}}$ & VMR of H$_2$O  & \nodata &  log & 10$^{-12}$ -- 10$^{-1}$
\\
$X_{\mathrm{CO_2}}$ & VMR of CO$_2$ & \nodata &  log & 10$^{-12}$ -- 10$^{-1}$
\\
$X_{\mathrm{SiO}}$ & VMR of SiO & \nodata & log & 10$^{-12}$ -- 10$^{-1}$
\\
$X_{\mathrm{Na}}$ & VMR of Na & \nodata & log & 10$^{-12}$ -- 10$^{-1}$
\\
$X_{\mathrm{K}}$ & VMR of K & \nodata &  log & 10$^{-12}$ -- 10$^{-1}$
\\
$X_{\mathrm{CO}}$* & VMR of CO & \nodata &  log & 10$^{-12}$ -- 10$^{-1}$
\\
$X_{\mathrm{N_2}}$* & VMR of N$_2$, representing heavy-inactive gases & \nodata & log & 10$^{-12}$ -- 10$^{-1}$
\\
$X_{\mathrm{Li}}$* & VMR of Li, representing light-inactive gases & \nodata & log & 10$^{-12}$ -- 10$^{-1}$
\\
\hline
$r_\mathrm{HPS}$ & HPS cloud particle size & m & log & 10$^{-7}$ -- 10$^{-4}$
\\
$N_\mathrm{HPS}$ & HPS cloud particle number density & m$^{-3}$ & log & 10$^{0}$ -- 10$^{8}$
\\
$p_\mathrm{HPS,deck}$ & HPS cloud deck pressure & Pa & log & 10$^{-1}$ -- 10$^{4}$
\\
$p_\mathrm{HPS,base}$ & HPS cloud base pressure & Pa & log & 10$^{-1}$ -- 10$^{4}$
\\
$p_\mathrm{grey}$ & Grey cloud deck pressure & Pa & log & 10$^{-1}$ -- 10$^{4}$
\enddata
\tablecomments{VMR represents volume mixing ratio. HPS represents homogeneous particle-size clouds. *Retrieved parameters in non-reference model experiments.}
\end{deluxetable*}

\subsubsection{Optimisation of the atmospheric thermal profile using retrieval techniques (Step~3)}

\begin{figure}[ht!]
\plotone{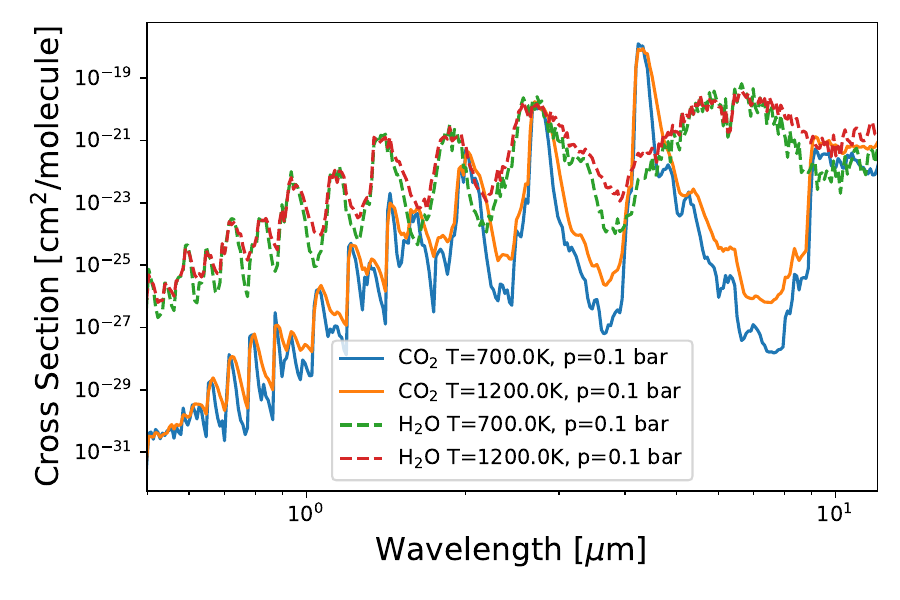}
\caption{Cross sections of H$_2$O \citep{Polyansky2018pokazatel} and CO$_2$ \citep{Yurchenko2020CO2} at two different temperatures at 0.1\,bar collected by the ExoMol database.
\label{fig:lineshape}}
\end{figure}

From the posteriors obtained in Step~2, we learned that H$_2$O and CO$_2$ are key species absorbing in the atmosphere of WASP-39\,b. These molecules have cross sections that are very sensitive to the temperature, as showcased in Fig.~\ref{fig:lineshape}, so we can use them to constrain the atmospheric $T$--$p$ profile.

\subsubsection{Optimisation of retrieved chemical abundances including clouds (Step~4)}

In Step~4, we ran the full equilibrium chemistry model using the $T$–$p$ profiles obtained in Step~3 as input. Condensation processes and charges were included in the simulations.
Approximately 600 chemical species were estimated by \texttt{GGChem} to be present in the atmosphere. However, only a small fraction contributes significantly to the transit spectrum. Others contribute to the atmospheric properties, such as the mean molecular weight. Through retrieval simulations, we converged to the set of molecular abundances and $T$–$p$ profiles that better fit the observations and compared them to the predictions of \texttt{GGChem} to verify their consistency and estimate the relative abundances of the muted species a posteriori. This last step is important to constrain the metallicity and elemental ratios of the atmosphere. 

Given the time gap between the MIRI observations and those obtained with NIRSpec, NIRCam, and NIRISS, we have also conducted a separate cloud particle size distribution optimisation that considers only the MIRI observations alongside the homogeneous analysis across the entire spectrum available. 
For the potential contamination longwards 10\,$\mu$m mentioned in Section~\ref{sec:intro}, we have prioritised the observations shortwards 10~$\mu$m in model optimisation.
Our iterative approach, moving from equilibrium chemistry modelling to retrievals and back to equilibrium chemistry modelling, aims to constrain molecular species -- in gaseous form or condensed -- with prominent spectral features, and then use their molecular abundances as baseline assumptions for deeper analysis with more complex atmospheric models. 

%%%%%%%%%%%%%%%
\subsection{Tests on cloud particles' properties and composition}
\label{tests}

Concerning the composition of silicate clouds, we have run equilibrium chemistry (\texttt{GGChem}) and retrieval simulations to assess whether MgSiO$_3$ (enstatite),  SiO$_2$ or Mg$_2$SiO$_4$ (forsterite) are expected to be the dominant species in the atmosphere of WASP-39\,b.

We also conducted sensitivity tests about our assumptions of cloud particles' size distributions and scattering/absorption properties in our simulations. 
More specifically, we tested the impact of 
different fractions of absorption/scattering contributions. Scattering and absorption estimates were taken from \citet{Jager2003RIMgSiO3}, assuming MgSiO$_3$ spherical amorphous particles.

In addition, we performed simulations using particle size distributions at each pressure layer adapted from  \citet{Deirmendjian1964distribution}:
\begin{equation}
f(n) = \begin{cases}(\frac{r_c}{r_\mathrm{fit}})^a e^{-b\frac{r_c}{r_\mathrm{fit}}} &\mbox{if } 0< r_\mathrm{fit} < 2.7  \\
(\frac{r_c - r_t}{r_0})^a e^{-b\frac{r_c - r_t}{r_0}} & \mbox{if }  r_\mathrm{fit} \geq 2.7
\label{eq:distribution}
\end{cases}
\end{equation}
where $r_\mathrm{fit}$ is the best-fit single particle size. We assumed $b=6$ when $0 < r_\mathrm{fit} < 2.7$ and $b=1.5, r_0=1, r_t = r_\mathrm{fit} \sigma - r_0$, when $r_\mathrm{fit} \geq 2.7$, where $\sigma=0.7$.
We compared the results of these simulations with the assumption of homogeneous particle sizes. 

Finally, to test the robustness of our retrieval simulations, we adopted the A–M cloud-microphysics approach available in \texttt{YunMa}. We adopted the free convective assumptions \citep{gierasch1985energy} to estimate the eddy diffusion from the mixing length, which is derived from the scaled scale height as described in the A–M approach. The saturation vapour pressure (SVP) is taken from \citet{Visscher2010SVPMgSiO3}.

%%%%%%%%%%%%%%%

\section{Results} \label{sec:result}
%__________ Results _____________
\subsection{Selection of chemical species for free retrievals} \label{sec:mol_select}
%% Selection result
The selection process described in Section~\ref{sec:method_chem_select} (Steps 1 and 2) identifies species that are expected to be abundant according to models assuming chemical equilibrium but also to have large cross sections in the wavelength range probed by JWST, making them the most likely contributors to the transit spectra of WASP-39\,b. The active gaseous species playing a major role in our retrieval experiments are: H$_2$O, CO$_2$, SiO, Na and K. A more comprehensive selection of active chemical species is reported in Appendix~\ref{appendix:venn}.
We also report there how molecular abundances change with metallicity and temperature at the relevant pressures probed in the exoplanet atmosphere.

%% SiS
Some gaseous species were excluded from the selection of priors during preliminary retrieval tests because their spectral features are completely masked by other species within a reasonable abundance range, despite being highlighted in the initial selections and showing features similar to those in observations. For example, SiS is predicted to be abundant in the atmosphere of WASP-39\,b under chemical equilibrium conditions. However, its spectral features are less prominent than those of other selected species. Only if SiS's abundance exceeded those of H$_2$O and CO$_2$ by an order of magnitude, an unlikely scenario, would its features become discernible in the transit depth rather than being obscured by other molecular signatures. However, in such a case, SiS would significantly increase the mean molecular weight, reducing the atmospheric scale height and compressing the overall molecular features in a way that is not compatible with the observations.

\subsection{Free retrieval results} \label{sec:optimized_retrieval_results}
\begin{figure*}[ht!]
\centering
\includegraphics[width=1\textwidth]{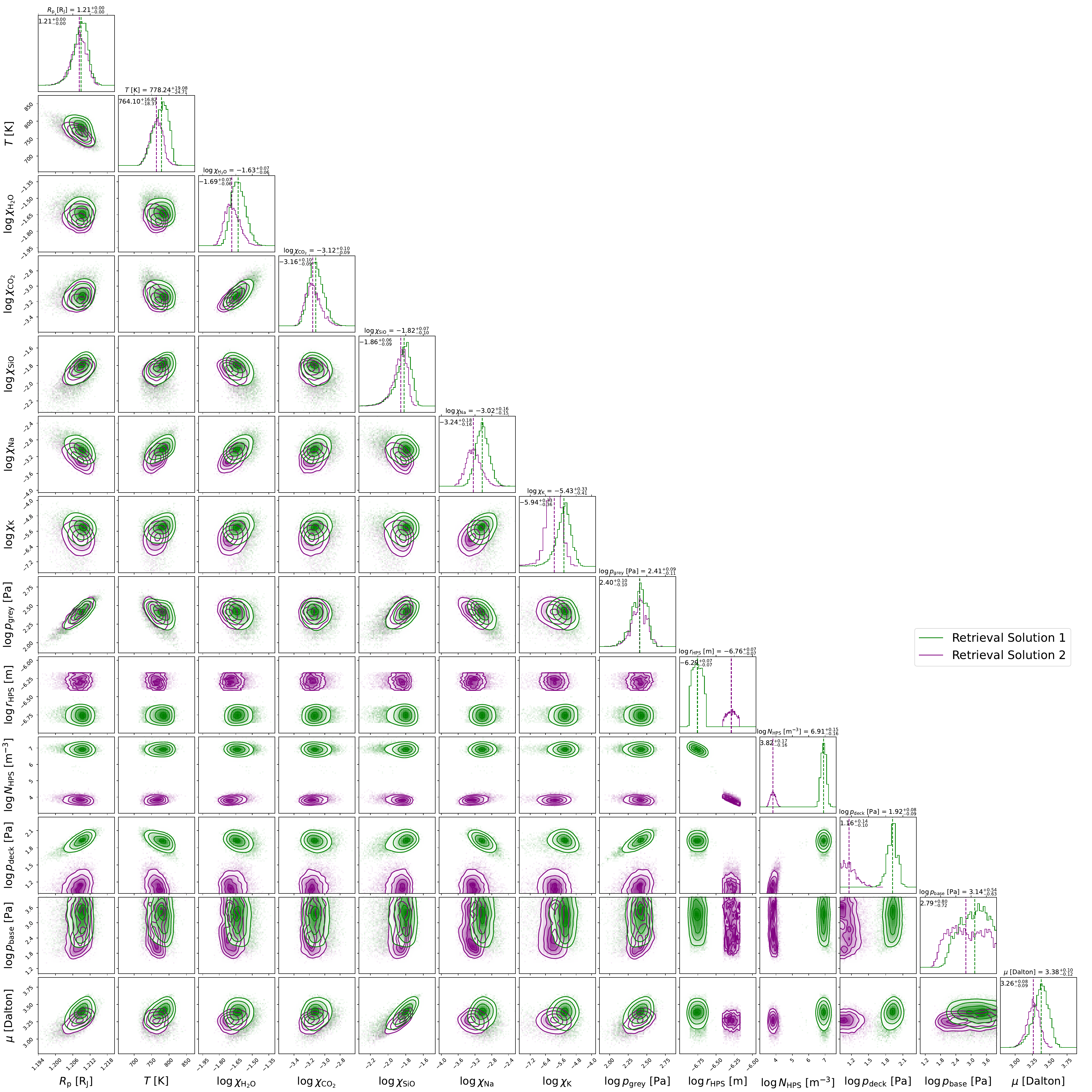}
\caption{Posterior distributions obtained with free retrievals of NIRISS, NIRSpec and NIRCam spectra (Step~2). We find two islands of solutions (green and purple) originating from the degeneracy of HPS clouds.
The numbers above each block represent the median value of the posteriors. The results suggest a significant amount of H$_2$O, SiO and CO$_2$ in the observable atmosphere. We plot the highest weighted $2\times10^4$ traces for visual ease. 
\label{fig:posteriors}}
\end{figure*}

In Steps 2 and 3, we performed free retrievals to constrain the gas abundances and thermal profiles. Fig.~\ref{fig:posteriors} shows the posterior distributions obtained in Step~2, where $T$--$p$ profiles are kept constant.
Here, two islands of solutions (Solution~1 and 2) are found, stemming from the degeneracy of the MgSiO$_3$ HPS clouds. Both retrieved solutions fit the observed spectrum with no statistical preference, as suggested by their Bayesian evidence. Despite the apparent degeneracy, the gas-phase species' abundances in the two cloud solutions are consistent. While the cloud particle radius in Solution~2 is larger than in Solution~1, the particle number density is three orders of magnitude lower. As a result, the opacity density of the HPS cloud in Solution~2 is lower, but it is still detectable in transit due to its high altitude. We show the spectral fit and optical depth details of the optimised retrieval in Appendix~\ref{appendix:MAP}. Test cases where CO or muted species (N$_2$ or Li) are displayed in Appendix~\ref{appendix:mmw}.

\begin{figure*}[ht!]
\includegraphics[width=1.\textwidth]{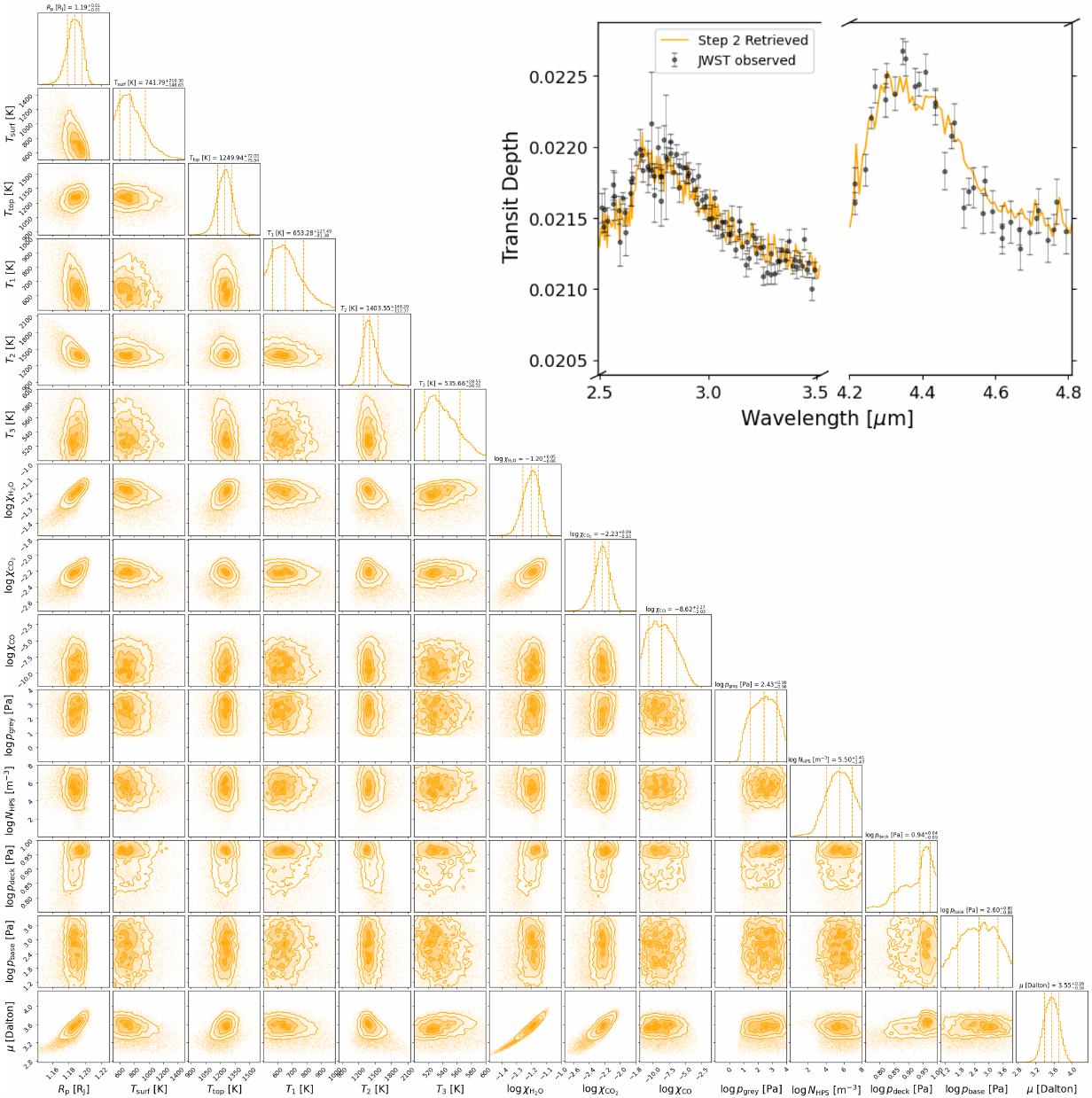}
\caption{Posteriors, the data points selected for retrieving the $T$--$p$ profile and the spectrum fitting in Step~3.
\label{fig:posteriors_Tp}}
\end{figure*}

\begin{figure}[ht!]
\plotone{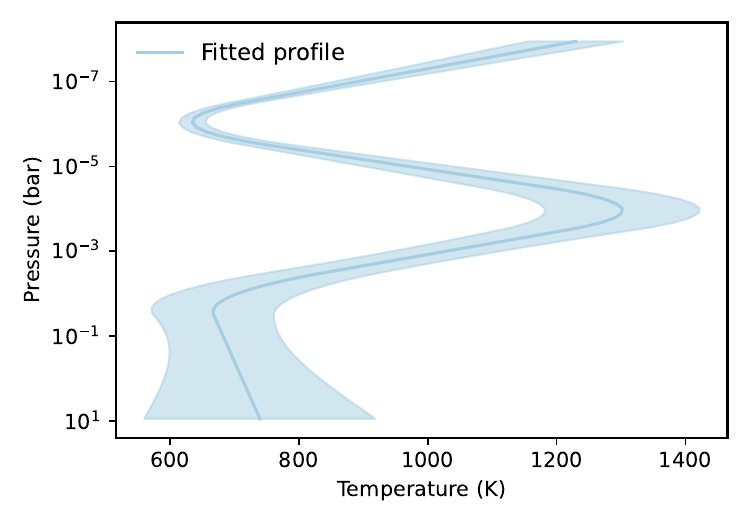}
\plotone{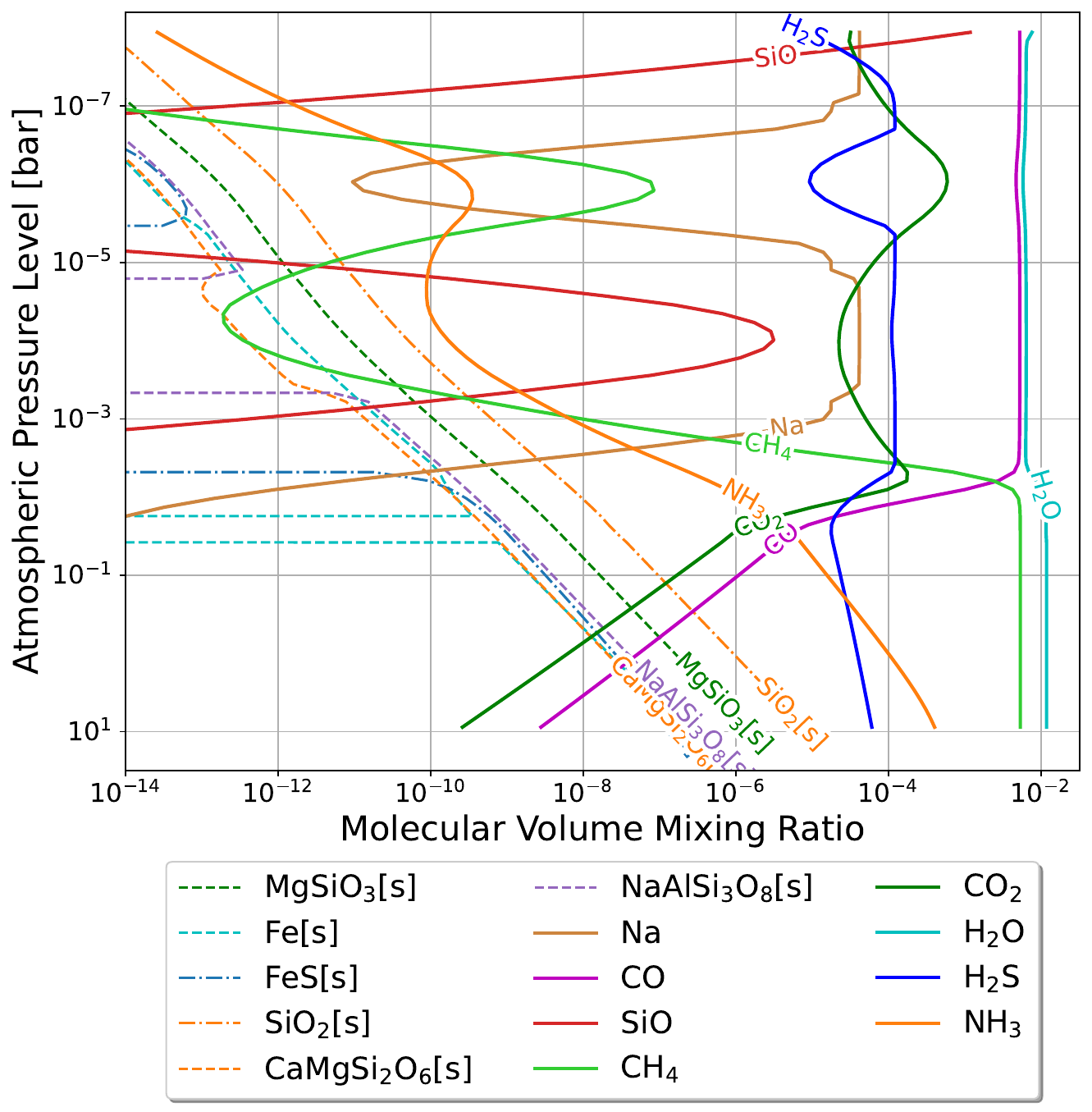}
\caption{Retrieved $T$--$p$ profile (Opt. $T$--$p$) and an example of the most abundant active gas (max. VMR $>10^{-5}$) and condensate species (max. VMR $>10^{-7}$) obtained with GGChem simulations using the median retrieved $T$--$p$ values as input. 
\label{fig:retrieved_tp}}
\end{figure}

From the free retrievals in Step~2, we determined that the 2.5--3.5 and 4.2--4.8~$\mu$m spectral windows—where H$_2$O and CO$_2$ exhibit strong absorption—are the most informative for further constraining the $T$--$p$ profile, because these are the dominant species shaping the overall spectrum and their spectral lines are highly temperature-sensitive.
We show in Fig.~\ref{fig:posteriors_Tp} the detailed retrieval results of this step and Fig.~\ref{fig:retrieved_tp} the optimal $T$--$p$ profile as retrieved in Step~3 and an example of \texttt{GGChem} simulations with condensates using the optimal $T$--$p$ profile as input.

\subsection{Optimal solutions to Webb data} \label{sec:convergence}
\begin{figure}[ht!]
\plotone{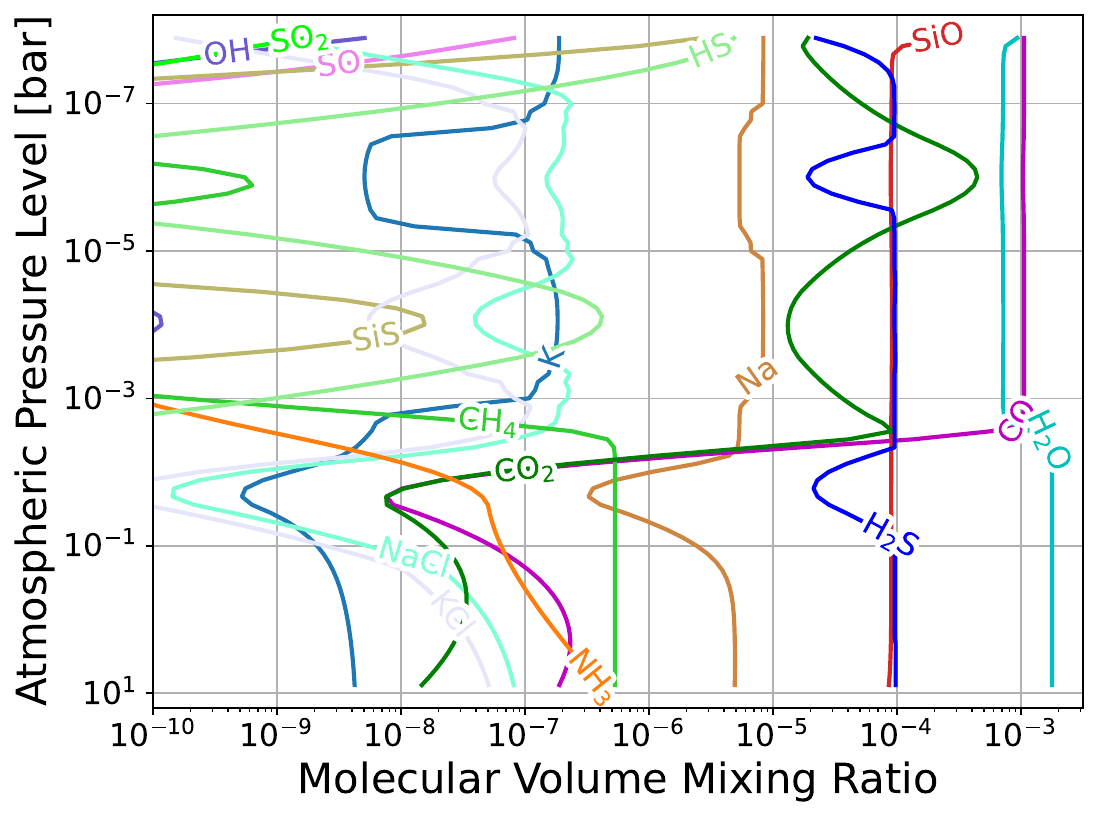}
\caption{Optimised VMRs of the active gases as a function of pressure: these were obtained by fine-tuning the results obtained with GGChem to better reproduce JWST's data. Only the most abundant gases are plotted here for visual ease.
\label{fig:vmr_opt}}
\end{figure}

Fig.~\ref{fig:vmr_opt} shows the abundances of the gaseous species as a function of pressure for the optimal solution: these were obtained by fine-tuning the results obtained with GGChem to better reproduce Webb's data. Only the VMRs of the most abundant active gases are shown in Fig.~\ref{fig:vmr_opt}. 
The optimal solution in the gas phase corresponds to the following elemental ratios: 
2.27\,$\times$\,O/H, 1.39\,$\times$\,C/H, 1.67\,$\times$\,Si/H, 3.35\,$\times$\,S/H, 2.06\,$\times$\,Na/H, 0.55\,$\times$\,K/H with respect to the solar abundance ratios from \citet{Lodders2021solar_photo_abund}.

In addition to the gaseous species, we have found that HPS clouds of multiple particle sizes are needed for an optimal fit. 
In our simulations, we have included three particle sizes for SiO$_2$, i.e. 0.01, 0.4 and 0.8\,$\mu$m, and one for MgSiO$_3$ clouds, i.e. 10\,$\mu$m. The fit to the MIRI data can be improved if we use 1\,$\mu$m particle-size for SiO$_2$ while keeping the MgSiO$_3$ particle size.

\begin{figure*}[ht!]
\includegraphics[width=1.\textwidth]{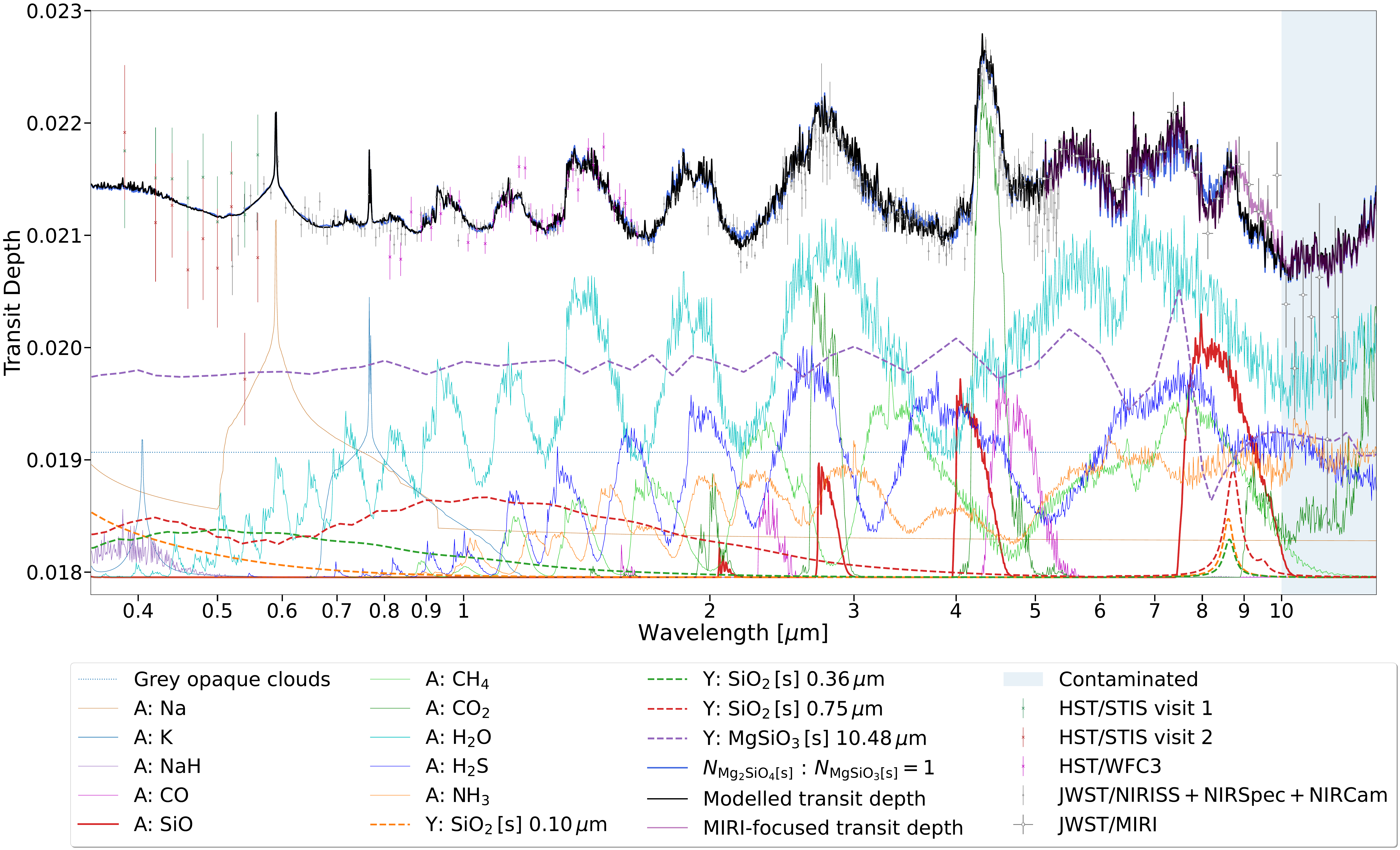}
\caption{Best fit to the JWST spectral data (black and purple lines at the top). $A$: individual gas-phase absorption, $Y$: cloud scattering and absorption (if taken), assuming specific cloud particle radii in the simulations. We only plotted the most significant opacity contributors for visual ease, where opacities of low-abundance species, including SO$_2$ are included in the calculation but are not shown in the plot. Rayleigh scattering and collision-induced absorption (CIA) are not shown for the same reason. 
NIRISS, NIRSpec and NIRCam spectral data are taken from \citet{Carter2024JWSTWASP39b}, MIRI data from \citet{Powell2024WASP39bMIRI} and
Hubble STIS and WFC3 data from \citet{Saba2024STIS}.
Details of the data used and offsets adopted here are described in Section~\ref{sec:specoffset}. 
Beyond 10\,$\mu$m, the MIRI spectrum is potentially contaminated according to \citet{Flagg2024WASP39b}. 
\label{fig:eq32}}
\end{figure*}

\begin{figure*}[ht!]
\plotone{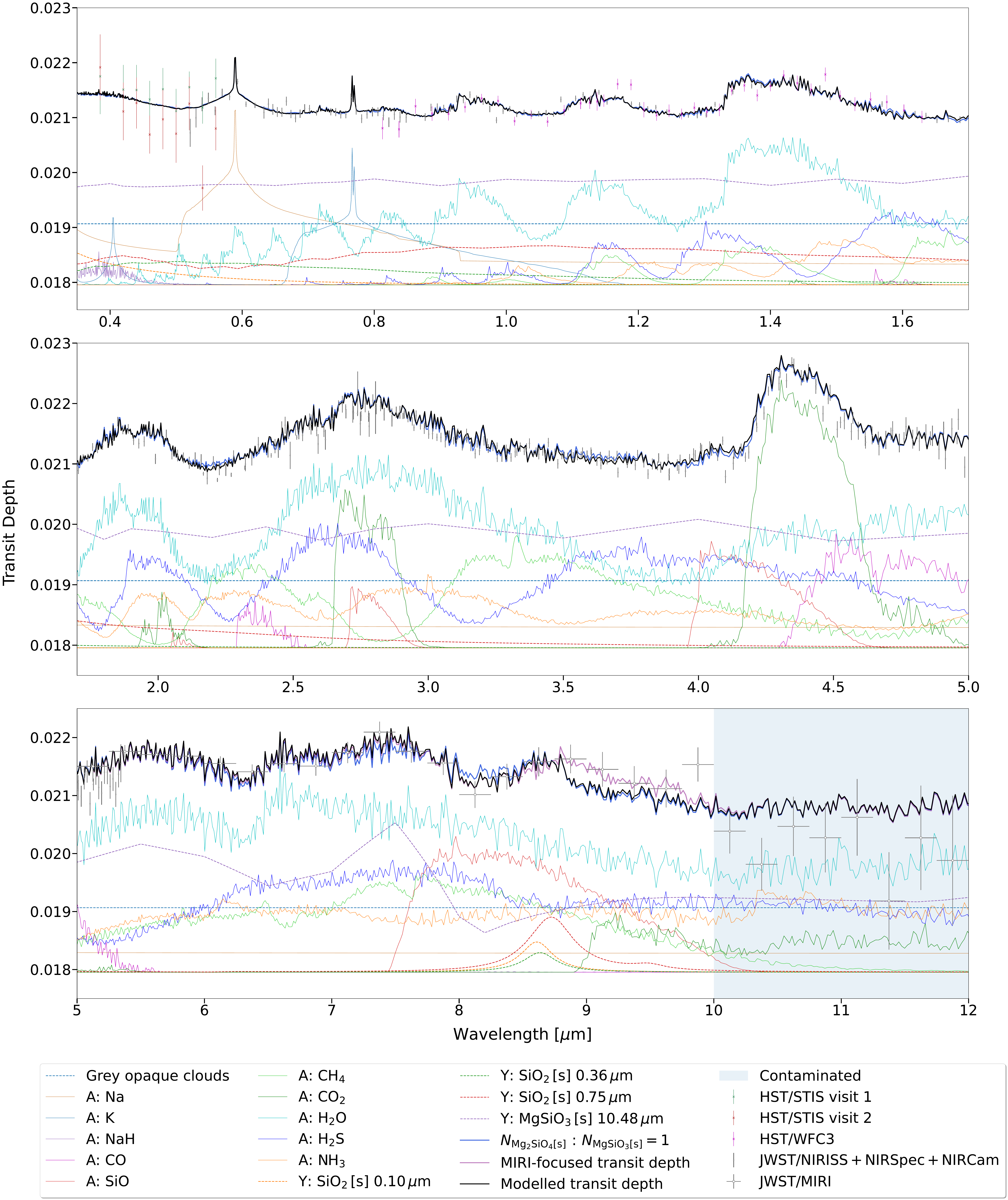}
\caption{Zoom in of Fig.~\ref{fig:eq32} in different spectral windows. Here, the x-axis of each subplot is displayed on a linear scale.\label{fig:eq32_zoomin}}
\end{figure*}

\begin{figure*}[ht!]
\plotone{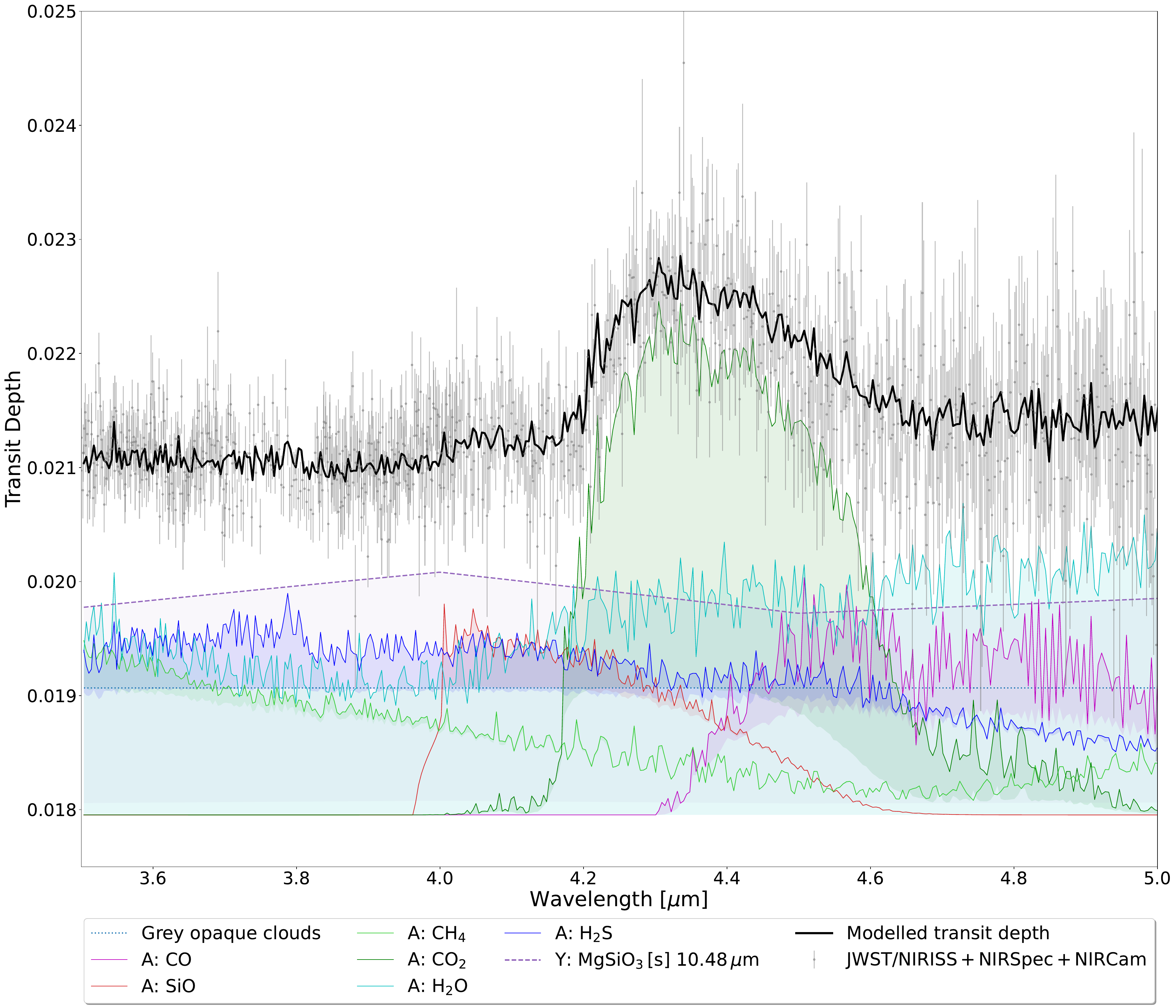}
\caption{Zoom in around 4.1~$\mu$m of Fig.~\ref{fig:eq32}. The model is compared with the native-resolution observations (grey data points with error bars). We show the individual molecular contributions (coloured) of the optimised case (black). To reveal their altitude dependence, we plot as shaded areas the gas-phase molecular absorptions integrated from 10$^{-1}$\,bar to the top of the atmosphere. \label{fig:41siofit}}
\end{figure*}

\begin{figure*}[ht!]
\plotone{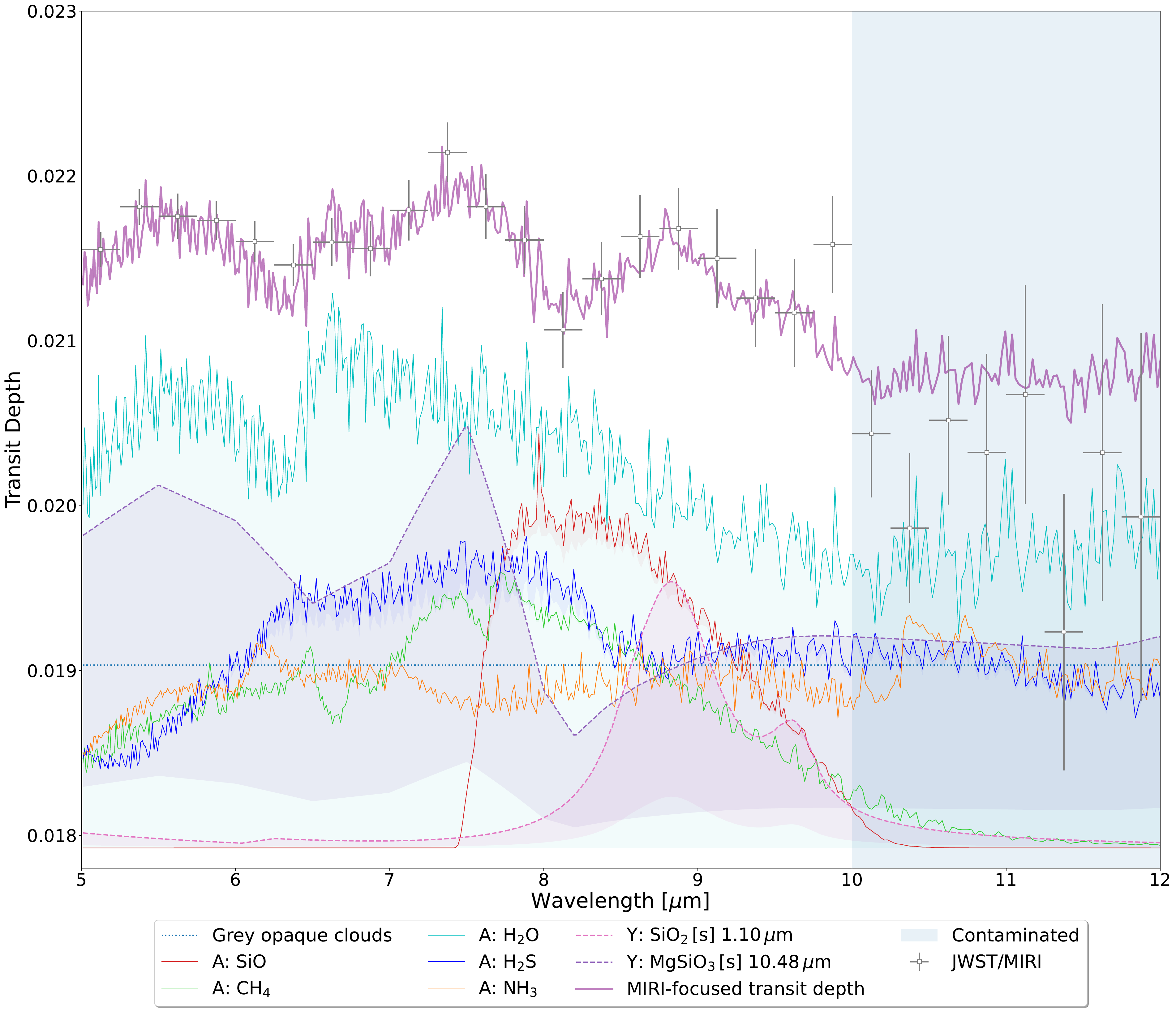}
\caption{Zoom in of the MIRI window of the MIRI-focused fit in Fig.~\ref{fig:eq32}. We plot as shaded areas the gas-phase molecular absorptions integrated from 10$^{-3}$\,bar to the top of the atmosphere.  \label{fig:mirifit}}
\end{figure*}

\begin{figure*}[ht!]
\plotone{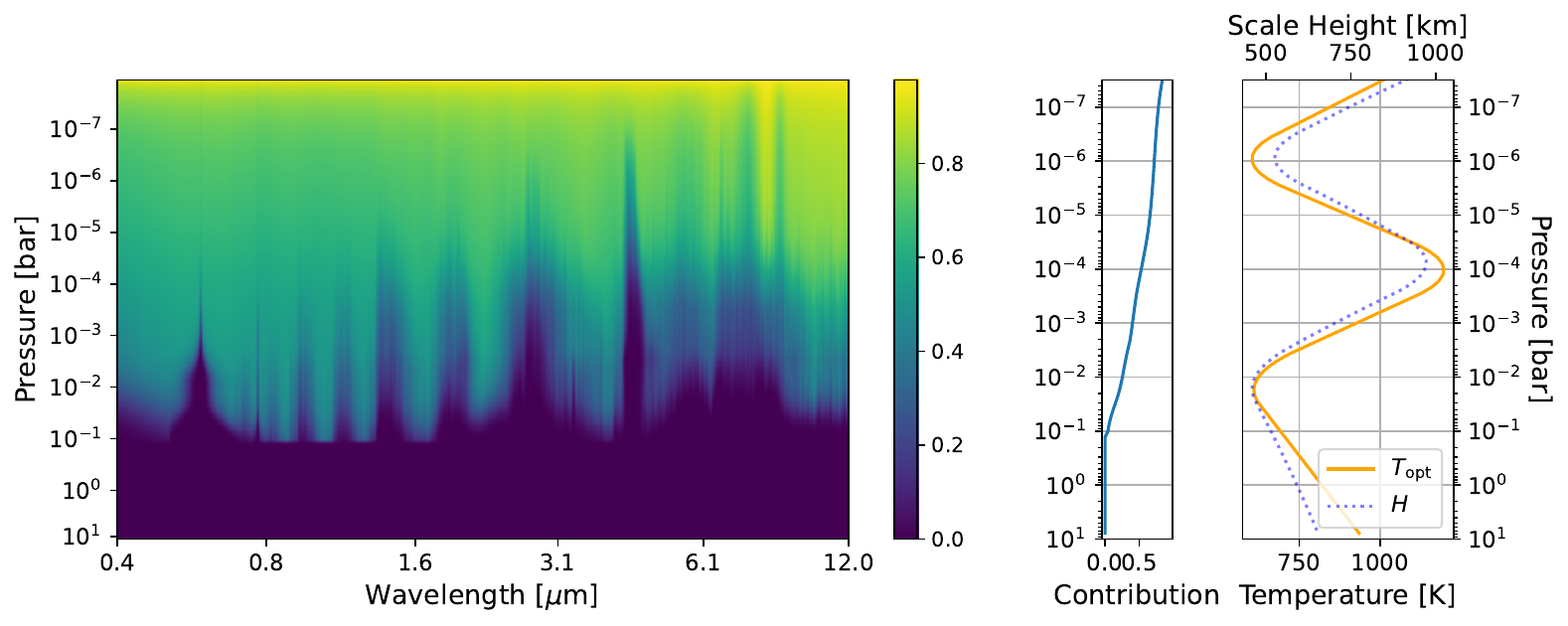}
\caption{Estimated optical depths through optical paths centred at each pressure level (left panel), wavelength-normalised opacity contributions (central panel), adopted temperature profile and estimated scale heights (right panel). 
\label{fig:tau_opt}}
\end{figure*}

In Figs.~\ref{fig:eq32},~\ref{fig:eq32_zoomin},~\ref{fig:41siofit} \&~\ref{fig:mirifit} the final simulated spectra are plotted against the data. Fig.~\ref{fig:tau_opt} presents the opacities along the optical paths, together with the atmospheric properties of each pressure layer in the observed atmosphere.

\subsection{Cloud particles' properties and composition: consistency with laboratory measurements and microphysics models}
\label{subsec:yunma}
GGChem chemical equilibrium simulations suggest a correlation between the Si:O ratio and the relative abundances of enstatite versus forsterite. 
Sub-solar Si/O prefers the formation of enstatite over forsterite. In addition, enstatite provides a better fit to the observations, as shown in Fig.~\ref{fig:eq32} \&~\ref{fig:eq32_zoomin}, in agreement with Si/O $\sim0.7$ as suggested in this work.

\begin{figure*}[ht]
\includegraphics[width=1.\textwidth]{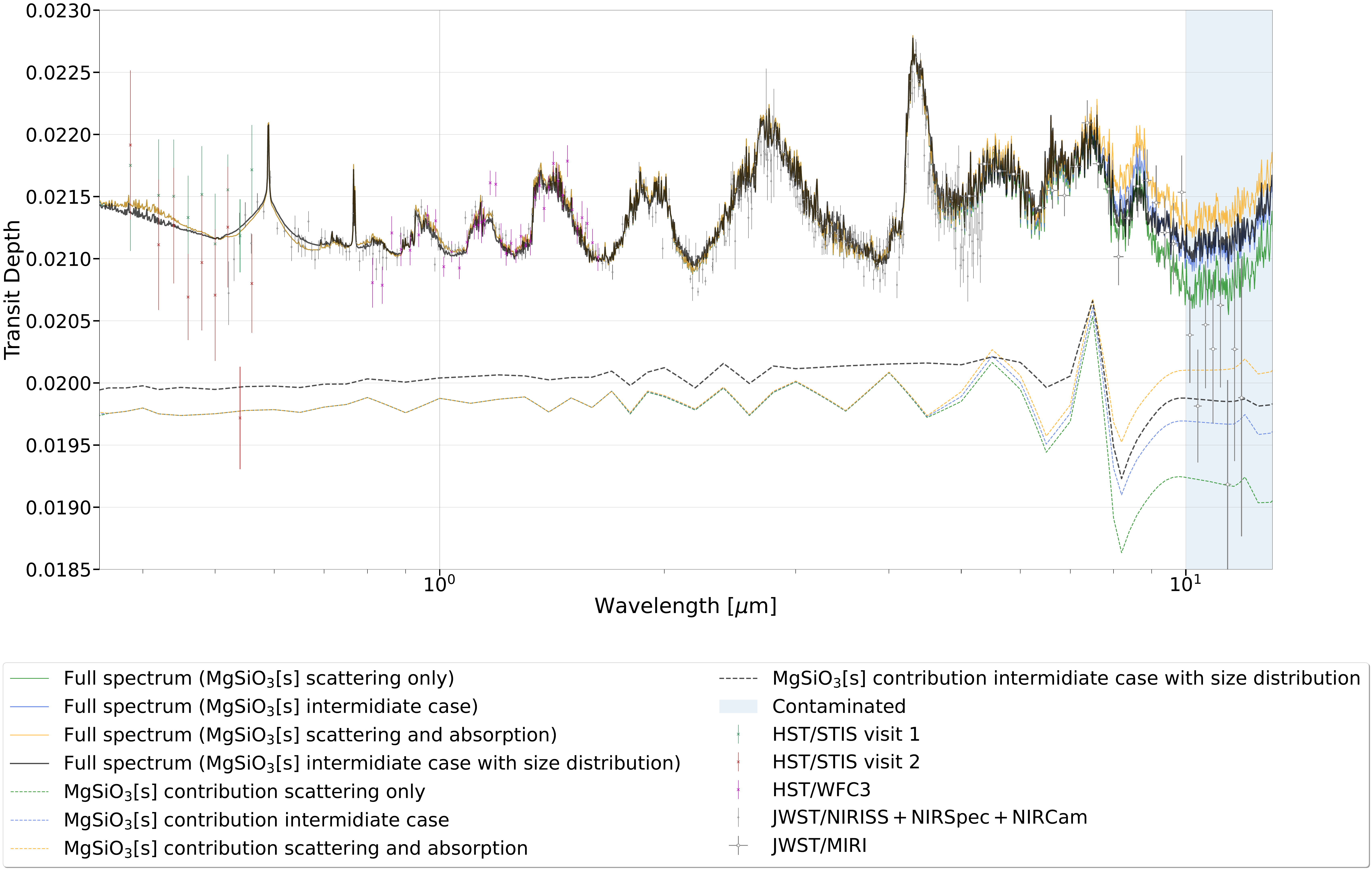}
\caption{Spectra estimated with different cloud absorption/scattering contribution and particle size distribution. Green: reference model as shown in Fig.~\ref{fig:eq32} with HPS cloud model, and only the scattering contribution is included.
Yellow: HPS clouds with scattering and absorption equally represented. 
Light blue: intermediate case between the yellow and green cases.
Black: similar to the light blue model but with particle size distribution following Equation~(\ref{eq:distribution}) adapted from \citet{Deirmendjian1964distribution} at each pressure layer. 
Scattering and absorption estimates were taken from \citet{Jager2003RIMgSiO3}.
}
\label{fig:absorption}
\end{figure*}

Furthermore, we tested the impact of different fractions of absorption/scattering contributions for enstatite clouds, as described in Section~\ref{tests}, and show the results in Fig.~\ref{fig:absorption} (green, yellow and light-blue spectra). In addition, we compared simulations using the particle size distributions described in Section~\ref{tests} with those assuming homogeneous particle sizes. The results are reported in Fig.~\ref{fig:absorption} (black and light blue spectra). These results show good agreement with the results using HPS clouds shown in Fig.~\ref{fig:vmr_opt} and~\ref{fig:eq32}.

% \section{\textbf{Cloud particles' retrieved properties and consistency with laboratory measurements and microphysics models}} \label{appendix:yunma}

\begin{figure}[ht]
\centering
\includegraphics[width=.48\textwidth]{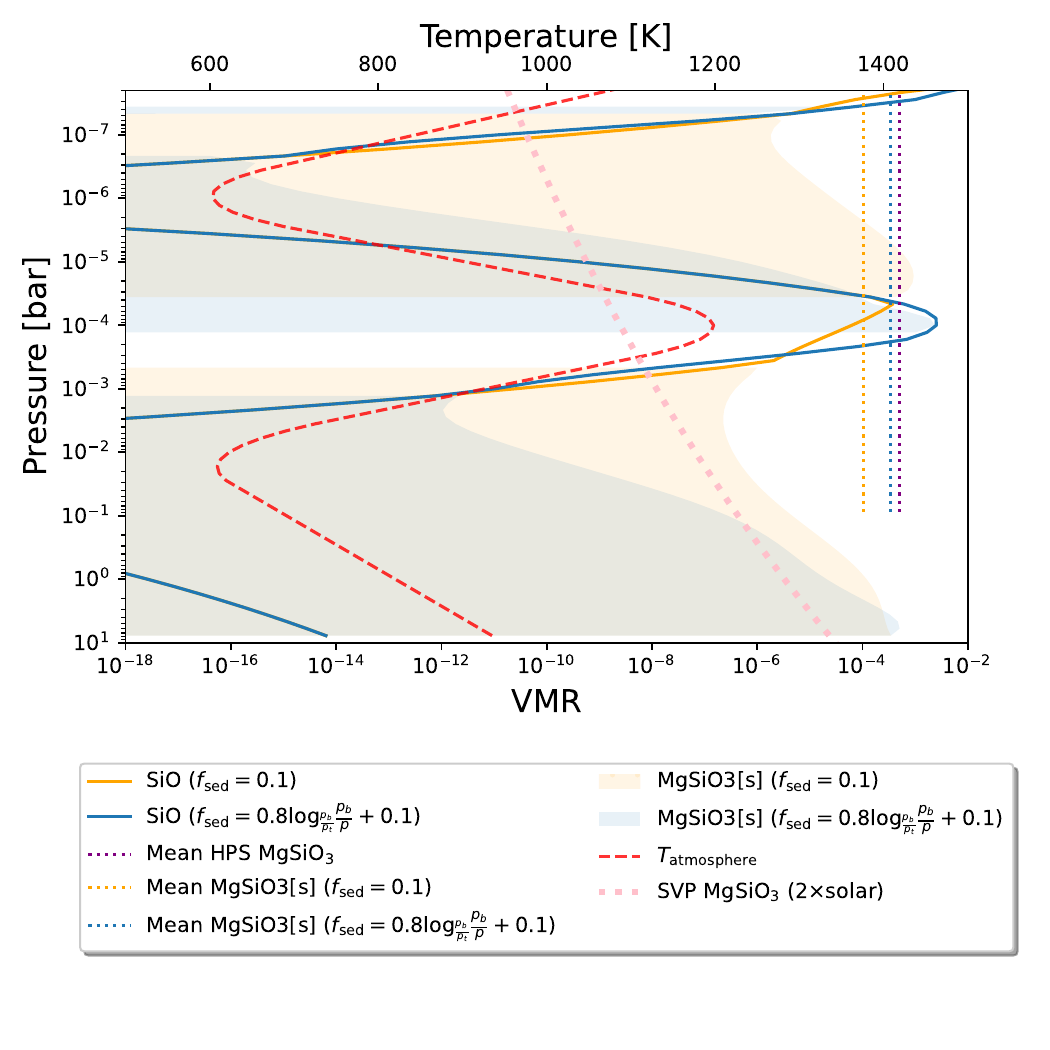}
\caption{SiO and cloud mixing ratios derived from the cloud microphysics approach as described in \citet{ackerman2001precipitating} and available in \texttt{YunMa}. We also show the mean VMR (dotted lines) of the MgSiO$_3$ clouds constrained by observations with the HPS model and by the molecular draft balance, with two assumptions of the sedimentation efficiency ($f_\mathrm{sed}$). The mean VMR are estimated above the grey cloud deck (refer to the saturated layer across all the wavelengths in Fig.~\ref{fig:tau_opt} left panel).}
\label{fig:amcloud}
\end{figure}

\begin{figure}[ht]
\plotone{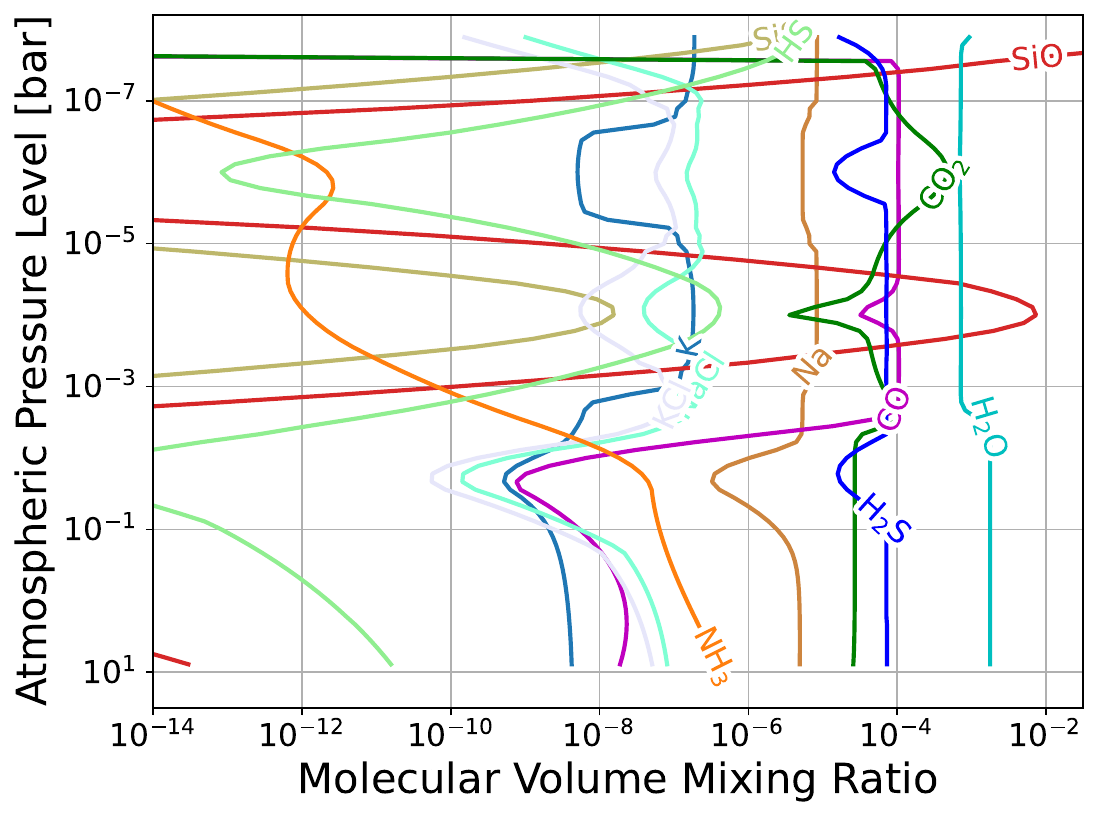}
\caption{Similar to Fig.~\ref{fig:vmr_opt} but optimised for the SiO profile shown in Fig.~\ref{fig:amcloud}.
\label{fig:am_VMR}}
\end{figure}

\begin{figure*}[ht]
\plotone{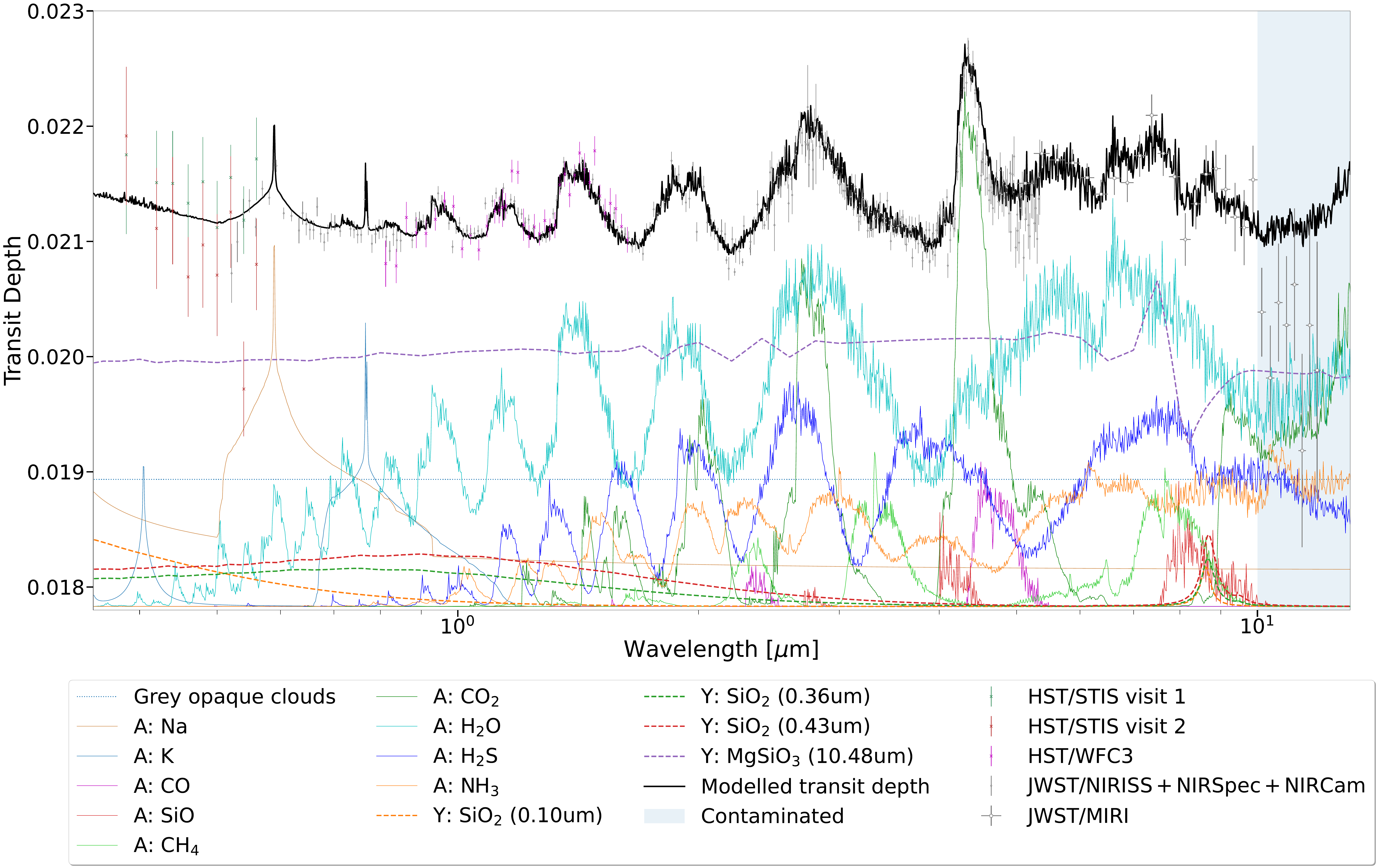}
\caption{Similar to Fig.~\ref{fig:eq32} but assuming the chemical profiles shown in Fig.~\ref{fig:am_VMR} and cloud particle size distribution same as the black spectrum in Fig.~\ref{fig:absorption}.
\label{fig:am_spec}}
\end{figure*}

To test the robustness of our retrieval simulations with respect to atmospheric dynamics, we adopted the cloud-microphysics approach described in Section~\ref{tests}. SiO and cloud mixing ratios derived from this approach are shown in Fig.~\ref{fig:amcloud}, and the other molecular mixing ratios are shown in Fig.~\ref{fig:am_VMR}. The simulations, with assumed values of the sedimentation coefficient $f_\mathrm{sed}$, align with the former observation-driven results. We compared the mean molecular mixing ratios of the clouds and found consistency between the HPS model constrained by observations and those simulated theoretically using the A–M approach, as shown in Fig.~\ref{fig:amcloud}.
We used the chemical mixing profile with the SiO volume mixing ratio from the A–M approach and plotted the simulated spectrum against the observations in Fig.~\ref{fig:am_spec}.

% \textbf{While complex cloud microphysics models provide a self-consistent way to simulate cloud properties, several challenges remain. In particular, the gas-phase chemistry—which provides the baseline for interpreting cloud formation from a chemical standpoint—is not constrained to a level that can be used to robustly infer cloud mixing ratios. Moreover, many of the fundamental assumptions regarding cloud formation remain difficult to validate in the context of exoplanet atmospheres; the complexity of atmospheric dynamics and cloud formation processes is not yet fully captured. Improved observations covering a broader range of orbital phases can help calibrate these models for exoplanetary conditions and to allow at least some of the microphysical parameters to be constrained by observations in the future. In the current situation, depending exclusively on such models makes it challenging to achieve good statistical consistency between observations and predictions from theoretical chemical and aerosol models.}

% %% This command is needed to show the entire author+affiliation list when
% %% the collaboration and author truncation commands are used. It has to
% %% go at the end of the manuscript.
% %\allauthors

% %% Include this line if you are using the \added, \replaced, \deleted
% %% commands to see a summary list of all changes at the end of the article.
% %\listofchanges

\section{Discussion} \label{sec:discussion}

\subsection{Silicon versus sulphur chemistry to explain JWST spectra of WASP-39\,b}
\begin{figure}[ht!]
\centering
\includegraphics[width=.48\textwidth]{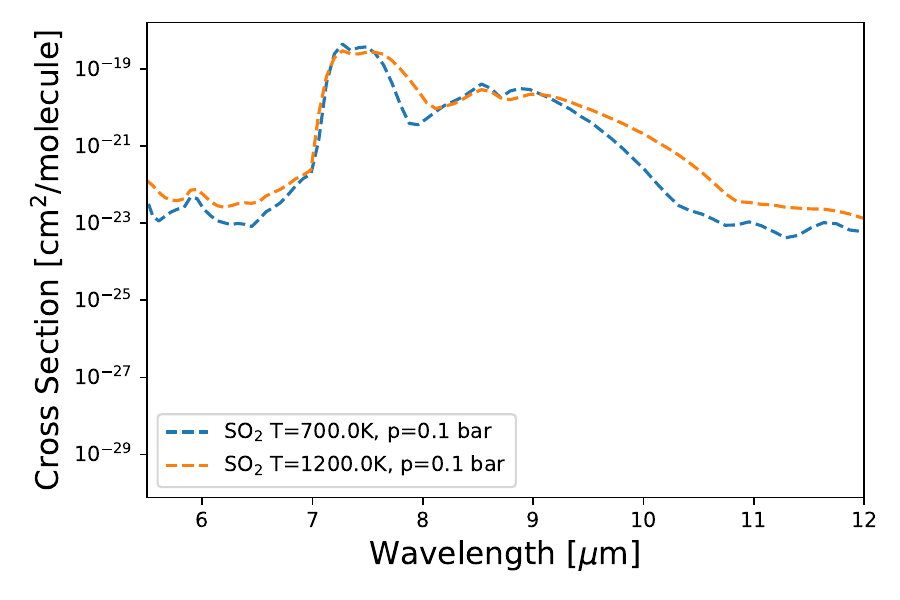}
\caption{Cross sections of SO$_2$ from \citet{Underwood2016SO2} collected by the ExoMol database.}
\label{fig:lineshape_SO2}
\end{figure}

\begin{figure}[ht!]
\centering
\includegraphics[width=.48\textwidth]{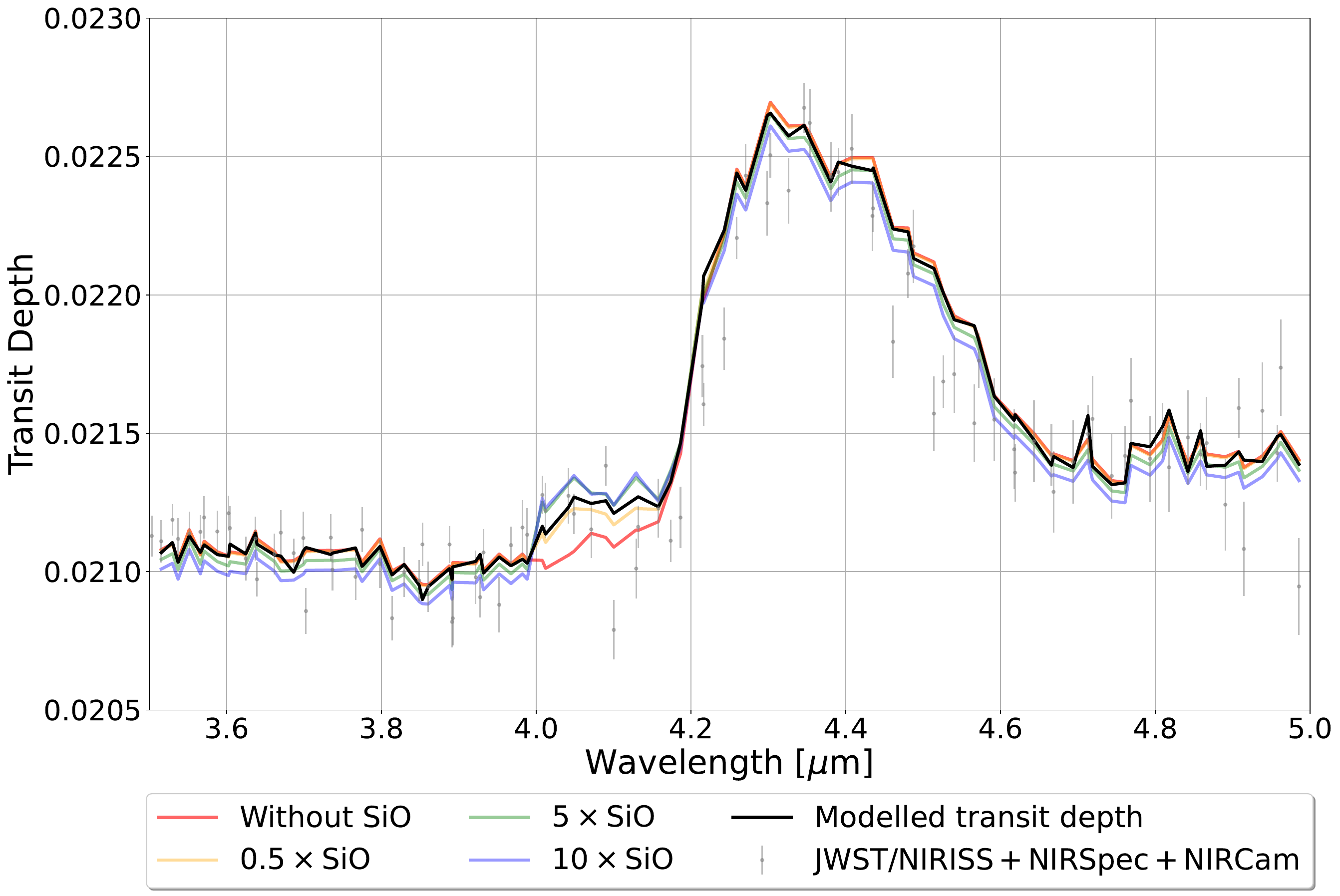}
\caption{Transit depth simulations with varying amounts of SiO indicating the importance of an optimised VMR of SiO for a good fit with the observed spectrum.}
\label{fig:noSiO}
\end{figure}

%% why silicate
The previous literature suggested SO$_2$ as an explanation of the features at 4.1, 7.7 and 8.5~$\mu$m (see Section~\ref{sec:intro_so2} and Fig.~\ref{fig:lineshape_SO2}). 
% Strong photochemical processes in the upper atmosphere are needed to explain the abundance of SO$_2$. 
While we are not excluding the presence of SO$_2$, we propose here that the spectral feature observed at $4.1$~$\mu$m is mainly caused by SiO (Fig.~\ref{fig:41siofit}) and the features at 7.7 and 8.5~$\mu$m can be explained by silicate clouds (Fig.~\ref{fig:mirifit}). To be clearer, we compared the transit depths simulated with different amounts of SiO in Fig.~\ref{fig:noSiO}.
Results from our retrieval analyses suggest H$_2$O, CO$_2$, SiO, silicate clouds, Na, and K can fit well NIRISS, NIRSpec, NIRCam and MIRI observations. Further optimisation with both shorter wavelengths and MIRI observations helped to constrain the particle sizes, composition and location of silicate clouds. The presence and abundances of both gas and condensate species are compatible with a scenario of chemical equilibrium at the temperatures expected for this planet.

\subsection{Self-consistency with equilibrium chemistry and condensation}
Chemical equilibrium simulations that do not account for condensation represent an extreme case, where the gas-phase molecules remain abundant in the atmosphere; in other words, gas-phase molecules supersaturate instead of condensing out. By contrast, chemical equilibrium simulations including condensation represent another extreme case, as they predict maximum depletion from condensation processes. In the simulations performed as part of this study with \texttt{GGChem} assuming condensation, multiple species of silicate form, with MgSiO$_3$ and SiO$_2$ having the highest abundance in the observable atmosphere. 

We have further investigated self-consistency using one-dimensional, physically motivated models (see Section~\ref{subsec:yunma}). These results demonstrate a high degree of consistency with those presented in Section~\ref{sec:optimized_retrieval_results}; however, they are limited by the assumption of the one-dimensional parameterisation.

Additional complexity arises from uncertainties in cloud particle materials, forms, shapes and size distributions. A more detailed understanding of the annealing and crystallisation processes might be needed to fully grasp the cloud microphysics and properties. However, we defer these aspects to future studies, pending more advanced laboratory and observational data from both within and beyond the solar system.

% Further studies on eclipse and phase curve observations of the same planet could further constrain the atmospheric thermal and cloud structures.

%\subsection{Transit spectroscopy study strategy optimisation}
%The approach adopted in this paper, can be extended to interpret other transit spectroscopic observations using the next-generation facilities like Webb:
%\begin{enumerate} \item Chemical species selection;    \item Retrievals with optimised freedom;  \item Data range selection to constrain the $T$--$p$ profile.    \item Optimisation of the chemistry using self-consistent models.
%\end{enumerate}

\subsubsection{Cloud properties in retrievals} \label{discuss:cloud}
%%% Retrieval parameterisation Optimisation
%Considering the information content of the observed data, state of the art statistical methods and computing facilities, we need to optimise retrieval model complexity with trade-offs between the model self-consistency and retrieval parameter freedom. Efforts have been made to find a balance between these two, e.g., \citet{constantinou2024viraexoplanetatmosphericretrieval} and \citet{ Constantinou2023WASP39b}. Specifically, the degree of retrieval freedom refers to how flexible observations determine a profile. This approach avoids biases from invalidated physical assumptions or a misunderstanding of the planet's environment that supports typical physical processes while obtaining reasonable retrieval sampling dimensionality.  

%%% Attempts to constrain clouds
Our retrieval simulations have indicated that gas-phase species alone may not account for the observed opacity in the optical and infrared wavelength range. While clouds can explain the missing opacity, they also represent a complex challenge for atmospheric characterisation, requiring significantly higher statistical and computing capabilities. Studies that include clouds need to balance the trade-offs between parametric/characteristic models and more complex microphysics models. There have been efforts to model cloud formation on WASP-39\,b, as discussed by \citet{Arfaux2024Cloud}, who compared cloud models of varying complexity, highlighting their respective strengths and limitations. However, due to the uncertainties in the underlying atmospheric chemistry and dynamics, key aspects of the cloud properties on WASP-39\,b, such as cloud composition and size distribution, remain ambiguous. Detailed parameters like particle shapes, contact angles and surface tension are beyond the reach of current observational capabilities. Therefore, an optimal strategy for interpreting the available data is to begin with simplified cloud models in retrievals, allowing optimised flexibility in particle size distributions. More advanced cloud retrieval methods, such as those proposed by \citet{Ma2023YunMa}, are theoretically feasible but are better suited for follow-up studies once a more precise understanding of the basic atmospheric chemistry and dynamics is established. By constraining the dimensionality of the models and refining the priors to fit the parameters, our approach allows for more efficient and tailored retrievals. Therefore, as an optimisation for the current JWST retrieval studies, we chose to use a characteristic HPS model plus a grey cloud model, as detailed in Section~\ref{sec:retrievalup}, to start with. 
We have tested the robustness of these simplistic assumptions by comparing the results in Section~\ref{sec:optimized_retrieval_results} with those obtained with cloud microphysical models (Section~\ref{tests} and \ref{subsec:yunma}). The high degree of consistency between the results obtained with the two different approaches is a good indicator of the plausibility of our assumptions.

%%Add T-p retrieval

\subsubsection{Optimised free retrieval in Step~2}
% Reconstruction
As explained in Section~\ref{sec:convergence}, the direct indication from retrieval results is the ratios between major tracers. The relative abundances between species determine the shape of the transit depth versus wavelengths. 
Indirectly, the retrieved results constrain the metallicity and temperature.

Parameters such as the absolute abundances of major tracers may shift due to the exclusion of most of the species (see Section~\ref{sec:method_chem_select} and~\ref{sec:mol_select}) during retrieval. These unselected species have two major impacts on the spectrum: their mass tunes the atmospheric scale height, and their cumulative opacities contribute directly to the spectrum's continuum. While the inclusion of a grey cloud model mitigates the latter impact to a certain extent, as a result, the retrieval algorithm adjusts other atmospheric properties, such as density and temperature, to optimise the integration path and maintain the strength of key spectral features.

\subsection{Uncertainties and suggestions}
%% Model uncertainties
Attention is needed to select the initial elemental ratios and condensation parameterisation. Although certainty cannot be guaranteed, our study relies on accepted knowledge, with sufficient margins to account for uncertainties in thermal profiles, elemental ratios, and condensation processes. Additionally, more accurate opacity data of varied temperature and pressure could help resolve the degeneracy between temperature, metallicity, and other parameters. However, conducting laboratory experiments under conditions that replicate those found in extraterrestrial environments remains challenging.

%% Metallicity calculation
As the reported values for the atmospheric properties are sensitive to and heavily dependent on atmospheric model assumptions -- including the species considered in the chemistry and opacity -- we highlight the importance of clarifying in models the molecular species assumed: 
\begin{itemize}
    \item in calculating the basic atmospheric properties, e.g., mean molecular weight;
    \item as opacity sources contributing to the transit depth;
    \item in deriving the elemental ratios to facilitate easier understanding and comparison of results in future work.
\end{itemize}

%% Metallicity definition
In addition, some misunderstanding of literature values may stem from different definitions of metallicity. In this paper, we followed the definitions summarised by \citet{Hinkel2022metallicity} \citep[see also the review by][]{Swain2024metallicitymass}. We report the values of the total volume fraction for all elements other than H$_2$ and He relative to solar and report the elemental ratios to H directly in the final results. For the benefit of future work in the field, we advocate for clarity in specifying the reference and which elemental abundance (e.g., solar outer convective zone or protostellar disk) is being used, whether in mass ratios or volume ratios, linear or log scales, solar units or absolute values when referring to the term ``metallicity''.

\section{Conclusions}

We have studied the chemical composition and cloud/haze formation in the atmosphere of WASP-39\,b, using recent JWST spectroscopic, transit observations of this planet to inform our hypotheses and models.
For this work, we have used a novel, hybrid approach that combines multistep, free retrieval simulations using the results of equilibrium chemistry models and cloud microphysics to constrain the priors. 
 
Our analysis suggests that, in addition to H$_2$O, CO$_2$, Na and K, already identified in previous works in the literature, silicon-based chemistry plays a major role in shaping the chemistry and condensates of the atmosphere
of WASP-39\,b. Our Bayesian retrievals strongly support the presence of the gas-phase SiO together with MgSiO$_3$ and SiO$_2$ clouds to explain the observed 
spectral features around 4.1, 7.5 and 8.8~$\mu$m. This scenario is compatible with results from equilibrium chemistry models.
H$_2$O and CO$_2$ absorptions in the 2.5--3.5 and 4.2--4.8~$\mu$m spectral windows were used to constrain the thermal profile as a function of pressure. 

Previous literature suggested SO$_2$ as an explanation of the features at 4.1, 7.7 and 8.5~$\mu$m, where the SO$_2$ formation was explained by potential photochemical processes. Meanwhile, it is important to explore alternative scenarios. As current models of chemical disequilibrium and photochemistry are inevitably limited by the elements considered and the size of the chemical network \citep[see e.g.,][]{Veillet2025phCHONS}, we adopt here equilibrium chemistry simulations to include a more comprehensive list of elements as initial conditions.

Our optimal solution in the gas phase corresponds to the following elemental ratios: 
2.27\,$\times$\,O/H, 1.39\,$\times$\,C/H, 1.67\,$\times$\,Si/H, 3.35\,$\times$\,S/H, 2.06\,$\times$\,Na/H, 0.55\,$\times$\,K/H with respect to the solar abundance ratios from \citet{Lodders2021solar_photo_abund}.
In addition to the gaseous species, our optimal solution includes three particle sizes for SiO$_2$, i.e. 0.01, 0.4 and 0.8\,$\mu$m, and one for MgSiO$_3$ clouds, i.e. 10\,$\mu$m. The fit to the MIRI data can be improved if we use 1\,$\mu$m particle-size for SiO$_2$ while keeping the MgSiO$_3$ particle size.

While the solution proposed here is able to faithfully match all JWST and HST data, we are not in a position to claim its uniqueness, as the method adopted does not comprehensively scan the parameter space of possible chemical species -- in gas or solid/liquid phase -- and thermodynamical conditions.
However, our hybrid approach that combines optimised free retrievals and complex forward models offers higher flexibility and robustness to interpret spectroscopic data compared to previous efforts. For instance, it leverages Bayesian statistics for the sampling algorithm to search for solutions based on observational data, while minimising biases from prior scientific knowledge.
In addition, self-consistent atmospheric models allow to limit the number of implausible scenarios considered by free retrievals, by constraining the priors with physically informed models.

To further investigate and confirm the chemical and elemental composition of the atmosphere of WASP-39\,b, additional spectroscopic data, such as very high-spectral resolution data from the ground aimed at detecting SO$_2$ and SiO, would be needed.

\begin{acknowledgments}
% __________ Acknowledgments __________
This work is based on observations made with the NASA/ESA/CSA James Webb Space Telescope. The data were obtained from the Mikulski Archive for Space Telescopes at the Space Telescope Science Institute, which is operated by the Association of Universities for Research in Astronomy, Inc., under NASA contract NAS 5-03127 for JWST. These observations are associated with program JWST-ERS-01366 and JWST-DD-2783. The specific observations presented can be accessed via \dataset[10.17909/qvjr-xx74]{https://doi.org/10.17909/qvjr-xx74}. This work was supported by the Science and Technology Facilities Council [grant number UKRI3305]. SM and GT thank the support from UKSA grants ST/X002616/1 and ST/W00254X/1. SY thanks the STFC Project No. ST/Y001508/1. This project has received funding from the European Research Council (ERC) under the European Union's Horizon 2020 research and innovation program through Advanced Investigator grant number 883830 (ExoMolHD). This work utilised the Cambridge Service for Data-Driven Discovery (CSD3), part of which is operated by the University of Cambridge Research Computing on behalf of the STFC DiRAC HPC Facility (dirac.ac.uk). The DiRAC component of CSD3 was funded by BEIS capital funding via STFC capital grants ST/P002307/1, ST/R002452/1, and STFC operations grant ST/R00689X/1. DiRAC is part of the National e-Infrastructure.
\end{acknowledgments}

% \facility{JWST (NIRISS, NIRCAM, NIRSpec, MIRI), HST (STIS, WFC3)}

\software{python 3.10.0, numpy 1.26.0 \citep{harris2020numpy}, scipy 1.14.0 \citep{Virtanen2020SciPy-NMeth}, pymultinest 2.12 \citep{Buchner2014pymultinest}, astropy 5.3.4 \citep{astropy:2022, astropy:2018, astropy:2013}, numba 0.59.0 \citep{lam2015numba}}

% \textit{Datasets:} The JWST data presented in this article were obtained from the Mikulski Archive for Space Telescopes (MAST) at the Space Telescope Science Institute. The specific observations presented can be accessed via \dataset[10.17909/qvjr-xx74]{https://doi.org/10.17909/qvjr-xx74}.

%% For this sample we use BibTeX plus aasjournals.bst to generate the
%% the bibliography. The sample631.bib file was populated from ADS. To
%% get the citations to show in the compiled file do the following:
%%
%% pdflatex sample631.tex
%% bibtext sample631
%% pdflatex sample631.tex
%% pdflatex sample631.tex
\newpage
\bibliography{GasExplanation}{}

\begin{thebibliography}{}
\expandafter\ifx\csname natexlab\endcsname\relax\def\natexlab#1{#1}\fi
\providecommand{\url}[1]{\href{#1}{#1}}
\providecommand{\dodoi}[1]{doi:~\href{http://doi.org/#1}{\nolinkurl{#1}}}
\providecommand{\doeprint}[1]{\href{http://ascl.net/#1}{\nolinkurl{http://ascl.net/#1}}}
\providecommand{\doarXiv}[1]{\href{https://arxiv.org/abs/#1}{\nolinkurl{https://arxiv.org/abs/#1}}}

\bibitem[{{Ackerman} \& {Marley}(2001)}]{ackerman2001precipitating}
{Ackerman}, A.~S., \& {Marley}, M.~S. 2001, \apj, 556, 872, \dodoi{10.1086/321540}

\bibitem[{{Ag{\'u}ndez} {et~al.}(2020){Ag{\'u}ndez}, {Mart{\'\i}nez}, {de Andres}, {Cernicharo}, \& {Mart{\'\i}n-Gago}}]{Agundez2020ACE}
{Ag{\'u}ndez}, M., {Mart{\'\i}nez}, J.~I., {de Andres}, P.~L., {Cernicharo}, J., \& {Mart{\'\i}n-Gago}, J.~A. 2020, \aap, 637, A59, \dodoi{10.1051/0004-6361/202037496}

\bibitem[{{Ag{\'u}ndez} {et~al.}(2012){Ag{\'u}ndez}, {Venot}, {Iro}, {Selsis}, {Hersant}, {H{\'e}brard}, \& {Dobrijevic}}]{Agundez2012HD209}
{Ag{\'u}ndez}, M., {Venot}, O., {Iro}, N., {et~al.} 2012, \aap, 548, A73, \dodoi{10.1051/0004-6361/201220365}

\bibitem[{{Ahrer} {et~al.}(2023){Ahrer}, {Stevenson}, {Mansfield}, {Moran}, {Brande}, {Morello}, {Murray}, {Nikolov}, {Petit dit de la Roche}, {Schlawin}, {Wheatley}, {Zieba}, {Batalha}, {Damiano}, {Goyal}, {Lendl}, {Lothringer}, {Mukherjee}, {Ohno}, {Batalha}, {Battley}, {Bean}, {Beatty}, {Benneke}, {Berta-Thompson}, {Carter}, {Cubillos}, {Daylan}, {Espinoza}, {Gao}, {Gibson}, {Gill}, {Harrington}, {Hu}, {Kreidberg}, {Lewis}, {Line}, {L{\'o}pez-Morales}, {Parmentier}, {Powell}, {Sing}, {Tsai}, {Wakeford}, {Welbanks}, {Alam}, {Alderson}, {Allen}, {Anderson}, {Barstow}, {Bayliss}, {Bell}, {Blecic}, {Bryant}, {Burleigh}, {Carone}, {Casewell}, {Changeat}, {Chubb}, {Crossfield}, {Crouzet}, {Decin}, {D{\'e}sert}, {Feinstein}, {Flagg}, {Fortney}, {Gizis}, {Heng}, {Iro}, {Kempton}, {Kendrew}, {Kirk}, {Knutson}, {Komacek}, {Lagage}, {Leconte}, {Lustig-Yaeger}, {MacDonald}, {Mancini}, {May}, {Mayne}, {Miguel}, {Mikal-Evans}, {Molaverdikhani}, {Palle}, {Piaulet}, {Rackham}, {Redfield}, {Rogers}, {Roy}, {Rustamkulov},
  {Shkolnik}, {Sotzen}, {Taylor}, {Tremblin}, {Tucker}, {Turner}, {de Val-Borro}, {Venot}, \& {Zhang}}]{Ahrer2023WASP39b}
{Ahrer}, E.-M., {Stevenson}, K.~B., {Mansfield}, M., {et~al.} 2023, \nat, 614, 653, \dodoi{10.1038/s41586-022-05590-4}

\bibitem[{Al-Refaie {et~al.}(2022)Al-Refaie, Changeat, Venot, Waldmann, \& Tinetti}]{Al-Refaie2022taurex}
Al-Refaie, A.~F., Changeat, Q., Venot, O., Waldmann, I.~P., \& Tinetti, G. 2022, \apj, 932, 123, \dodoi{10.3847/1538-4357/ac6dcd}

\bibitem[{{Al-Refaie} {et~al.}(2021){Al-Refaie}, Changeat, Waldmann, \& Tinetti}]{alrefaie2021taurex31}
{Al-Refaie}, A.~F., Changeat, Q., Waldmann, I.~P., \& Tinetti, G. 2021, \apj, 917, 37, \dodoi{10.3847/1538-4357/ac0252}

\bibitem[{Alderson {et~al.}(2022)Alderson, Grant, \& Wakeford}]{lili_alderson_2022_7185855}
Alderson, L., Grant, D., \& Wakeford, H. 2022, Exo-TiC/ExoTiC-JEDI: v0.1-beta-release, v0.1,  Zenodo, \dodoi{10.5281/zenodo.7185855}

\bibitem[{{Alderson} {et~al.}(2023){Alderson}, {Wakeford}, {Alam}, {Batalha}, {Lothringer}, {Adams Redai}, {Barat}, {Brande}, {Damiano}, {Daylan}, {Espinoza}, {Flagg}, {Goyal}, {Grant}, {Hu}, {Inglis}, {Lee}, {Mikal-Evans}, {Ramos-Rosado}, {Roy}, {Wallack}, {Batalha}, {Bean}, {Benneke}, {Berta-Thompson}, {Carter}, {Changeat}, {Col{\'o}n}, {Crossfield}, {D{\'e}sert}, {Foreman-Mackey}, {Gibson}, {Kreidberg}, {Line}, {L{\'o}pez-Morales}, {Molaverdikhani}, {Moran}, {Morello}, {Moses}, {Mukherjee}, {Schlawin}, {Sing}, {Stevenson}, {Taylor}, {Aggarwal}, {Ahrer}, {Allen}, {Barstow}, {Bell}, {Blecic}, {Casewell}, {Chubb}, {Crouzet}, {Cubillos}, {Decin}, {Feinstein}, {Fortney}, {Harrington}, {Heng}, {Iro}, {Kempton}, {Kirk}, {Knutson}, {Krick}, {Leconte}, {Lendl}, {MacDonald}, {Mancini}, {Mansfield}, {May}, {Mayne}, {Miguel}, {Nikolov}, {Ohno}, {Palle}, {Parmentier}, {Petit dit de la Roche}, {Piaulet}, {Powell}, {Rackham}, {Redfield}, {Rogers}, {Rustamkulov}, {Tan}, {Tremblin}, {Tsai}, {Turner}, {de Val-Borro},
  {Venot}, {Welbanks}, {Wheatley}, \& {Zhang}}]{Alderson2023WASP39b}
{Alderson}, L., {Wakeford}, H.~R., {Alam}, M.~K., {et~al.} 2023, \nat, 614, 664, \dodoi{10.1038/s41586-022-05591-3}

\bibitem[{{Allard} {et~al.}(2016){Allard}, {Spiegelman}, \& {Kielkopf}}]{Allard2016K}
{Allard}, N.~F., {Spiegelman}, F., \& {Kielkopf}, J.~F. 2016, \aap, 589, A21, \dodoi{10.1051/0004-6361/201628270}

\bibitem[{{Allard} {et~al.}(2019){Allard}, {Spiegelman}, {Leininger}, \& {Molliere}}]{Allard2019Na}
{Allard}, N.~F., {Spiegelman}, F., {Leininger}, T., \& {Molliere}, P. 2019, \aap, 628, A120, \dodoi{10.1051/0004-6361/201935593}

\bibitem[{{Arfaux} \& {Lavvas}(2024)}]{Arfaux2024Cloud}
{Arfaux}, A., \& {Lavvas}, P. 2024, \mnras, 530, 482, \dodoi{10.1093/mnras/stae826}

\bibitem[{{Asplund} {et~al.}(2009){Asplund}, {Grevesse}, {Sauval}, \& {Scott}}]{Asplund2009solarabundance}
{Asplund}, M., {Grevesse}, N., {Sauval}, A.~J., \& {Scott}, P. 2009, \araa, 47, 481, \dodoi{10.1146/annurev.astro.46.060407.145222}

\bibitem[{{Astropy Collaboration} {et~al.}(2013){Astropy Collaboration}, {Robitaille}, {Tollerud}, {Greenfield}, {Droettboom}, {Bray}, {Aldcroft}, {Davis}, {Ginsburg}, {Price-Whelan}, {Kerzendorf}, {Conley}, {Crighton}, {Barbary}, {Muna}, {Ferguson}, {Grollier}, {Parikh}, {Nair}, {Unther}, {Deil}, {Woillez}, {Conseil}, {Kramer}, {Turner}, {Singer}, {Fox}, {Weaver}, {Zabalza}, {Edwards}, {Azalee Bostroem}, {Burke}, {Casey}, {Crawford}, {Dencheva}, {Ely}, {Jenness}, {Labrie}, {Lim}, {Pierfederici}, {Pontzen}, {Ptak}, {Refsdal}, {Servillat}, \& {Streicher}}]{astropy:2013}
{Astropy Collaboration}, {Robitaille}, T.~P., {Tollerud}, E.~J., {et~al.} 2013, \aap, 558, A33, \dodoi{10.1051/0004-6361/201322068}

\bibitem[{{Astropy Collaboration} {et~al.}(2018){Astropy Collaboration}, {Price-Whelan}, {Sip{\H{o}}cz}, {G{\"u}nther}, {Lim}, {Crawford}, {Conseil}, {Shupe}, {Craig}, {Dencheva}, {Ginsburg}, {Vand erPlas}, {Bradley}, {P{\'e}rez-Su{\'a}rez}, {de Val-Borro}, {Aldcroft}, {Cruz}, {Robitaille}, {Tollerud}, {Ardelean}, {Babej}, {Bach}, {Bachetti}, {Bakanov}, {Bamford}, {Barentsen}, {Barmby}, {Baumbach}, {Berry}, {Biscani}, {Boquien}, {Bostroem}, {Bouma}, {Brammer}, {Bray}, {Breytenbach}, {Buddelmeijer}, {Burke}, {Calderone}, {Cano Rodr{\'\i}guez}, {Cara}, {Cardoso}, {Cheedella}, {Copin}, {Corrales}, {Crichton}, {D'Avella}, {Deil}, {Depagne}, {Dietrich}, {Donath}, {Droettboom}, {Earl}, {Erben}, {Fabbro}, {Ferreira}, {Finethy}, {Fox}, {Garrison}, {Gibbons}, {Goldstein}, {Gommers}, {Greco}, {Greenfield}, {Groener}, {Grollier}, {Hagen}, {Hirst}, {Homeier}, {Horton}, {Hosseinzadeh}, {Hu}, {Hunkeler}, {Ivezi{\'c}}, {Jain}, {Jenness}, {Kanarek}, {Kendrew}, {Kern}, {Kerzendorf}, {Khvalko}, {King}, {Kirkby}, {Kulkarni},
  {Kumar}, {Lee}, {Lenz}, {Littlefair}, {Ma}, {Macleod}, {Mastropietro}, {McCully}, {Montagnac}, {Morris}, {Mueller}, {Mumford}, {Muna}, {Murphy}, {Nelson}, {Nguyen}, {Ninan}, {N{\"o}the}, {Ogaz}, {Oh}, {Parejko}, {Parley}, {Pascual}, {Patil}, {Patil}, {Plunkett}, {Prochaska}, {Rastogi}, {Reddy Janga}, {Sabater}, {Sakurikar}, {Seifert}, {Sherbert}, {Sherwood-Taylor}, {Shih}, {Sick}, {Silbiger}, {Singanamalla}, {Singer}, {Sladen}, {Sooley}, {Sornarajah}, {Streicher}, {Teuben}, {Thomas}, {Tremblay}, {Turner}, {Terr{\'o}n}, {van Kerkwijk}, {de la Vega}, {Watkins}, {Weaver}, {Whitmore}, {Woillez}, {Zabalza}, \& {Astropy Contributors}}]{astropy:2018}
{Astropy Collaboration}, {Price-Whelan}, A.~M., {Sip{\H{o}}cz}, B.~M., {et~al.} 2018, \aj, 156, 123, \dodoi{10.3847/1538-3881/aabc4f}

\bibitem[{{Astropy Collaboration} {et~al.}(2022){Astropy Collaboration}, {Price-Whelan}, {Lim}, {Earl}, {Starkman}, {Bradley}, {Shupe}, {Patil}, {Corrales}, {Brasseur}, {N{"o}the}, {Donath}, {Tollerud}, {Morris}, {Ginsburg}, {Vaher}, {Weaver}, {Tocknell}, {Jamieson}, {van Kerkwijk}, {Robitaille}, {Merry}, {Bachetti}, {G{"u}nther}, {Aldcroft}, {Alvarado-Montes}, {Archibald}, {B{'o}di}, {Bapat}, {Barentsen}, {Baz{'a}n}, {Biswas}, {Boquien}, {Burke}, {Cara}, {Cara}, {Conroy}, {Conseil}, {Craig}, {Cross}, {Cruz}, {D'Eugenio}, {Dencheva}, {Devillepoix}, {Dietrich}, {Eigenbrot}, {Erben}, {Ferreira}, {Foreman-Mackey}, {Fox}, {Freij}, {Garg}, {Geda}, {Glattly}, {Gondhalekar}, {Gordon}, {Grant}, {Greenfield}, {Groener}, {Guest}, {Gurovich}, {Handberg}, {Hart}, {Hatfield-Dodds}, {Homeier}, {Hosseinzadeh}, {Jenness}, {Jones}, {Joseph}, {Kalmbach}, {Karamehmetoglu}, {Ka{l}uszy{'n}ski}, {Kelley}, {Kern}, {Kerzendorf}, {Koch}, {Kulumani}, {Lee}, {Ly}, {Ma}, {MacBride}, {Maljaars}, {Muna}, {Murphy}, {Norman}, {O'Steen},
  {Oman}, {Pacifici}, {Pascual}, {Pascual-Granado}, {Patil}, {Perren}, {Pickering}, {Rastogi}, {Roulston}, {Ryan}, {Rykoff}, {Sabater}, {Sakurikar}, {Salgado}, {Sanghi}, {Saunders}, {Savchenko}, {Schwardt}, {Seifert-Eckert}, {Shih}, {Jain}, {Shukla}, {Sick}, {Simpson}, {Singanamalla}, {Singer}, {Singhal}, {Sinha}, {Sip{H{o}}cz}, {Spitler}, {Stansby}, {Streicher}, {{{S}}umak}, {Swinbank}, {Taranu}, {Tewary}, {Tremblay}, {Val-Borro}, {Van Kooten}, {Vasovi{'c}}, {Verma}, {de Miranda Cardoso}, {Williams}, {Wilson}, {Winkel}, {Wood-Vasey}, {Xue}, {Yoachim}, {Zhang}, {Zonca}, \& {Astropy Project Contributors}}]{astropy:2022}
{Astropy Collaboration}, {Price-Whelan}, A.~M., {Lim}, P.~L., {et~al.} 2022, \apj, 935, 167, \dodoi{10.3847/1538-4357/ac7c74}

\bibitem[{{Azzam} {et~al.}(2016){Azzam}, {Tennyson}, {Yurchenko}, \& {Naumenko}}]{Azzam2016H2S}
{Azzam}, A. A.~A., {Tennyson}, J., {Yurchenko}, S.~N., \& {Naumenko}, O.~V. 2016, \mnras, 460, 4063, \dodoi{10.1093/mnras/stw1133}

\bibitem[{{Bell} {et~al.}(2022){Bell}, {Ahrer}, {Brande}, {Carter}, {Feinstein}, {Caloca}, {Mansfield}, {Zieba}, {Piaulet}, {Benneke}, {Filippazzo}, {May}, {Roy}, {Kreidberg}, \& {Stevenson}}]{Bell2022Eureka}
{Bell}, T., {Ahrer}, E.-M., {Brande}, J., {et~al.} 2022, The Journal of Open Source Software, 7, 4503, \dodoi{10.21105/joss.04503}

\bibitem[{{Benneke} {et~al.}(2019){Benneke}, {Wong}, {Piaulet}, {Knutson}, {Lothringer}, {Morley}, {Crossfield}, {Gao}, {Greene}, {Dressing}, {Dragomir}, {Howard}, {McCullough}, {Kempton}, {Fortney}, \& {Fraine}}]{2019ApJ...887L..14B}
{Benneke}, B., {Wong}, I., {Piaulet}, C., {et~al.} 2019, \apjl, 887, L14, \dodoi{10.3847/2041-8213/ab59dc}

\bibitem[{Bohren \& Huffman(2008)}]{bohren2008absorption}
Bohren, C.~F., \& Huffman, D.~R. 2008, Absorption and scattering of light by small particles (New York: Wiley)

\bibitem[{{Bonomo} {et~al.}(2017){Bonomo}, {Desidera}, {Benatti}, {Borsa}, {Crespi, S.}, {Damasso, M.}, {Lanza, A. F.}, {Sozzetti, A.}, {Lodato, G.}, {Marzari, F.}, {Boccato, C.}, {Claudi, R. U.}, {Cosentino, R.}, {Covino, E.}, {Gratton, R.}, {Maggio, A.}, {Micela, G.}, {Molinari, E.}, {Pagano, I.}, {Piotto, G.}, {Poretti, E.}, {Smareglia, R.}, {Affer, L.}, {Biazzo, K.}, {Bignamini, A.}, {Esposito, M.}, {Giacobbe, P.}, {HÃ©brard, G.}, {Malavolta, L.}, {Maldonado, J.}, {Mancini, L.}, {Martinez Fiorenzano, A.}, {Masiero, S.}, {Nascimbeni, V.}, {Pedani, M.}, {Rainer, M.}, \& {Scandariato, G.}}]{Bonomo2017WASP39b}
{Bonomo}, A.~S., {Desidera}, S., {Benatti}, S., {et~al.} 2017, A\&A, 602, A107, \dodoi{10.1051/0004-6361/201629882}

\bibitem[{{Bourrier} {et~al.}(2020){Bourrier}, {Kitzmann}, {Kuntzer}, {Nascimbeni}, {Lendl}, {Lavie}, {Hoeijmakers}, {Pino}, {Ehrenreich}, {Heng}, {Allart}, {Cegla}, {Dumusque}, {Melo}, {Astudillo-Defru}, {Caldwell}, {Cretignier}, {Giles}, {Henze}, {Jenkins}, {Lovis}, {Murgas}, {Pepe}, {Ricker}, {Rose}, {Seager}, {Segransan}, {Su{\'a}rez-Mascare{\~n}o}, {Udry}, {Vanderspek}, \& {Wyttenbach}}]{Bourrier2020wasp121b}
{Bourrier}, V., {Kitzmann}, D., {Kuntzer}, T., {et~al.} 2020, \aap, 637, A36, \dodoi{10.1051/0004-6361/201936647}

\bibitem[{{Buchner} {et~al.}(2014){Buchner}, {Georgakakis}, {Nandra}, {Hsu}, {Rangel}, {Brightman}, {Merloni}, {Salvato}, {Donley}, \& {Kocevski}}]{Buchner2014pymultinest}
{Buchner}, J., {Georgakakis}, A., {Nandra}, K., {et~al.} 2014, \aap, 564, A125, \dodoi{10.1051/0004-6361/201322971}

\bibitem[{{Carter} {et~al.}(2024){Carter}, {May}, {Espinoza}, {Welbanks}, {Ahrer}, {Alderson}, {Brahm}, {Feinstein}, {Grant}, {Line}, {Morello}, {O'Steen}, {Radica}, {Rustamkulov}, {Stevenson}, {Turner}, {Alam}, {Anderson}, {Batalha}, {Battley}, {Bayliss}, {Bean}, {Benneke}, {Berta-Thompson}, {Brande}, {Bryant}, {Burleigh}, {Coulombe}, {Crossfield}, {Damiano}, {D{\'e}sert}, {Flagg}, {Gill}, {Inglis}, {Kirk}, {Knutson}, {Kreidberg}, {L{\'o}pez Morales}, {Mansfield}, {Moran}, {Murray}, {Nixon}, {Petit dit de la Roche}, {Rackham}, {Schlawin}, {Sing}, {Wakeford}, {Wallack}, {Wheatley}, {Zieba}, {Aggarwal}, {Barstow}, {Bell}, {Blecic}, {Caceres}, {Crouzet}, {Cubillos}, {Daylan}, {de Val-Borro}, {Decin}, {Fortney}, {Gibson}, {Heng}, {Hu}, {Kempton}, {Lagage}, {Lothringer}, {Lustig-Yaeger}, {Mancini}, {Mayne}, {Mayorga}, {Molaverdikhani}, {Nasedkin}, {Ohno}, {Parmentier}, {Powell}, {Redfield}, {Roy}, {Taylor}, \& {Zhang}}]{Carter2024JWSTWASP39b}
{Carter}, A.~L., {May}, E.~M., {Espinoza}, N., {et~al.} 2024, Nature Astronomy, \dodoi{10.1038/s41550-024-02292-x}

\bibitem[{{Changeat} {et~al.}(2025){Changeat}, {Bardet}, {Chubb}, {Dyrek}, {Edwards}, {Ohno}, \& {Venot}}]{Changeat2025titan}
{Changeat}, Q., {Bardet}, D., {Chubb}, K., {et~al.} 2025, arXiv e-prints, arXiv:2505.18715, \dodoi{10.48550/arXiv.2505.18715}

\bibitem[{{Changeat} {et~al.}(2024){Changeat}, {Ito}, {Al-Refaie}, {Yip}, \& {Lueftinger}}]{Changeat2024}
{Changeat}, Q., {Ito}, Y., {Al-Refaie}, A.~F., {Yip}, K.~H., \& {Lueftinger}, T. 2024, \aj, 167, 195, \dodoi{10.3847/1538-3881/ad3032}

\bibitem[{{{Changeat}} {et~al.}(2022){{Changeat}}, {Edwards}, {Al-Refaie}, {Tsiaras}, {Skinner}, {Cho}, {Yip}, {Anisman}, {Ikoma}, {Bieger}, {Venot}, {Shibata}, {Waldmann}, \& {Tinetti}}]{changeat2022five}
{{Changeat}}, Q., {Edwards}, B., {Al-Refaie}, A.~F., {et~al.} 2022, \apjs, 260, 3, \dodoi{10.3847/1538-4365/ac5cc2}

\bibitem[{{Chen} {et~al.}(2025){Chen}, {Ji}, {Chen}, {Yan}, \& {Tan}}]{Chen2025WASP39b}
{Chen}, Z., {Ji}, J., {Chen}, G., {Yan}, F., \& {Tan}, X. 2025, \aj, 169, 294, \dodoi{10.3847/1538-3881/adc803}

\bibitem[{{Chubb} {et~al.}(2021){Chubb}, {Rocchetto}, {Yurchenko}, {Min}, {Waldmann}, {Barstow}, {Molli{\`e}re}, {Al-Refaie}, {Phillips}, \& {Tennyson}}]{chubb2021exomol}
{Chubb}, K.~L., {Rocchetto}, M., {Yurchenko}, S.~N., {et~al.} 2021, \aap, 646, A21, \dodoi{10.1051/0004-6361/202038350}

\bibitem[{{Chubb} {et~al.}(2024){Chubb}, {Robert}, {Sousa-Silva}, {Yurchenko}, {Allard}, {Boudon}, {Buldyreva}, {Bultel}, {Coustenis}, {Foltynowicz}, {Gordon}, {Hargreaves}, {Helling}, {Hill}, {Rafn Hrodmarsson}, {Karman}, {Lecoq-Molinos}, {Migliorini}, {Rey}, {Richard}, {Sadiek}, {Schmidt}, {Sokolov}, {Stefani}, {Tennyson}, {Venot}, {Wright}, {Arenales-Lope}, {Barstow}, {Bocchieri}, {Carrasco}, {Dubey}, {Egorov}, {Garc{\'\i}a Mu{\~n}oz}, {Ehsan}, {Gharib-Nezhad}, {Gkouvelis}, {Gr{\"u}bel}, {Irwin}, {Kn{\'\i}{\v{z}}ek}, {Lewis}, {Lodge}, {Ma}, {Martins}, {Molaverdikhani}, {Morello}, {Nikitin}, {Panek}, {Rengel}, {Rinaldi}, {Skinner}, {Tinetti}, {van Kempen}, {Yang}, \& {Zingales}}]{Chubb2024Database_Whitepaper_ph}
{Chubb}, K.~L., {Robert}, S., {Sousa-Silva}, C., {et~al.} 2024, RAS Tech. Instr., \dodoi{10.1093/rasti/rzae039}

\bibitem[{{Coles} {et~al.}(2019){Coles}, {Yurchenko}, \& {Tennyson}}]{Coles2019NH3}
{Coles}, P.~A., {Yurchenko}, S.~N., \& {Tennyson}, J. 2019, \mnras, 490, 4638, \dodoi{10.1093/mnras/stz2778}

\bibitem[{Constantinou \& Madhusudhan(2024)}]{constantinou2024viraexoplanetatmosphericretrieval}
Constantinou, S., \& Madhusudhan, N. 2024, VIRA: An Exoplanet Atmospheric Retrieval Framework for JWST Transmission Spectroscopy.
\newblock \doarXiv{2403.04825}

\bibitem[{Cox(2015)}]{Cox2015allen}
Cox, A.~N. 2015, Allen’s Astrophysical Quantities, 4th edn. (New York: Springer)

\bibitem[{{Darveau-Bernier} {et~al.}(2022){Darveau-Bernier}, {Albert}, {Talens}, {Lafreni{\`e}re}, {Radica}, {Doyon}, {Cook}, {Rowe}, {Allart}, {Artigau}, {Benneke}, {Cowan}, {Dang}, {Espinoza}, {Johnstone}, {Kaltenegger}, {Lim}, {Pauly}, {Pelletier}, {Piaulet}, {Roy}, {Roy}, {Splinter}, {Taylor}, \& {Turner}}]{2022PASP..134i4502D}
{Darveau-Bernier}, A., {Albert}, L., {Talens}, G.~J., {et~al.} 2022, \pasp, 134, 094502, \dodoi{10.1088/1538-3873/ac8a77}

\bibitem[{Davey {et~al.}(2024)Davey, Yip, Al-Refaie, \& Waldmann}]{davey2024binning}
Davey, J.~J., Yip, K.~H., Al-Refaie, A.~F., \& Waldmann, I.~P. 2024, Monthly Notices of the Royal Astronomical Society, stae2731, \dodoi{10.1093/mnras/stae2731}

\bibitem[{Deirmendjian(1964)}]{Deirmendjian1964distribution}
Deirmendjian, D. 1964, Appl. Opt., 3, 187, \dodoi{10.1364/AO.3.000187}

\bibitem[{{Dorschner} {et~al.}(1995){Dorschner}, {Begemann}, {Henning}, {Jaeger}, \& {Mutschke}}]{Dorschner1995RIMgSiO3}
{Dorschner}, J., {Begemann}, B., {Henning}, T., {Jaeger}, C., \& {Mutschke}, H. 1995, \aap, 300, 503

\bibitem[{{Esparza-Borges} {et~al.}(2023){Esparza-Borges}, {L{\'o}pez-Morales}, {Adams Redai}, {Pall{\'e}}, {Kirk}, {Casasayas-Barris}, {Batalha}, {Rackham}, {Bean}, {Casewell}, {Decin}, {Dos Santos}, {Garc{\'\i}a Mu{\~n}oz}, {Harrington}, {Heng}, {Hu}, {Mancini}, {Molaverdikhani}, {Morello}, {Nikolov}, {Nixon}, {Redfield}, {Stevenson}, {Wakeford}, {Alam}, {Benneke}, {Blecic}, {Crouzet}, {Daylan}, {Inglis}, {Kreidberg}, {Petit dit de la Roche}, \& {Turner}}]{Esparza-Borges2023WASP39b}
{Esparza-Borges}, E., {L{\'o}pez-Morales}, M., {Adams Redai}, J.~I., {et~al.} 2023, \apjl, 955, L19, \dodoi{10.3847/2041-8213/acf27b}

\bibitem[{Espinoza(2022)}]{espinoza_2022_6960924}
Espinoza, N. 2022, TransitSpectroscopy, 0.3.11,  Zenodo, \dodoi{10.5281/zenodo.6960924}

\bibitem[{{Espinoza} {et~al.}(2019){Espinoza}, {Kossakowski}, \& {Brahm}}]{2019MNRAS.490.2262E}
{Espinoza}, N., {Kossakowski}, D., \& {Brahm}, R. 2019, \mnras, 490, 2262, \dodoi{10.1093/mnras/stz2688}

\bibitem[{{Fabian} {et~al.}(2000){Fabian}, {J{\"a}ger}, {Henning}, {Dorschner}, \& {Mutschke}}]{Fabian2000RIMgSiO3}
{Fabian}, D., {J{\"a}ger}, C., {Henning}, T., {Dorschner}, J., \& {Mutschke}, H. 2000, \aap, 364, 282

\bibitem[{{Faedi} {et~al.}(2011){Faedi}, {Barros}, {Anderson}, {Brown}, {Collier Cameron}, {Pollacco}, {Boisse}, {H{\'e}brard}, {Lendl}, {Lister}, {Smalley}, {Street}, {Triaud}, {Bento}, {Bouchy}, {Butters}, {Enoch}, {Haswell}, {Hellier}, {Keenan}, {Miller}, {Moulds}, {Moutou}, {Norton}, {Queloz}, {Santerne}, {Simpson}, {Skillen}, {Smith}, {Udry}, {Watson}, {West}, \& {Wheatley}}]{Faedi2011WASP39b}
{Faedi}, F., {Barros}, S.~C.~C., {Anderson}, D.~R., {et~al.} 2011, \aap, 531, A40, \dodoi{10.1051/0004-6361/201116671}

\bibitem[{{Feinstein} {et~al.}(2023){Feinstein}, {Radica}, {Welbanks}, {Murray}, {Ohno}, {Coulombe}, {Espinoza}, {Bean}, {Teske}, {Benneke}, {Line}, {Rustamkulov}, {Saba}, {Tsiaras}, {Barstow}, {Fortney}, {Gao}, {Knutson}, {MacDonald}, {Mikal-Evans}, {Rackham}, {Taylor}, {Parmentier}, {Batalha}, {Berta-Thompson}, {Carter}, {Changeat}, {dos Santos}, {Gibson}, {Goyal}, {Kreidberg}, {L{\'o}pez-Morales}, {Lothringer}, {Miguel}, {Molaverdikhani}, {Moran}, {Morello}, {Mukherjee}, {Sing}, {Stevenson}, {Wakeford}, {Ahrer}, {Alam}, {Alderson}, {Allen}, {Batalha}, {Bell}, {Blecic}, {Brande}, {Caceres}, {Casewell}, {Chubb}, {Crossfield}, {Crouzet}, {Cubillos}, {Decin}, {D{\'e}sert}, {Harrington}, {Heng}, {Henning}, {Iro}, {Kempton}, {Kendrew}, {Kirk}, {Krick}, {Lagage}, {Lendl}, {Mancini}, {Mansfield}, {May}, {Mayne}, {Nikolov}, {Palle}, {Petit dit de la Roche}, {Piaulet}, {Powell}, {Redfield}, {Rogers}, {Roman}, {Roy}, {Nixon}, {Schlawin}, {Tan}, {Tremblin}, {Turner}, {Venot}, {Waalkes}, {Wheatley}, \&
  {Zhang}}]{Feinstein2023WASP39b}
{Feinstein}, A.~D., {Radica}, M., {Welbanks}, L., {et~al.} 2023, \nat, 614, 670, \dodoi{10.1038/s41586-022-05674-1}

\bibitem[{Feroz \& Hobson(2008)}]{Feroz2008multinest}
Feroz, F., \& Hobson, M.~P. 2008, Monthly Notices of the Royal Astronomical Society, 384, 449, \dodoi{10.1111/j.1365-2966.2007.12353.x}

\bibitem[{Feroz {et~al.}(2009)Feroz, Hobson, \& Bridges}]{Feroz2009Multinest}
Feroz, F., Hobson, M.~P., \& Bridges, M. 2009, Monthly Notices of the Royal Astronomical Society, 398, 1601, \dodoi{10.1111/j.1365-2966.2009.14548.x}

\bibitem[{{Feroz} {et~al.}(2019){Feroz}, {Hobson}, {Cameron}, \& {Pettitt}}]{Feroz2019multinest}
{Feroz}, F., {Hobson}, M.~P., {Cameron}, E., \& {Pettitt}, A.~N. 2019, The Open Journal of Astrophysics, 2, 10, \dodoi{10.21105/astro.1306.2144}

\bibitem[{{Fischer} {et~al.}(2016){Fischer}, {Knutson}, {Sing}, {Henry}, {Williamson}, {Fortney}, {Burrows}, {Kataria}, {Nikolov}, {Showman}, {Ballester}, {D{\'e}sert}, {Aigrain}, {Deming}, {Lecavelier des Etangs}, \& {Vidal-Madjar}}]{Fischer2016WASP39b}
{Fischer}, P.~D., {Knutson}, H.~A., {Sing}, D.~K., {et~al.} 2016, \apj, 827, 19, \dodoi{10.3847/0004-637X/827/1/19}

\bibitem[{{Fisher} \& {Heng}(2018)}]{FisherHeng2018WFC3retrieval}
{Fisher}, C., \& {Heng}, K. 2018, \mnras, 481, 4698, \dodoi{10.1093/mnras/sty2550}

\bibitem[{Fisher {et~al.}(2024)Fisher, Taylor, Parmentier, Kitzmann, Birkby, Radica, Barstow, Yang, \& Morello}]{Fisher2024jwst}
Fisher, C., Taylor, J., Parmentier, V., {et~al.} 2024, Monthly Notices of the Royal Astronomical Society, 535, 27, \dodoi{10.1093/mnras/stae2240}

\bibitem[{{Flagg} {et~al.}(2024){Flagg}, {Weinberger}, {Bell}, {Welbanks}, {Morello}, {Powell}, {Bean}, {Blecic}, {Crouzet}, {Gao}, {Inglis}, {Kirk}, {Lopez-Morales}, {Molaverdikhani}, {Nikolov}, {Oza}, {Rackham}, {Redfield}, {Tsai}, {Jayawardhana}, {Kreidberg}, {Nixon}, {Stevenson}, \& {Turner}}]{Flagg2024WASP39b}
{Flagg}, L., {Weinberger}, A.~J., {Bell}, T.~J., {et~al.} 2024, arXiv e-prints, arXiv:2406.02305, \dodoi{10.48550/arXiv.2406.02305}

\bibitem[{Gierasch \& Conrath(1985)}]{gierasch1985energy}
Gierasch, P., \& Conrath, B. 1985, Recent Advances in Planetary Meteorology, 121

\bibitem[{{Gorman} {et~al.}(2019){Gorman}, {Yurchenko}, \& {Tennyson}}]{Gorman2019HS}
{Gorman}, M.~N., {Yurchenko}, S.~N., \& {Tennyson}, J. 2019, \mnras, 490, 1652, \dodoi{10.1093/mnras/stz2517}

\bibitem[{{Grant} {et~al.}(2023){Grant}, {Lothringer}, {Wakeford}, {Alam}, {Alderson}, {Bean}, {Benneke}, {D{\'e}sert}, {Daylan}, {Flagg}, {Hu}, {Inglis}, {Kirk}, {Kreidberg}, {L{\'o}pez-Morales}, {Mancini}, {Mikal-Evans}, {Molaverdikhani}, {Palle}, {Rackham}, {Redfield}, {Stevenson}, {Valenti}, {Wallack}, {Aggarwal}, {Ahrer}, {Crossfield}, {Crouzet}, {Iro}, {Nikolov}, {Wheatley}, \& {JWST Transiting Exoplanet Community ERS Team}}]{2023ApJ...949L..15G}
{Grant}, D., {Lothringer}, J.~D., {Wakeford}, H.~R., {et~al.} 2023, \apjl, 949, L15, \dodoi{10.3847/2041-8213/acd544}

\bibitem[{Harris {et~al.}(2020)Harris, Millman, van~der Walt, Gommers, Virtanen, Cournapeau, Wieser, Taylor, Berg, Smith, Kern, Picus, Hoyer, van Kerkwijk, Brett, Haldane, del R{\'{i}}o, Wiebe, Peterson, G{\'{e}}rard-Marchant, Sheppard, Reddy, Weckesser, Abbasi, Gohlke, \& Oliphant}]{harris2020numpy}
Harris, C.~R., Millman, K.~J., van~der Walt, S.~J., {et~al.} 2020, Nature, 585, 357, \dodoi{10.1038/s41586-020-2649-2}

\bibitem[{{Hinkel} {et~al.}(2022){Hinkel}, {Young}, \& {Wheeler}}]{Hinkel2022metallicity}
{Hinkel}, N.~R., {Young}, P.~A., \& {Wheeler}, Caleb~H., I. 2022, \aj, 164, 256, \dodoi{10.3847/1538-3881/ac9bfa}

\bibitem[{{Holmberg} \& {Madhusudhan}(2023)}]{Holmberg2023_NIRISS}
{Holmberg}, M., \& {Madhusudhan}, N. 2023, \mnras, 524, 377, \dodoi{10.1093/mnras/stad1580}

\bibitem[{{Jaeger} {et~al.}(1998){Jaeger}, {Molster}, {Dorschner}, {Henning}, {Mutschke}, \& {Waters}}]{Jaeger1998RIMgSiO3}
{Jaeger}, C., {Molster}, F.~J., {Dorschner}, J., {et~al.} 1998, \aap, 339, 904

\bibitem[{{Jaeger} {et~al.}(1994){Jaeger}, {Mutschke}, {Begemann}, {Dorschner}, \& {Henning}}]{Jaeger1994RIMgSiO3}
{Jaeger}, C., {Mutschke}, H., {Begemann}, B., {Dorschner}, J., \& {Henning}, T. 1994, \aap, 292, 641

\bibitem[{{J{\"a}ger} {et~al.}(2003){J{\"a}ger}, {Dorschner}, {Mutschke}, {Posch}, \& {Henning}}]{Jager2003RIMgSiO3}
{J{\"a}ger}, C., {Dorschner}, J., {Mutschke}, H., {Posch}, T., \& {Henning}, T. 2003, \aap, 408, 193, \dodoi{10.1051/0004-6361:20030916}

\bibitem[{{JWST Transiting Exoplanet Community Early Release Science Team} {et~al.}(2023){JWST Transiting Exoplanet Community Early Release Science Team}, {Ahrer}, {Alderson}, {Batalha}, {Batalha}, {Bean}, {Beatty}, {Bell}, {Benneke}, {Berta-Thompson}, {Carter}, {Crossfield}, {Espinoza}, {Feinstein}, {Fortney}, {Gibson}, {Goyal}, {Kempton}, {Kirk}, {Kreidberg}, {L{\'o}pez-Morales}, {Line}, {Lothringer}, {Moran}, {Mukherjee}, {Ohno}, {Parmentier}, {Piaulet}, {Rustamkulov}, {Schlawin}, {Sing}, {Stevenson}, {Wakeford}, {Allen}, {Birkmann}, {Brande}, {Crouzet}, {Cubillos}, {Damiano}, {D{\'e}sert}, {Gao}, {Harrington}, {Hu}, {Kendrew}, {Knutson}, {Lagage}, {Leconte}, {Lendl}, {MacDonald}, {May}, {Miguel}, {Molaverdikhani}, {Moses}, {Murray}, {Nehring}, {Nikolov}, {Petit dit de la Roche}, {Radica}, {Roy}, {Stassun}, {Taylor}, {Waalkes}, {Wachiraphan}, {Welbanks}, {Wheatley}, {Aggarwal}, {Alam}, {Banerjee}, {Barstow}, {Blecic}, {Casewell}, {Changeat}, {Chubb}, {Col{\'o}n}, {Coulombe}, {Daylan}, {de Val-Borro},
  {Decin}, {Dos Santos}, {Flagg}, {France}, {Fu}, {Garc{\'\i}a Mu{\~n}oz}, {Gizis}, {Glidden}, {Grant}, {Heng}, {Henning}, {Hong}, {Inglis}, {Iro}, {Kataria}, {Komacek}, {Krick}, {Lee}, {Lewis}, {Lillo-Box}, {Lustig-Yaeger}, {Mancini}, {Mandell}, {Mansfield}, {Marley}, {Mikal-Evans}, {Morello}, {Nixon}, {Ortiz Ceballos}, {Piette}, {Powell}, {Rackham}, {Ramos-Rosado}, {Rauscher}, {Redfield}, {Rogers}, {Roman}, {Roudier}, {Scarsdale}, {Shkolnik}, {Southworth}, {Spake}, {Steinrueck}, {Tan}, {Teske}, {Tremblin}, {Tsai}, {Tucker}, {Turner}, {Valenti}, {Venot}, {Waldmann}, {Wallack}, {Zhang}, \& {Zieba}}]{2023Natur.614..649J}
{JWST Transiting Exoplanet Community Early Release Science Team}, {Ahrer}, E.-M., {Alderson}, L., {et~al.} 2023, \nat, 614, 649, \dodoi{10.1038/s41586-022-05269-w}

\bibitem[{{Karman} {et~al.}(2019){Karman}, {Gordon}, {van der Avoird}, {Baranov}, {Boulet}, {Drouin}, {Groenenboom}, {Gustafsson}, {Hartmann}, {Kurucz}, {Rothman}, {Sun}, {Sung}, {Thalman}, {Tran}, {Wishnow}, {Wordsworth}, {Vigasin}, {Volkamer}, \& {van der Zande}}]{Karman2019hitran}
{Karman}, T., {Gordon}, I.~E., {van der Avoird}, A., {et~al.} 2019, \icarus, 328, 160, \dodoi{10.1016/j.icarus.2019.02.034}

\bibitem[{{Kempton} {et~al.}(2023){Kempton}, {Zhang}, {Bean}, {Steinrueck}, {Piette}, {Parmentier}, {Malsky}, {Roman}, {Rauscher}, {Gao}, {Bell}, {Xue}, {Taylor}, {Savel}, {Arnold}, {Nixon}, {Stevenson}, {Mansfield}, {Kendrew}, {Zieba}, {Ducrot}, {Dyrek}, {Lagage}, {Stassun}, {Henry}, {Barman}, {Lupu}, {Malik}, {Kataria}, {Ih}, {Fu}, {Welbanks}, \& {McGill}}]{2023Natur.620...67K}
{Kempton}, E. M.~R., {Zhang}, M., {Bean}, J.~L., {et~al.} 2023, \nat, 620, 67, \dodoi{10.1038/s41586-023-06159-5}

\bibitem[{{Khorshid} {et~al.}(2024){Khorshid}, {Min}, {Polman}, \& {Waters}}]{2024A&A...685A..64K}
{Khorshid}, N., {Min}, M., {Polman}, J., \& {Waters}, L.~B.~F.~M. 2024, \aap, 685, A64, \dodoi{10.1051/0004-6361/202347124}

\bibitem[{{Kirk} {et~al.}(2019){Kirk}, {L{\'o}pez-Morales}, {Wheatley}, {Weaver}, {Skillen}, {Louden}, {McCormac}, \& {Espinoza}}]{Kirk2019WASP39b}
{Kirk}, J., {L{\'o}pez-Morales}, M., {Wheatley}, P.~J., {et~al.} 2019, \aj, 158, 144, \dodoi{10.3847/1538-3881/ab397d}

\bibitem[{{Kitzmann} {et~al.}(2020){Kitzmann}, {Heng}, {Oreshenko}, {Grimm}, {Apai}, {Bowler}, {Burgasser}, \& {Marley}}]{Kitzmann2020helios}
{Kitzmann}, D., {Heng}, K., {Oreshenko}, M., {et~al.} 2020, \apj, 890, 174, \dodoi{10.3847/1538-4357/ab6d71}

\bibitem[{Lam {et~al.}(2015)Lam, Pitrou, \& Seibert}]{lam2015numba}
Lam, S.~K., Pitrou, A., \& Seibert, S. 2015, in Proceedings of the Second Workshop on the LLVM Compiler Infrastructure in HPC, 1--6

\bibitem[{{Lendl} {et~al.}(2017){Lendl}, {Cubillos}, {Hagelberg}, {M{\"u}ller}, {Juvan}, \& {Fossati}}]{2017A&A...606A..18L}
{Lendl}, M., {Cubillos}, P.~E., {Hagelberg}, J., {et~al.} 2017, \aap, 606, A18, \dodoi{10.1051/0004-6361/201731242}

\bibitem[{{Lendl} {et~al.}(2016){Lendl}, {Delrez}, {Gillon}, {Madhusudhan}, {Jehin}, {Queloz}, {Anderson}, {Demory}, \& {Hellier}}]{2016A&A...587A..67L}
{Lendl}, M., {Delrez}, L., {Gillon}, M., {et~al.} 2016, \aap, 587, A67, \dodoi{10.1051/0004-6361/201527594}

\bibitem[{{Lodders}(2021)}]{Lodders2021solar_photo_abund}
{Lodders}, K. 2021, \ssr, 217, 44, \dodoi{10.1007/s11214-021-00825-8}

\bibitem[{{Louca} {et~al.}(2023){Louca}, {Miguel}, \& {Kubyshkina}}]{2023ApJ...956L..19L}
{Louca}, A.~J., {Miguel}, Y., \& {Kubyshkina}, D. 2023, \apjl, 956, L19, \dodoi{10.3847/2041-8213/acfaec}

\bibitem[{{Lueber} {et~al.}(2024){Lueber}, {Novais}, {Fisher}, \& {Heng}}]{Lueber2024WASP39b}
{Lueber}, A., {Novais}, A., {Fisher}, C., \& {Heng}, K. 2024, arXiv e-prints, arXiv:2405.02656, \dodoi{10.48550/arXiv.2405.02656}

\bibitem[{{Ma} {et~al.}(2023){Ma}, {Ito}, {Al-Refaie}, {Changeat}, {Edwards}, \& {Tinetti}}]{Ma2023YunMa}
{Ma}, S., {Ito}, Y., {Al-Refaie}, A.~F., {et~al.} 2023, \apj, 957, 104, \dodoi{10.3847/1538-4357/acf8ca}

\bibitem[{{Maciejewski} {et~al.}(2016){Maciejewski}, {Dimitrov}, {Mancini}, {Southworth}, {Ciceri}, {D'Ago}, {Bruni}, {Raetz}, {Nowak}, {Ohlert}, {Puchalski}, {Saral}, {Derman}, {Petrucci}, {Jofre}, {Seeliger}, \& {Henning}}]{Maciejewski2016WASP39b}
{Maciejewski}, G., {Dimitrov}, D., {Mancini}, L., {et~al.} 2016, \actaa, 66, 55, \dodoi{10.48550/arXiv.1603.03268}

\bibitem[{{Mancini} {et~al.}(2018){Mancini}, {Esposito}, {Covino}, {Southworth}, {Biazzo}, {Bruni}, {Ciceri}, {Evans}, {Lanza}, {Poretti}, {Sarkis}, {Smith}, {Brogi}, {Affer}, {Benatti}, {Bignamini}, {Boccato}, {Bonomo}, {Borsa}, {Carleo}, {Claudi}, {Cosentino}, {Damasso}, {Desidera}, {Giacobbe}, {Gonz{\'a}lez-{\'A}lvarez}, {Gratton}, {Harutyunyan}, {Leto}, {Maggio}, {Malavolta}, {Maldonado}, {Martinez-Fiorenzano}, {Masiero}, {Micela}, {Molinari}, {Nascimbeni}, {Pagano}, {Pedani}, {Piotto}, {Rainer}, {Scandariato}, {Smareglia}, {Sozzetti}, {Andreuzzi}, \& {Henning}}]{Mancini2018WASP39b}
{Mancini}, L., {Esposito}, M., {Covino}, E., {et~al.} 2018, \aap, 613, A41, \dodoi{10.1051/0004-6361/201732234}

\bibitem[{{McKemmish} {et~al.}(2019){McKemmish}, {Masseron}, {Hoeijmakers}, {P{\'e}rez-Mesa}, {Grimm}, {Yurchenko}, \& {Tennyson}}]{McKemmish2019TiO}
{McKemmish}, L.~K., {Masseron}, T., {Hoeijmakers}, H.~J., {et~al.} 2019, \mnras, 488, 2836, \dodoi{10.1093/mnras/stz1818}

\bibitem[{{McKemmish} {et~al.}(2016){McKemmish}, {Yurchenko}, \& {Tennyson}}]{McKemmish2016VO}
{McKemmish}, L.~K., {Yurchenko}, S.~N., \& {Tennyson}, J. 2016, \mnras, 463, 771, \dodoi{10.1093/mnras/stw1969}

\bibitem[{{Mugnai} {et~al.}(2024){Mugnai}, {Swain}, {Estrela}, \& {Roudier}}]{Mugnai2024population}
{Mugnai}, L.~V., {Swain}, M.~R., {Estrela}, R., \& {Roudier}, G.~M. 2024, \mnras, \dodoi{10.1093/mnras/stae1073}

\bibitem[{{Nikolov} {et~al.}(2016){Nikolov}, {Sing}, {Gibson}, {Fortney}, {Evans}, {Barstow}, {Kataria}, \& {Wilson}}]{Nikolov2016WASP39b}
{Nikolov}, N., {Sing}, D.~K., {Gibson}, N.~P., {et~al.} 2016, \apj, 832, 191, \dodoi{10.3847/0004-637X/832/2/191}

\bibitem[{Niraula {et~al.}(2023)Niraula, de~Wit, Gordon, Hargreaves, \& Sousa-Silva}]{Niraula2023WASP39b}
Niraula, P., de~Wit, J., Gordon, I.~E., Hargreaves, R.~J., \& Sousa-Silva, C. 2023, The Astrophysical Journal Letters, 950, L17, \dodoi{10.3847/2041-8213/acd6f8}

\bibitem[{Novais {et~al.}(2025)Novais, Fisher, Ghezzi, Kitzmann, Thorsbro, \& Heng}]{Novais2025paramdegene}
Novais, A., Fisher, C., Ghezzi, L., {et~al.} 2025, Monthly Notices of the Royal Astronomical Society, 538, 2521, \dodoi{10.1093/mnras/staf397}

\bibitem[{{Owens} {et~al.}(2021){Owens}, {Tennyson}, \& {Yurchenko}}]{Owen2021NaOHKOH}
{Owens}, A., {Tennyson}, J., \& {Yurchenko}, S.~N. 2021, \mnras, 502, 1128, \dodoi{10.1093/mnras/staa4041}

\bibitem[{{Pinhas} {et~al.}(2019){Pinhas}, {Madhusudhan}, {Gandhi}, \& {MacDonald}}]{Pinhas2019hot-jupiterH2O}
{Pinhas}, A., {Madhusudhan}, N., {Gandhi}, S., \& {MacDonald}, R. 2019, \mnras, 482, 1485, \dodoi{10.1093/mnras/sty2544}

\bibitem[{{Pinhas} {et~al.}(2018){Pinhas}, {Rackham}, {Madhusudhan}, \& {Apai}}]{Pinhas2018WASP39b}
{Pinhas}, A., {Rackham}, B.~V., {Madhusudhan}, N., \& {Apai}, D. 2018, \mnras, 480, 5314, \dodoi{10.1093/mnras/sty2209}

\bibitem[{{Polanski} {et~al.}(2022){Polanski}, {Crossfield}, {Howard}, {Isaacson}, \& {Rice}}]{Polanski2022CO}
{Polanski}, A.~S., {Crossfield}, I. J.~M., {Howard}, A.~W., {Isaacson}, H., \& {Rice}, M. 2022, Research Notes of the American Astronomical Society, 6, 155, \dodoi{10.3847/2515-5172/ac8676}

\bibitem[{Polyansky {et~al.}(2018)Polyansky, Kyuberis, Zobov, Tennyson, Yurchenko, \& Lodi}]{Polyansky2018pokazatel}
Polyansky, O.~L., Kyuberis, A.~A., Zobov, N.~F., {et~al.} 2018, \mnras, 480, 2597, \dodoi{10.1093/mnras/sty1877}

\bibitem[{{Powell} {et~al.}(2024){Powell}, {Feinstein}, {Lee}, {Zhang}, {Tsai}, {Taylor}, {Kirk}, {Bell}, {Barstow}, {Gao}, {Bean}, {Blecic}, {Chubb}, {Crossfield}, {Jordan}, {Kitzmann}, {Moran}, {Morello}, {Moses}, {Welbanks}, {Yang}, {Zhang}, {Ahrer}, {Bello-Arufe}, {Brande}, {Casewell}, {Crouzet}, {Cubillos}, {Demory}, {Dyrek}, {Flagg}, {Hu}, {Inglis}, {Jones}, {Kreidberg}, {L{\'o}pez-Morales}, {Lagage}, {Meier Vald{\'e}s}, {Miguel}, {Parmentier}, {Piette}, {Rackham}, {Radica}, {Redfield}, {Stevenson}, {Wakeford}, {Aggarwal}, {Alam}, {Batalha}, {Batalha}, {Benneke}, {Berta-Thompson}, {Brady}, {Caceres}, {Carter}, {D{\'e}sert}, {Harrington}, {Iro}, {Line}, {Lothringer}, {MacDonald}, {Mancini}, {Molaverdikhani}, {Mukherjee}, {Nixon}, {Oza}, {Palle}, {Rustamkulov}, {Sing}, {Steinrueck}, {Venot}, {Wheatley}, \& {Yurchenko}}]{Powell2024WASP39bMIRI}
{Powell}, D., {Feinstein}, A.~D., {Lee}, E. K.~H., {et~al.} 2024, \nat, 626, 979, \dodoi{10.1038/s41586-024-07040-9}

\bibitem[{{Prajapat} {et~al.}(2017){Prajapat}, {Jagoda}, {Lodi}, {Gorman}, {Yurchenko}, \& {Tennyson}}]{Prajapat2017POPS}
{Prajapat}, L., {Jagoda}, P., {Lodi}, L., {et~al.} 2017, \mnras, 472, 3648, \dodoi{10.1093/mnras/stx2229}

\bibitem[{{Radica}(2024)}]{Radica2024exoTEDR}
{Radica}, M. 2024, The Journal of Open Source Software, 9, 6898, \dodoi{10.21105/joss.06898}

\bibitem[{{Radica} {et~al.}(2023){Radica}, {Welbanks}, {Espinoza}, {Taylor}, {Coulombe}, {Feinstein}, {Goyal}, {Scarsdale}, {Albert}, {Baghel}, {Bean}, {Blecic}, {Lafreni{\`e}re}, {MacDonald}, {Zamyatina}, {Allart1}, {Artigau}, {Batalha}, {Cook}, {Cowan}, {Dang}, {Doyon}, {Fournier-Tondreau}, {Johnstone}, {Line}, {Moran}, {Mukherjee}, {Pelletier}, {Roy}, {Talens}, {Filippazzo}, {Pontoppidan}, \& {Volk}}]{Radica2023exoTEDRF}
{Radica}, M., {Welbanks}, L., {Espinoza}, N., {et~al.} 2023, \mnras, 524, 835, \dodoi{10.1093/mnras/stad1762}

\bibitem[{{Rigby} {et~al.}(2023){Rigby}, {Perrin}, {McElwain}, {Kimble}, {Friedman}, {Lallo}, {Doyon}, {Feinberg}, {Ferruit}, {Glasse}, {Rieke}, {Rieke}, {Wright}, {Willott}, {Colon}, {Milam}, {Neff}, {Stark}, {Valenti}, {Abell}, {Abney}, {Abul-Huda}, {Acton}, {Adams}, {Adler}, {Aguilar}, {Ahmed}, {Albert}, {Alberts}, {Aldridge}, {Allen}, {Altenburg}, {{\'A}lvarez-M{\'a}rquez}, {Alves de Oliveira}, {Andersen}, {Anderson}, {Anderson}, {Argyriou}, {Armstrong}, {Arribas}, {Artigau}, {Arvai}, {Atkinson}, {Bacon}, {Bair}, {Banks}, {Barrientes}, {Barringer}, {Bartosik}, {Bast}, {Baudoz}, {Beatty}, {Bechtold}, {Beck}, {Bergeron}, {Bergkoetter}, {Bhatawdekar}, {Birkmann}, {Blazek}, {Blome}, {Boccaletti}, {B{\"o}ker}, {Boia}, {Bonaventura}, {Bond}, {Bosley}, {Boucarut}, {Bourque}, {Bouwman}, {Bower}, {Bowers}, {Boyer}, {Bradley}, {Brady}, {Braun}, {Breda}, {Bresnahan}, {Bright}, {Britt}, {Bromenschenkel}, {Brooks}, {Brooks}, {Brown}, {Brown}, {Brown}, {Bunker}, {Burger}, {Bushouse}, {Cale}, {Cameron}, {Cameron},
  {Canipe}, {Caplinger}, {Caputo}, {Cara}, {Carey}, {Carniani}, {Carrasquilla}, {Carruthers}, {Case}, {Catherine}, {Chance}, {Chapman}, {Charlot}, {Charlow}, {Chayer}, {Chen}, {Cherinka}, {Chichester}, {Chilton}, {Chonis}, {Clampin}, {Clark}, {Clark}, {Coe}, {Coleman}, {Comber}, {Comeau}, {Connolly}, {Cooper}, {Cooper}, {Coppock}, {Correnti}, {Cossou}, {Coulais}, {Coyle}, {Cracraft}, {Curti}, {Cuturic}, {Davis}, {Davis}, {Dean}, {DeLisa}, {deMeester}, {Dencheva}, {Dencheva}, {DePasquale}, {Deschenes}, {Hunor Detre}, {Diaz}, {Dicken}, {DiFelice}, {Dillman}, {Dixon}, {Doggett}, {Donaldson}, {Douglas}, {DuPrie}, {Dupuis}, {Durning}, {Easmin}, {Eck}, {Edeani}, {Egami}, {Ehrenwinkler}, {Eisenhamer}, {Eisenhower}, {Elie}, {Elliott}, {Elliott}, {Ellis}, {Engesser}, {Espinoza}, {Etienne}, {Etxaluze}, {Falini}, {Feeney}, {Ferry}, {Filippazzo}, {Fincham}, {Fix}, {Flagey}, {Florian}, {Flynn}, {Fontanella}, {Ford}, {Forshay}, {Fox}, {Franz}, {Fu}, {Fullerton}, {Galkin}, {Galyer}, {Garc{\'\i}a Mar{\'\i}n}, {Gardner},
  {Gardner}, {Garland}, {Garrett}, {Gasman}, {Gaspar}, {Gaudreau}, {Gauthier}, {Geers}, {Geithner}, {Gennaro}, {Giardino}, {Girard}, {Giuliano}, {Glassmire}, {Glauser}, {Glazer}, {Godfrey}, {Golimowski}, {Gollnitz}, {Gong}, {Gonzaga}, {Gordon}, {Gordon}, {Goudfrooij}, {Greene}, {Greenhouse}, {Grimaldi}, {Groebner}, {Grundy}, {Guillard}, {Gutman}, {Ha}, {Haderlein}, {Hagedorn}, {Hainline}, {Haley}, {Hami}, {Hamilton}, {Hammel}, {Hansen}, {Harkins}, {Harr}, {Hart}, {Hart}, {Hartig}, {Hashimoto}, {Haskins}, {Hathaway}, {Havey}, {Hayden}, {Hecht}, {Heller-Boyer}, {Henriques}, {Henry}, {Hermann}, {Hernandez}, {Hesman}, {Hicks}, {Hilbert}, {Hines}, {Hoffman}, {Holfeltz}, {Holler}, {Hoppa}, {Hott}, {Howard}, {Howard}, {Hunter}, {Hunter}, {Hurst}, {Husemann}, {Hustak}, {Ilinca Ignat}, {Illingworth}, {Irish}, {Jackson}, {Jahromi}, {Jakobsen}, {James}, {James}, {Januszewski}, {Jenkins}, {Jirdeh}, {Johnson}, {Johnson}, {Jones}, {Jones}, {Jones}, {Jones}, {Jordan}, {Jordan}, {Jurczyk}, {Jurling}, {Kaleida}, {Kalmanson},
  {Kammerer}, {Kang}, {Kao}, {Karakla}, {Kavanagh}, {Kelly}, {Kendrew}, {Kennedy}, {Kenny}, {Keski-kuha}, {Keyes}, {Kidwell}, {Kinzel}, {Kirk}, {Kirkpatrick}, {Kirshenblat}, {Klaassen}, {Knapp}, {Knight}, {Knollenberg}, {Koehler}, {Koekemoer}, {Kovacs}, {Kulp}, {Kumari}, {Kyprianou}, {La Massa}, {Labador}, {Labiano}, {Lagage}, {Lajoie}, {Lallo}, {Lam}, {Lamb}, {Lambros}, {Lampenfield}, {Langston}, {Larson}, {Law}, {Lawrence}, {Lee}, {Leisenring}, {Lepo}, {Leveille}, {Levenson}, {Levine}, {Levy}, {Lewis}, {Lewis}, {Libralato}, {Lightsey}, {Link}, {Liu}, {Lo}, {Lockwood}, {Logue}, {Long}, {Long}, {Loomis}, {Lopez-Caniego}, {Lorenzo Alvarez}, {Love-Pruitt}, {Lucy}, {Luetzgendorf}, {Maghami}, {Maiolino}, {Major}, {Malla}, {Malumuth}, {Manjavacas}, {Mannfolk}, {Marrione}, {Marston}, {Martel}, {Maschmann}, {Masci}, {Masciarelli}, {Maszkiewicz}, {Mather}, {McKenzie}, {McLean}, {McMaster}, {Melbourne}, {Mel{\'e}ndez}, {Menzel}, {Merz}, {Meyett}, {Meza}, {Miskey}, {Misselt}, {Moller}, {Morrison}, {Morse}, {Moseley},
  {Mosier}, {Mountain}, {Mueckay}, {Mueller}, {Mullally}, {Murphy}, {Murray}, {Murray}, {Mustelier}, {Muzerolle}, {Mycroft}, {Myers}, {Myrick}, {Nanavati}, {Nance}, {Nayak}, {Naylor}, {Nelan}, {Nickson}, {Nielson}, {Nieto-Santisteban}, {Nikolov}, {Noriega-Crespo}, {O'Shaughnessy}, {O'Sullivan}, {Ochs}, {Ogle}, {Oleszczuk}, {Olmsted}, {Osborne}, {Ottens}, {Owens}, {Pacifici}, {Pagan}, {Page}, {Park}, {Parrish}, {Patapis}, {Paul}, {Pauly}, {Pavlovsky}, {Pedder}, {Peek}, {Pena-Guerrero}, {Penanen}, {Perez}, {Perna}, {Perriello}, {Phillips}, {Pietraszkiewicz}, {Pinaud}, {Pirzkal}, {Pitman}, {Piwowar}, {Platais}, {Player}, {Plesha}, {Pollizi}, {Polster}, {Pontoppidan}, {Porterfield}, {Proffitt}, {Pueyo}, {Pulliam}, {Quirt}, {Quispe Neira}, {Ramos Alarcon}, {Ramsay}, {Rapp}, {Rapp}, {Rauscher}, {Ravindranath}, {Rawle}, {Regan}, {Reichard}, {Reis}, {Ressler}, {Rest}, {Reynolds}, {Rhue}, {Richon}, {Rickman}, {Ridgaway}, {Ritchie}, {Rix}, {Robberto}, {Robinson}, {Robinson}, {Robinson}, {Rock}, {Rodriguez}, {Rodriguez
  Del Pino}, {Roellig}, {Rohrbach}, {Roman}, {Romelfanger}, {Rose}, {Roteliuk}, {Roth}, {Rothwell}, {Rowlands}, {Roy}, {Royer}, {Royle}, {Rui}, {Rumler}, {Runnels}, {Russ}, {Rustamkulov}, {Ryden}, {Ryer}, {Sabata}, {Sabatke}, {Sabbi}, {Samuelson}, {Sapp}, {Sappington}, {Sargent}, {Sauer}, {Scheithauer}, {Schlawin}, {Schlitz}, {Schmitz}, {Schneider}, {Schreiber}, {Schulze}, {Schwab}, {Scott}, {Sembach}, {Shanahan}, {Shaughnessy}, {Shaw}, {Shawger}, {Shay}, {Sheehan}, {Shen}, {Sherman}, {Shiao}, {Shih}, {Shivaei}, {Sienkiewicz}, {Sing}, {Sirianni}, {Sivaramakrishnan}, {Skipper}, {Sloan}, {Slocum}, {Slowinski}, {Smith}, {Smith}, {Smith}, {Smith}, {Snyder}, {Soh}, {Sohn}, {Soto}, {Spencer}, {Stallcup}, {Stansberry}, {Starr}, {Starr}, {Stewart}, {Stiavelli}, {Straughn}, {Strickland}, {Stys}, {Summers}, {Sun}, {Sunnquist}, {Swade}, {Swam}, {Swaters}, {Swoish}, {Taylor}, {Taylor}, {Te Plate}, {Tea}, {Teague}, {Telfer}, {Temim}, {Thatte}, {Thompson}, {Thompson}, {Thomson}, {Tikkanen}, {Tippet}, {Todd}, {Toolan},
  {Tran}, {Trejo}, {Truong}, {Tsukamoto}, {Tustain}, {Tyra}, {Ubeda}, {Underwood}, {Uzzo}, {Van Campen}, {Vandal}, {Vandenbussche}, {Vila}, {Volk}, {Wahlgren}, {Waldman}, {Walker}, {Wander}, {Warfield}, {Warner}, {Wasiak}, {Watkins}, {Weaver}, {Weilert}, {Weiser}, {Weiss}, {Weissman}, {Welty}, {West}, {Wheate}, {Wheatley}, {Wheeler}, {White}, {Whiteaker}, {Whitehouse}, {Whiteleather}, {Whitman}, {Williams}, {Willmer}, {Willoughby}, {Wilson}, {Wirth}, {Wislowski}, {Wolf}, {Wolfe}, {Wolff}, {Workman}, {Wright}, {Wu}, {Wu}, {Wymer}, {Yates}, {Yeager}, {Yeates}, {Yerger}, {Yoon}, {Young}, {Yu}, {Zak}, {Zeidler}, {Zhou}, {Zielinski}, {Zincke}, \& {Zonak}}]{2023PASP..135d8001R}
{Rigby}, J., {Perrin}, M., {McElwain}, M., {et~al.} 2023, \pasp, 135, 048001, \dodoi{10.1088/1538-3873/acb293}

\bibitem[{{Rivlin} {et~al.}(2015){Rivlin}, {Lodi}, {Yurchenko}, {Tennyson}, \& {Le Roy}}]{Rivlin2015NaH}
{Rivlin}, T., {Lodi}, L., {Yurchenko}, S.~N., {Tennyson}, J., \& {Le Roy}, R.~J. 2015, \mnras, 451, 634, \dodoi{10.1093/mnras/stv979}

\bibitem[{{Rocchetto} {et~al.}(2016){Rocchetto}, {Waldmann}, {Venot}, {Lagage}, \& {Tinetti}}]{Rocchetto2016Tp}
{Rocchetto}, M., {Waldmann}, I.~P., {Venot}, O., {Lagage}, P.~O., \& {Tinetti}, G. 2016, \apj, 833, 120, \dodoi{10.3847/1538-4357/833/1/120}

\bibitem[{{Roy-Perez} {et~al.}(2025){Roy-Perez}, {P{\'e}rez-Hoyos}, {Barrado-Izagirre}, \& {Chen-Chen}}]{Roy-Perez2025phAerosol_R}
{Roy-Perez}, J., {P{\'e}rez-Hoyos}, S., {Barrado-Izagirre}, N., \& {Chen-Chen}, H. 2025, arXiv e-prints, arXiv:2501.17728.
\newblock \doarXiv{2501.17728}

\bibitem[{{Rustamkulov} {et~al.}(2022){Rustamkulov}, {Sing}, {Liu}, \& {Wang}}]{2022ApJ...928L...7R}
{Rustamkulov}, Z., {Sing}, D.~K., {Liu}, R., \& {Wang}, A. 2022, \apjl, 928, L7, \dodoi{10.3847/2041-8213/ac5b6f}

\bibitem[{{Rustamkulov} {et~al.}(2023){Rustamkulov}, {Sing}, {Mukherjee}, {May}, {Kirk}, {Schlawin}, {Line}, {Piaulet}, {Carter}, {Batalha}, {Goyal}, {L{\'o}pez-Morales}, {Lothringer}, {MacDonald}, {Moran}, {Stevenson}, {Wakeford}, {Espinoza}, {Bean}, {Batalha}, {Benneke}, {Berta-Thompson}, {Crossfield}, {Gao}, {Kreidberg}, {Powell}, {Cubillos}, {Gibson}, {Leconte}, {Molaverdikhani}, {Nikolov}, {Parmentier}, {Roy}, {Taylor}, {Turner}, {Wheatley}, {Aggarwal}, {Ahrer}, {Alam}, {Alderson}, {Allen}, {Banerjee}, {Barat}, {Barrado}, {Barstow}, {Bell}, {Blecic}, {Brande}, {Casewell}, {Changeat}, {Chubb}, {Crouzet}, {Daylan}, {Decin}, {D{\'e}sert}, {Mikal-Evans}, {Feinstein}, {Flagg}, {Fortney}, {Harrington}, {Heng}, {Hong}, {Hu}, {Iro}, {Kataria}, {Kempton}, {Krick}, {Lendl}, {Lillo-Box}, {Louca}, {Lustig-Yaeger}, {Mancini}, {Mansfield}, {Mayne}, {Miguel}, {Morello}, {Ohno}, {Palle}, {Petit dit de la Roche}, {Rackham}, {Radica}, {Ramos-Rosado}, {Redfield}, {Rogers}, {Shkolnik}, {Southworth}, {Teske}, {Tremblin},
  {Tucker}, {Venot}, {Waalkes}, {Welbanks}, {Zhang}, \& {Zieba}}]{Rustamkulov2023WASP39b}
{Rustamkulov}, Z., {Sing}, D.~K., {Mukherjee}, S., {et~al.} 2023, \nat, 614, 659, \dodoi{10.1038/s41586-022-05677-y}

\bibitem[{{Saba} {et~al.}(2024){Saba}, {Thompson}, {Hou Yip}, {Ma}, {Tsiaras}, {Faris Al-Refaie}, \& {Tinetti}}]{Saba2024STIS}
{Saba}, A., {Thompson}, A., {Hou Yip}, K., {et~al.} 2024, arXiv e-prints, arXiv:2404.15505, \dodoi{10.48550/arXiv.2404.15505}

\bibitem[{{Sarkar} {et~al.}(2024){Sarkar}, {Madhusudhan}, {Constantinou}, \& {Holmberg}}]{Sarkar2024WASP39b}
{Sarkar}, S., {Madhusudhan}, N., {Constantinou}, S., \& {Holmberg}, M. 2024, \mnras, 531, 2731, \dodoi{10.1093/mnras/stae1230}

\bibitem[{{Schleich} {et~al.}(2024){Schleich}, {Boro Saikia}, {Changeat}, {G{\"u}del}, {Voigt}, \& {Waldmann}}]{Schleich2024Tp}
{Schleich}, S., {Boro Saikia}, S., {Changeat}, Q., {et~al.} 2024, \aap, 690, A336, \dodoi{10.1051/0004-6361/202451845}

\bibitem[{{Scott} \& {Duley}(1996)}]{Scott1996RIMgSiO3}
{Scott}, A., \& {Duley}, W.~W. 1996, \apjs, 105, 401, \dodoi{10.1086/192321}

\bibitem[{{Semenov} {et~al.}(2025){Semenov}, {El-Kork}, {Yurchenko}, \& {Tennyson}}]{Semenov2025PN}
{Semenov}, M., {El-Kork}, N., {Yurchenko}, S.~N., \& {Tennyson}, J. 2025, \mnras, 536, 714, \dodoi{10.1093/mnras/stae2610}

\bibitem[{{Sing} {et~al.}(2016){Sing}, {Fortney}, {Nikolov}, {Wakeford}, {Kataria}, {Evans}, {Aigrain}, {Ballester}, {Burrows}, {Deming}, {D{\'e}sert}, {Gibson}, {Henry}, {Huitson}, {Knutson}, {Lecavelier Des Etangs}, {Pont}, {Showman}, {Vidal-Madjar}, {Williamson}, \& {Wilson}}]{Sing2016WASP39b}
{Sing}, D.~K., {Fortney}, J.~J., {Nikolov}, N., {et~al.} 2016, \nat, 529, 59, \dodoi{10.1038/nature16068}

\bibitem[{{Skinner} \& {Wei}(2025)}]{Skinner2025WASP39b}
{Skinner}, J.~W., \& {Wei}, S. 2025, arXiv e-prints, arXiv:2505.01397, \dodoi{10.48550/arXiv.2505.01397}

\bibitem[{{Sousa-Silva} {et~al.}(2014){Sousa-Silva}, {Hesketh}, {Yurchenko}, {Hill}, \& {Tennyson}}]{Sousa-Silva2014PH3}
{Sousa-Silva}, C., {Hesketh}, N., {Yurchenko}, S.~N., {Hill}, C., \& {Tennyson}, J. 2014, \jqsrt, 142, 66, \dodoi{10.1016/j.jqsrt.2014.03.012}

\bibitem[{{Stock} {et~al.}(2018){Stock}, {Kitzmann}, {Patzer}, \& {Sedlmayr}}]{Stock2018Fastchem}
{Stock}, J.~W., {Kitzmann}, D., {Patzer}, A. B.~C., \& {Sedlmayr}, E. 2018, \mnras, 479, 865, \dodoi{10.1093/mnras/sty1531}

\bibitem[{{Swain} {et~al.}(2024){Swain}, {Hasegawa}, {Thorngren}, \& {Roudier}}]{Swain2024metallicitymass}
{Swain}, M.~R., {Hasegawa}, Y., {Thorngren}, D.~P., \& {Roudier}, G.~M. 2024, \ssr, 220, 61, \dodoi{10.1007/s11214-024-01098-7}

\bibitem[{Tada {et~al.}(2025)Tada, Kawahara, Kawashima, Kotani, \& Masuda}]{Tada2025WASP39b}
Tada, S., Kawahara, H., Kawashima, Y., Kotani, T., \& Masuda, K. 2025, The Astronomical Journal, 169, 255, \dodoi{10.3847/1538-3881/adbe65}

\bibitem[{Tennyson {et~al.}(2024)Tennyson, Yurchenko, Zhang, Bowesman, Brady, Buldyreva, Chubb, Gamache, Gorman, Guest, Hill, Kefala, Lynas-Gray, Mellor, McKemmish, Mitev, Mizus, Owens, Peng, Perri, Pezzella, Polyansky, Qu, Semenov, Smola, Solokov, Somogyi, Upadhyay, Wright, \& Zobov}]{jt939}
Tennyson, J., Yurchenko, S.~N., Zhang, J., {et~al.} 2024, J. Quant. Spectrosc. Radiat. Transf., 326, 109083, \dodoi{10.1016/j.jqsrt.2024.109083}

\bibitem[{{Thompson} {et~al.}(2024){Thompson}, {Biagini}, {Cracchiolo}, {Petralia}, {Changeat}, {Saba}, {Morello}, {Morvan}, {Micela}, \& {Tinetti}}]{Thompson2024stellar}
{Thompson}, A., {Biagini}, A., {Cracchiolo}, G., {et~al.} 2024, \apj, 960, 107, \dodoi{10.3847/1538-4357/ad0369}

\bibitem[{{Thorngren} \& {Fortney}(2019)}]{ThorngrenFortney2019WASP39b}
{Thorngren}, D., \& {Fortney}, J.~J. 2019, \apjl, 874, L31, \dodoi{10.3847/2041-8213/ab1137}

\bibitem[{{Tsai} {et~al.}(2023{\natexlab{a}}){Tsai}, {Moses}, {Powell}, \& {Lee}}]{2023ApJ...959L..30T}
{Tsai}, S.-M., {Moses}, J.~I., {Powell}, D., \& {Lee}, E. K.~H. 2023{\natexlab{a}}, \apjl, 959, L30, \dodoi{10.3847/2041-8213/ad1405}

\bibitem[{{Tsai} {et~al.}(2023{\natexlab{b}}){Tsai}, {Lee}, {Powell}, {Gao}, {Zhang}, {Moses}, {H{\'e}brard}, {Venot}, {Parmentier}, {Jordan}, {Hu}, {Alam}, {Alderson}, {Batalha}, {Bean}, {Benneke}, {Bierson}, {Brady}, {Carone}, {Carter}, {Chubb}, {Inglis}, {Leconte}, {Line}, {L{\'o}pez-Morales}, {Miguel}, {Molaverdikhani}, {Rustamkulov}, {Sing}, {Stevenson}, {Wakeford}, {Yang}, {Aggarwal}, {Baeyens}, {Barat}, {de Val-Borro}, {Daylan}, {Fortney}, {France}, {Goyal}, {Grant}, {Kirk}, {Kreidberg}, {Louca}, {Moran}, {Mukherjee}, {Nasedkin}, {Ohno}, {Rackham}, {Redfield}, {Taylor}, {Tremblin}, {Visscher}, {Wallack}, {Welbanks}, {Youngblood}, {Ahrer}, {Batalha}, {Behr}, {Berta-Thompson}, {Blecic}, {Casewell}, {Crossfield}, {Crouzet}, {Cubillos}, {Decin}, {D{\'e}sert}, {Feinstein}, {Gibson}, {Harrington}, {Heng}, {Henning}, {Kempton}, {Krick}, {Lagage}, {Lendl}, {Lothringer}, {Mansfield}, {Mayne}, {Mikal-Evans}, {Palle}, {Schlawin}, {Shorttle}, {Wheatley}, \& {Yurchenko}}]{Tsai2023WASP39b}
{Tsai}, S.-M., {Lee}, E. K.~H., {Powell}, D., {et~al.} 2023{\natexlab{b}}, \nat, 617, 483, \dodoi{10.1038/s41586-023-05902-2}

\bibitem[{{Tsiaras} {et~al.}(2016){Tsiaras}, {Waldmann}, {Rocchetto}, {Varley}, {Morello}, {Damiano}, \& {Tinetti}}]{2016ApJ...832..202T}
{Tsiaras}, A., {Waldmann}, I.~P., {Rocchetto}, M., {et~al.} 2016, \apj, 832, 202, \dodoi{10.3847/0004-637X/832/2/202}

\bibitem[{{Tsiaras} {et~al.}(2018){Tsiaras}, {Waldmann}, {Zingales}, {Rocchetto}, {Morello}, {Damiano}, {Karpouzas}, {Tinetti}, {McKemmish}, {Tennyson}, \& {Yurchenko}}]{Tsiaras2018hot-jupiter}
{Tsiaras}, A., {Waldmann}, I.~P., {Zingales}, T., {et~al.} 2018, \aj, 155, 156, \dodoi{10.3847/1538-3881/aaaf75}

\bibitem[{{Underwood} {et~al.}(2016){Underwood}, {Tennyson}, {Yurchenko}, {Huang}, {Schwenke}, {Lee}, {Clausen}, \& {Fateev}}]{Underwood2016SO2}
{Underwood}, D.~S., {Tennyson}, J., {Yurchenko}, S.~N., {et~al.} 2016, \mnras, 459, 3890, \dodoi{10.1093/mnras/stw849}

\bibitem[{{Upadhyay} {et~al.}(2018){Upadhyay}, {Conway}, {Tennyson}, \& {Yurchenko}}]{Upadhyay2018SiS}
{Upadhyay}, A., {Conway}, E.~K., {Tennyson}, J., \& {Yurchenko}, S.~N. 2018, \mnras, 477, 1520, \dodoi{10.1093/mnras/sty998}

\bibitem[{{Veillet} {et~al.}(2025){Veillet}, {Venot}, {Sirjean}, {Citrangolo Destro}, {Fournet}, {Al-Refaie}, {H{\'e}brard}, {Glaude}, \& {Bounaceur}}]{Veillet2025phCHONS}
{Veillet}, R., {Venot}, O., {Sirjean}, B., {et~al.} 2025, arXiv e-prints, arXiv:2505.12152, \dodoi{10.48550/arXiv.2505.12152}

\bibitem[{Virtanen {et~al.}(2020)Virtanen, Gommers, Oliphant, Haberland, Reddy, Cournapeau, Burovski, Peterson, Weckesser, Bright, {van der Walt}, Brett, Wilson, Millman, Mayorov, Nelson, Jones, Kern, Larson, Carey, Polat, Feng, Moore, {VanderPlas}, Laxalde, Perktold, Cimrman, Henriksen, Quintero, Harris, Archibald, Ribeiro, Pedregosa, {van Mulbregt}, \& {SciPy 1.0 Contributors}}]{Virtanen2020SciPy-NMeth}
Virtanen, P., Gommers, R., Oliphant, T.~E., {et~al.} 2020, Nature Methods, 17, 261, \dodoi{10.1038/s41592-019-0686-2}

\bibitem[{{Visscher} {et~al.}(2010){Visscher}, {Lodders}, \& {Fegley}}]{Visscher2010SVPMgSiO3}
{Visscher}, C., {Lodders}, K., \& {Fegley}, Bruce, J. 2010, \apj, 716, 1060, \dodoi{10.1088/0004-637X/716/2/1060}

\bibitem[{{Wakeford} {et~al.}(2018){Wakeford}, {Sing}, {Deming}, {Lewis}, {Goyal}, {Wilson}, {Barstow}, {Kataria}, {Drummond}, {Evans}, {Carter}, {Nikolov}, {Knutson}, {Ballester}, \& {Mandell}}]{wakeford2018cloudretrieval}
{Wakeford}, H.~R., {Sing}, D.~K., {Deming}, D., {et~al.} 2018, \aj, 155, 29, \dodoi{10.3847/1538-3881/aa9e4e}

\bibitem[{Woitke {et~al.}(2018)Woitke, Helling, Hunter, Millard, Turner, Worters, Blecic, \& Stock}]{Woitke_2018}
Woitke, P., Helling, C., Hunter, G.~H., {et~al.} 2018, Astronomy \& Astrophysics, 614, A1, \dodoi{10.1051/0004-6361/201732193}

\bibitem[{{Yip} {et~al.}(2021){Yip}, {Changeat}, {Edwards}, {Morvan}, {Chubb}, {Tsiaras}, {Waldmann}, \& {Tinetti}}]{yip2021wasp96b}
{Yip}, K.~H., {Changeat}, Q., {Edwards}, B., {et~al.} 2021, \aj, 161, 4, \dodoi{10.3847/1538-3881/abc179}

\bibitem[{{Yurchenko} {et~al.}(2020){Yurchenko}, {Mellor}, {Freedman}, \& {Tennyson}}]{Yurchenko2020CO2}
{Yurchenko}, S.~N., {Mellor}, T.~M., {Freedman}, R.~S., \& {Tennyson}, J. 2020, \mnras, 496, 5282, \dodoi{10.1093/mnras/staa1874}

\bibitem[{{Yurchenko} {et~al.}(2024){Yurchenko}, {Owens}, {Kefala}, \& {Tennyson}}]{Yurchenko2024CH4}
{Yurchenko}, S.~N., {Owens}, A., {Kefala}, K., \& {Tennyson}, J. 2024, \mnras, 528, 3719, \dodoi{10.1093/mnras/stae148}

\bibitem[{{Yurchenko} {et~al.}(2022){Yurchenko}, {Tennyson}, {Syme}, {Adam}, {Clark}, {Cooper}, {Dobney}, {Donnelly}, {Gorman}, {Lynas-Gray}, {Meltzer}, {Owens}, {Qu}, {Semenov}, {Somogyi}, {Upadhyay}, {Wright}, \& {Zapata Trujillo}}]{Yurchenko2022SiO}
{Yurchenko}, S.~N., {Tennyson}, J., {Syme}, A.-M., {et~al.} 2022, \mnras, 510, 903, \dodoi{10.1093/mnras/stab3267}

\bibitem[{{Zeidler} {et~al.}(2015){Zeidler}, {Mutschke}, \& {Posch}}]{Zeidler2015Database}
{Zeidler}, S., {Mutschke}, H., \& {Posch}, T. 2015, \apj, 798, 125, \dodoi{10.1088/0004-637X/798/2/125}

\bibitem[{{Zeidler} {et~al.}(2013){Zeidler}, {Posch}, \& {Mutschke}}]{Zeidler2013SiO2}
{Zeidler}, S., {Posch}, T., \& {Mutschke}, H. 2013, \aap, 553, A81, \dodoi{10.1051/0004-6361/201220459}

\end{thebibliography}

\bibliographystyle{aasjournal}

%% Appendix material should be preceded with a single \appendix command.
%% There should be a \section command for each appendix. Mark appendix
%% subsections with the same markup you use in the main body of the paper.

%% Each Appendix (indicated with \section) will be lettered A, B, C, etc.
%% The equation counter will reset when it encounters the \appendix
%% command and will number appendix equations (A1), (A2), etc. The
%% Figure and Table counter will not reset.
\appendix

\section{Chemical species selection in wide parameter space} \label{appendix:venn}

\begin{figure*}[ht]
\centering
\includegraphics[width=.30\textwidth]{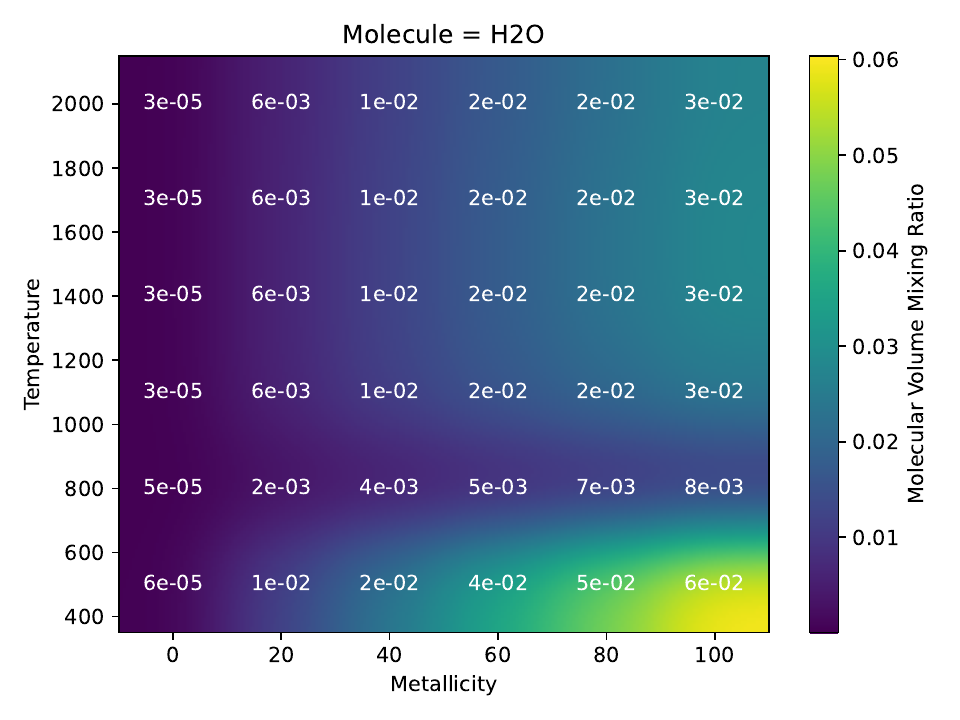}
\includegraphics[width=.30\textwidth]{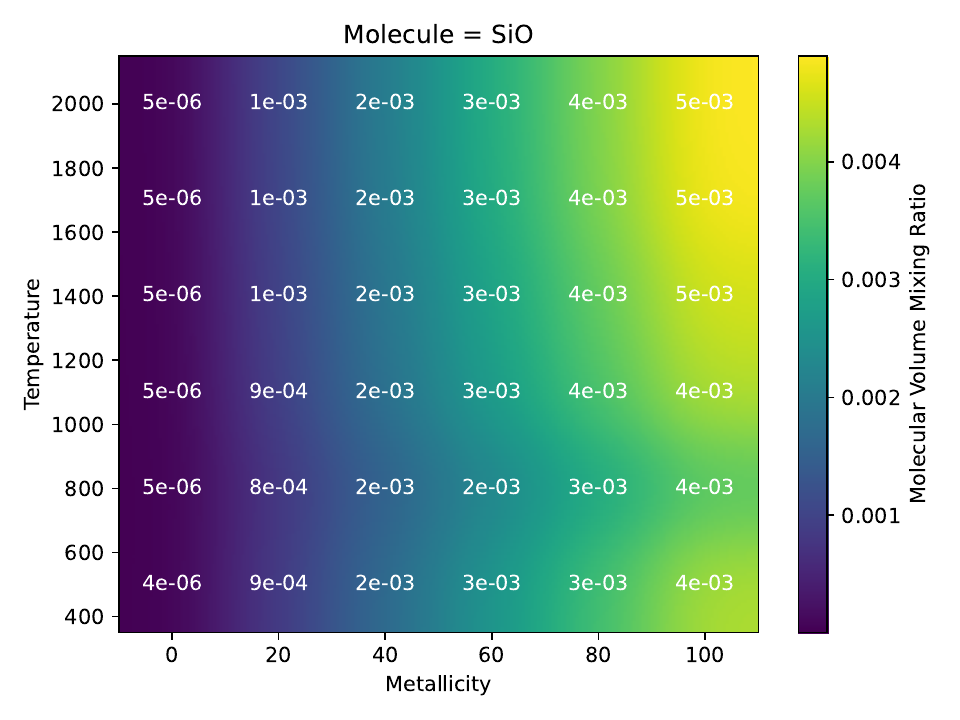}
\includegraphics[width=.30\textwidth]{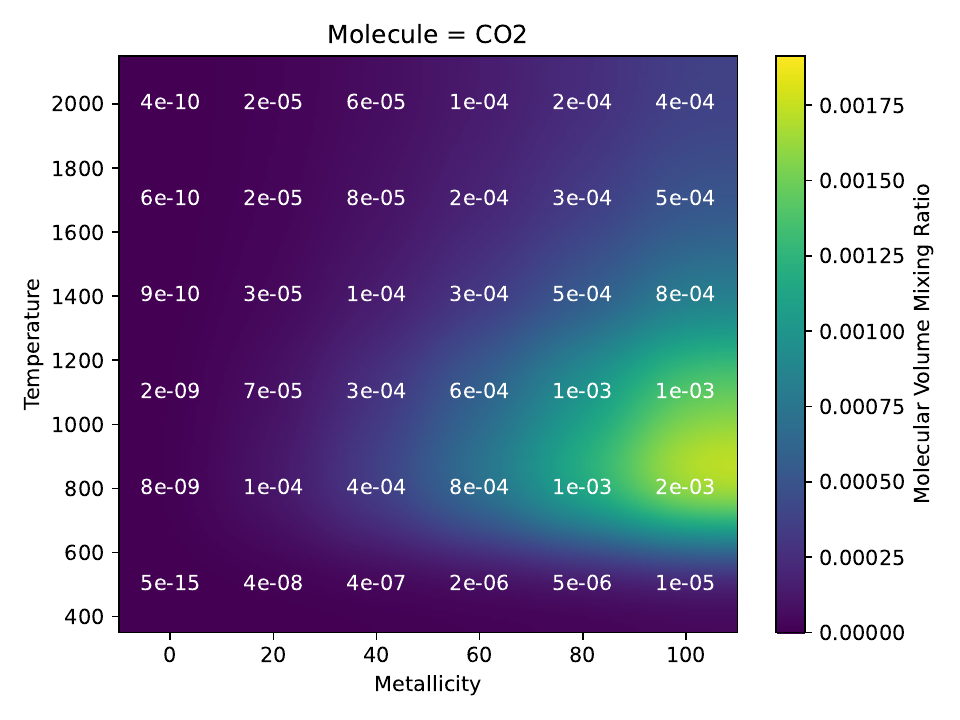}
\\
\includegraphics[width=.30\textwidth]{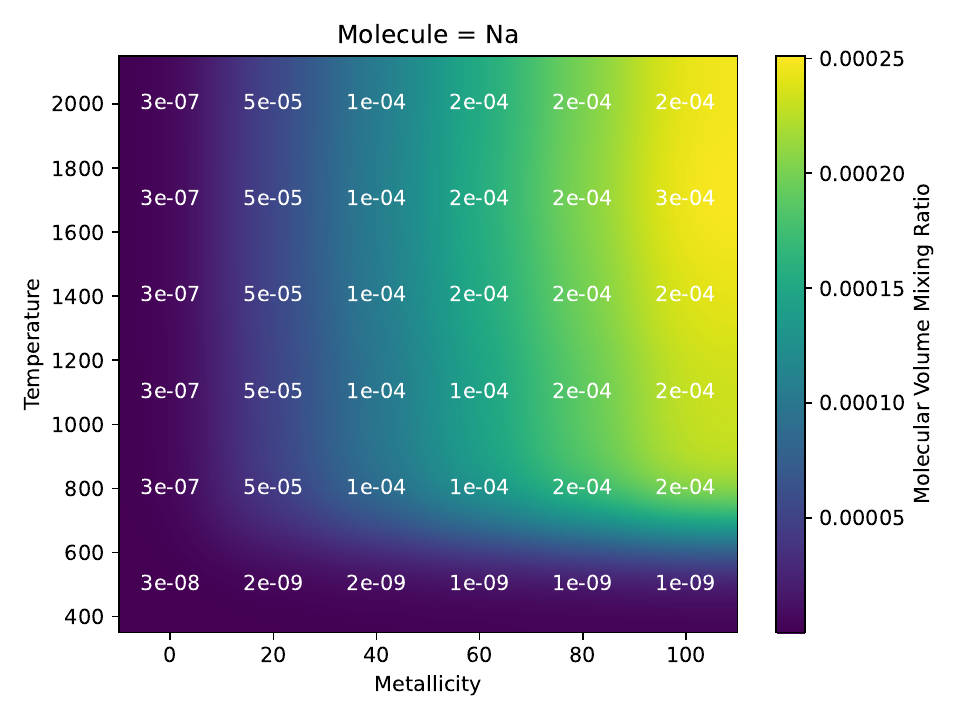}
\includegraphics[width=.30\textwidth]{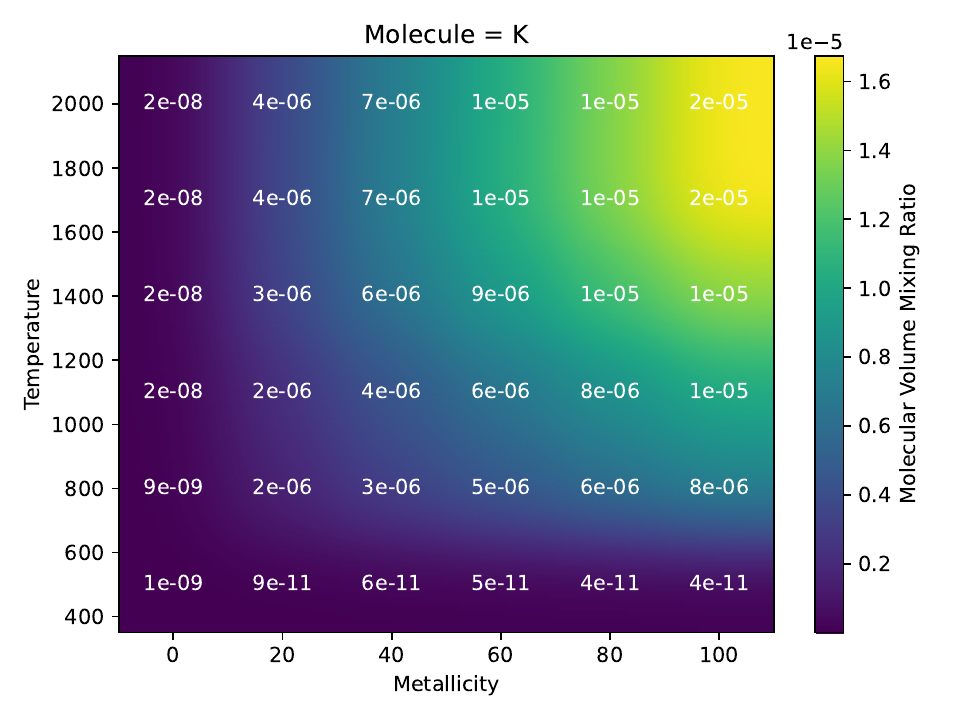}
\includegraphics[width=.30\textwidth]{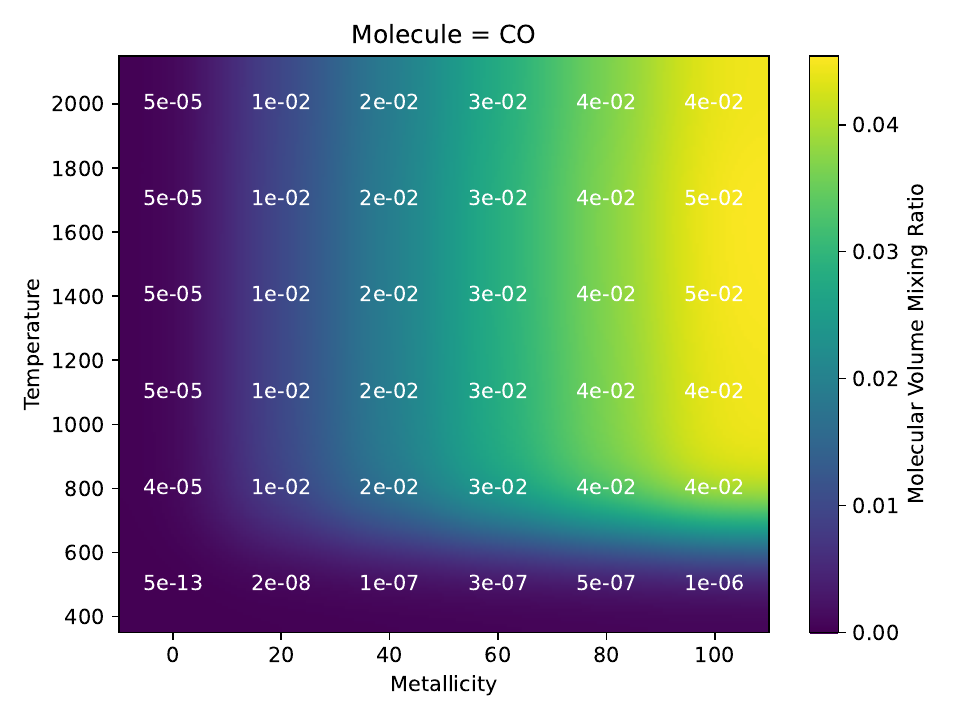}
\\
\includegraphics[width=.30\textwidth]{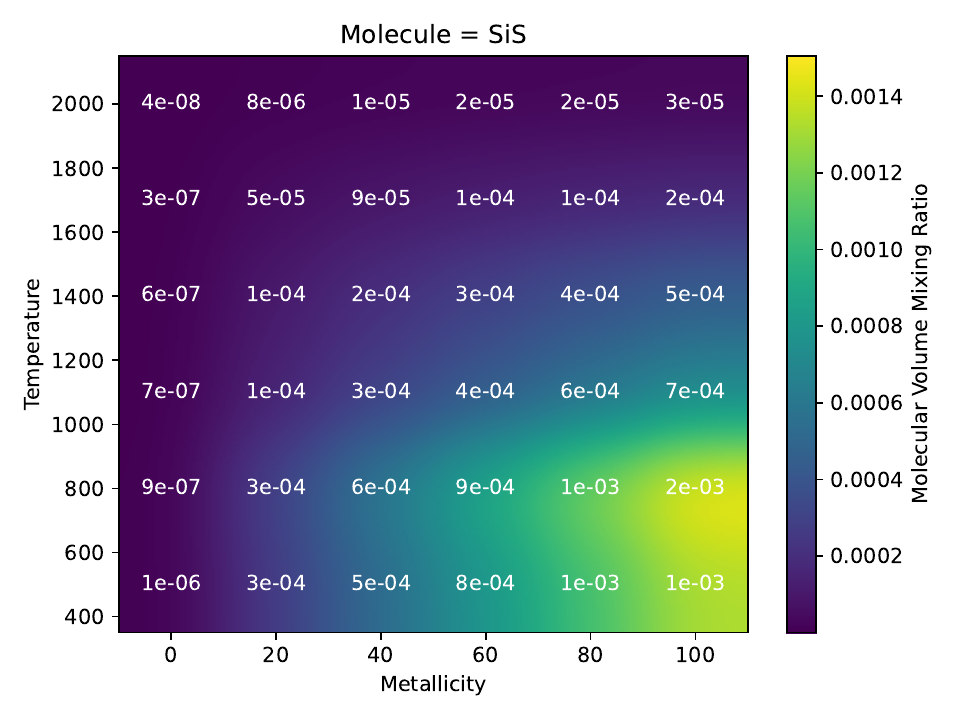}
\includegraphics[width=.30\textwidth]{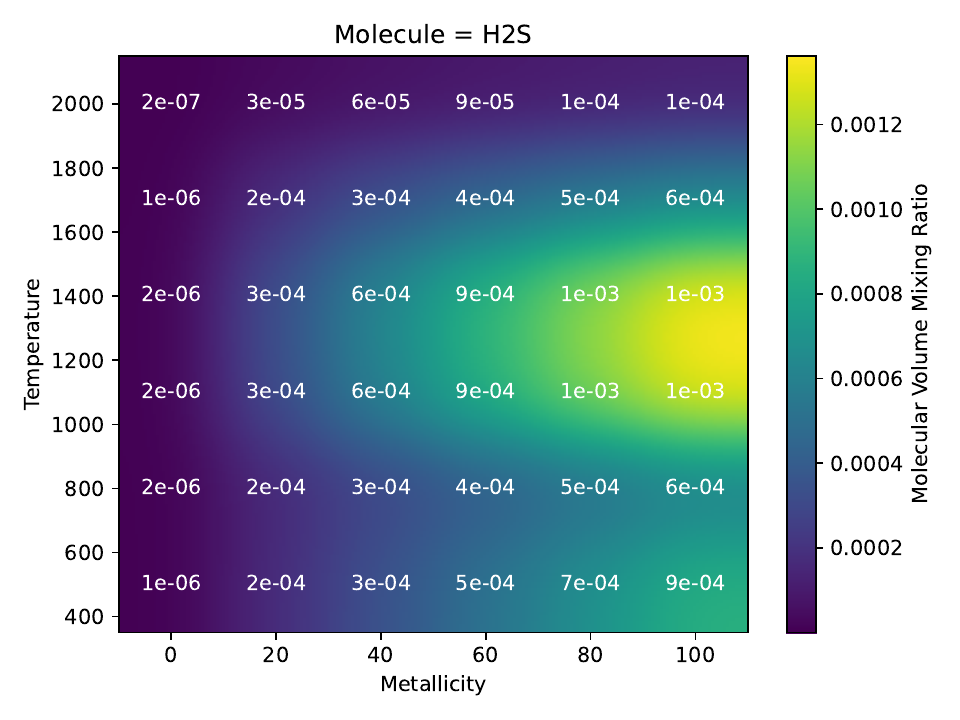}
\includegraphics[width=.30\textwidth]{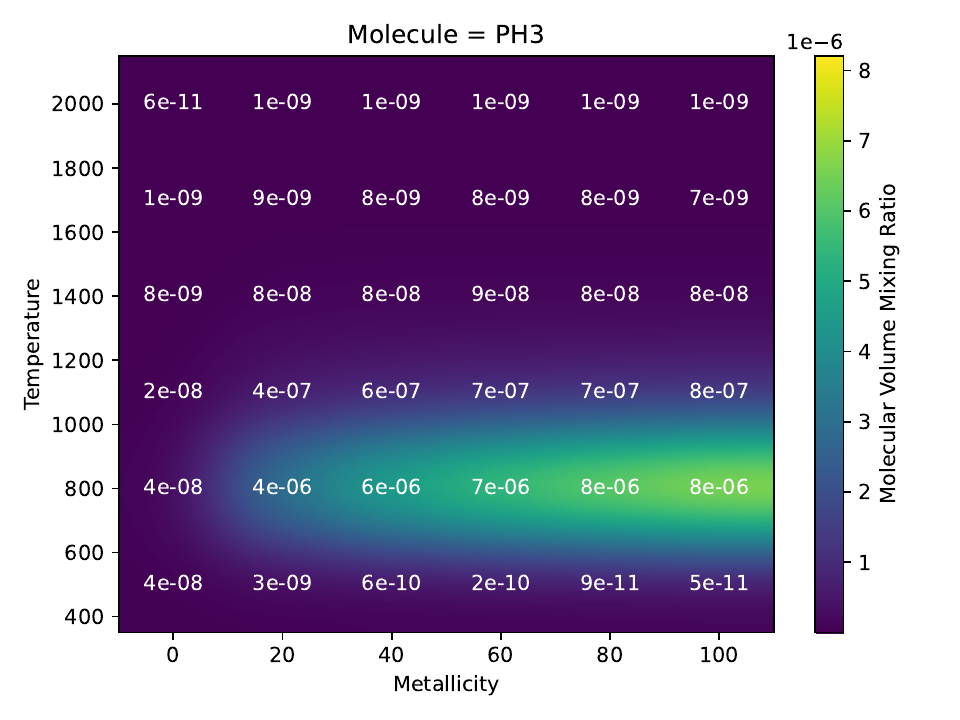}
\\
\includegraphics[width=.30\textwidth]{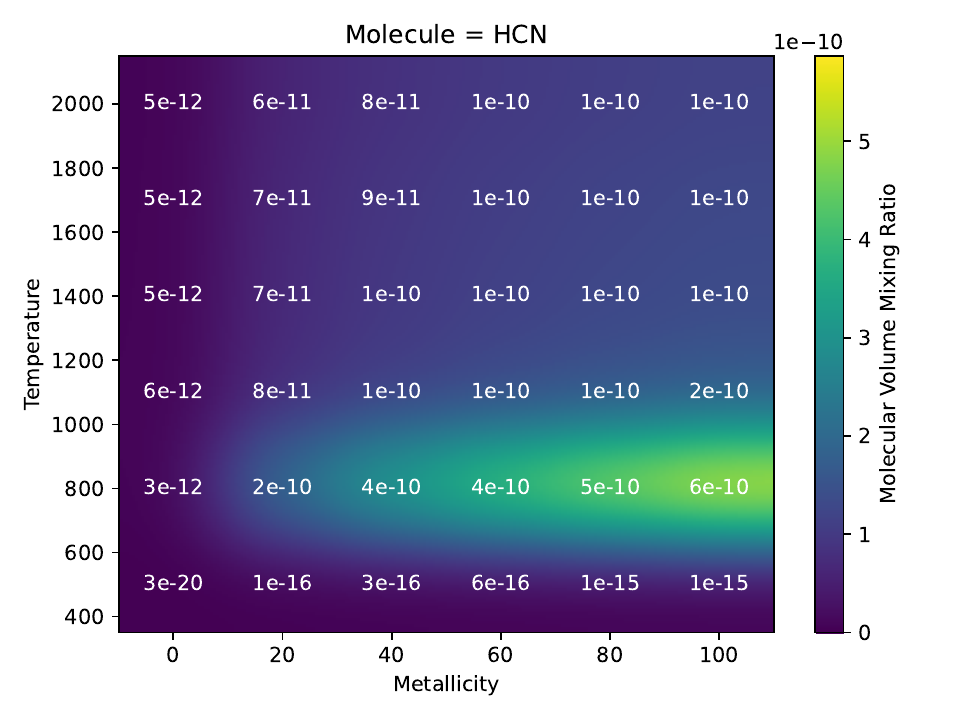}
\includegraphics[width=.30\textwidth]{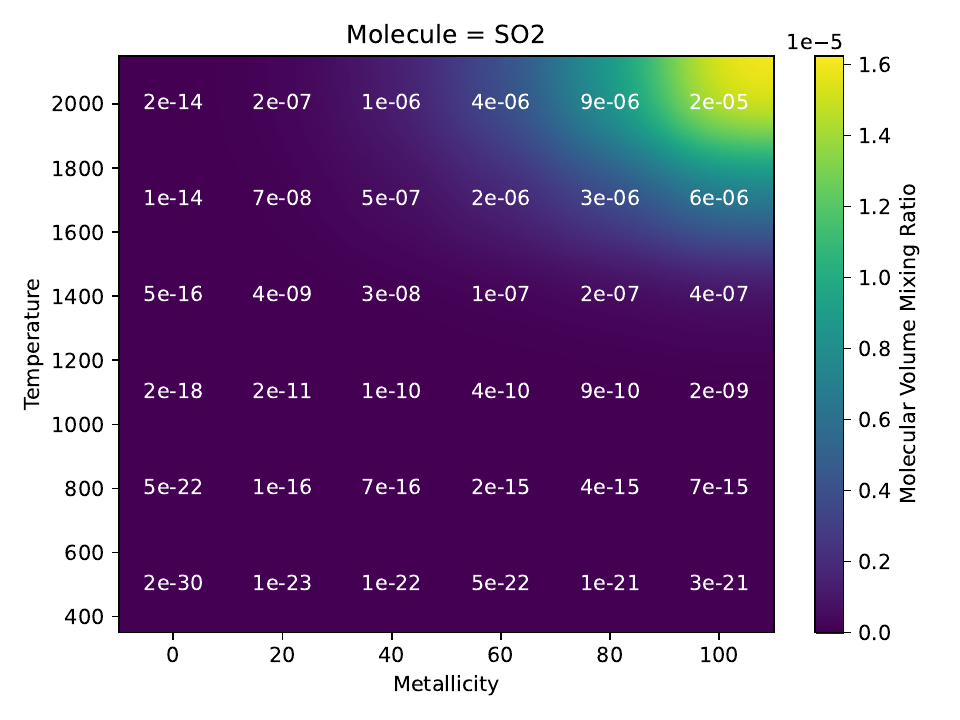}
\caption{Examples of molecular abundances from equilibrium chemistry in the temperatures and metallicities regime of WASP-39\,b's atmosphere. These plots showcase results at a pressure of ${\sim}10^{-3} \;\text{bar}$. Note that the colour scales vary for each plot. 
\label{fig:abundance}}
\end{figure*}

We conducted the chemical species selection following the methodology described in Section~\ref{sec:method_chem_select} covering the parameter space expected for WASP-39\,b, i.e. assuming as temperatures 700, 900, 1100, 1300 and 1500 K, and as volume metallicities 0.4, 4, 29 and 67\,$\times$ solar volume metallicity. The metallicity values are estimated from the constant species' abundance cases with molecular volume mixing ratios (VMRs) of $10^{-6}$, $10^{-5}$, $10^{-4}$, and $10^{-3}$. To maintain consistency between the equilibrium and homogeneous abundance simulations, we aim to keep the atmospheric metallicity and mean molecular weight within 20\,\% difference between the two sets of simulations. We list the highlighted species in Table~\ref{tab:selection}. Examples of key molecular abundances as a function of metallicity and temperature are given in Fig.~\ref{fig:abundance}. 

\begin{longrotatetable}
\begin{deluxetable*}{llll}
\tablecaption{Highlighted chemical species by the methodology in Section~\ref{sec:method_chem_select}. \label{tab:selection}}
\tablewidth{700pt}
\tabletypesize{\scriptsize}
\tablehead{
\colhead{VMR*} & \colhead{Met.$^\triangle$} & \colhead{$T_\mathrm{eq}$ [K]} & \colhead{Highlighted Species}}
\startdata
\textit{Eq.} Profiles & 0.4 & 700 & CH$_4$, Na, H$_2$O, CrH, PH$_3$, NH$_3$, SiH$_4$, H$_2$S, SiO \\
Constant Profiles 10$^{-6}$ & 0.4 & 700 & SiH$_4$, PH$_3$, CO$_2$, CrH, Na, CH$_4$, NH$_3$, CO, H$_2$O \\
\hline
\textit{Eq.} Profiles& 0.4 &  900 & H$_2$O, Na, CH$_4$, CrH, PH$_3$, CO, H$_2$S, SiO, NH$_3$, SiH$_4$ \\
Constant Profiles 10$^{-6}$ & 0.4 &  900 & SiH$_4$, PH$_3$, CrH, CH$_4$, Na, NH$_3$, H$_2$O, CO, H$_2$S, SiO \\
\hline
\textit{Eq.} Profiles& 0.4 &  1100 & H$_2$O, Na, CrH, CO, CH$_4$, H$_2$S, SiO, NH$_3$ \\
Constant Profiles 10$^{-6}$ & 0.4 &  1100 & CrH, CH$_4$, Na, NH$_3$, H$_2$O, CO, H$_2$S, SiO \\
\hline
\textit{Eq.} Profiles& 0.4 &  1300 & H$_2$O, Na, CO, CrH, H$_2$S, CH$_4$, SiO, NH$_3$ \\
Constant Profiles 10$^{-6}$ & 0.4 &  1300 & Cr, CH, N, NH, H$_2$, C, H$_2$, Si, SH \\
\hline
\textit{Eq.} Profiles& 0.4 &  1500 & H$_2$O, Na, CO, H$_2$S, SiO, CH$_4$, NH$_3$\\
Constant Profiles 10$^{-6}$ & 0.4 &  1500 & CH$_4$, Na, NH$_3$, H$_2$O, CO, H$_2$S, SiO, SH \\
\hline
Eq. Profiles & 4 & 700 & CO$_2$, CH$_4$, Na, PH$_3$, H$_2$O, CrH, K, NaOH, H$_2$S, NH$_3$, SiO, SiH$_4$, CO \\
Constant Profiles 10$^{-5}$ & 4 & 700 & SiH$_4$, PH$_3$, CO$_2$, NaOH, CH$_4$, CrH, K, Na, NH$_3$, H$_2$O, CO, H$_2$S, SiO \\
\hline
\textit{Eq.} Profiles& 4 & 900 & CaOH, CO$_2$, H$_2$O Na, CrH, PH$_3$, CO, CH$_4$, K, H$_2$S, SiO, NaOH, NH$_3$, SiH$_4$ \\
Constant Profiles 10$^{-5}$ & 4 & 900 & CaOH, SiH$_4$, PH$_3$, CO$_2$, NaOH, CH$_4$, CrH, Na, K, NH$_3$, H$_2$O, CO, H$_2$S, SiO \\
\hline
\textit{Eq.} Profiles& 4 & 1100 & CaOH, CO$_2$, H$_2$O Na, CrH, CO, PH$_3$, K, H$_2$S, MgH, CH$_4$, SiO, NH$_3$, SiS \\
Constant Profiles 10$^{-5}$ & 4 & 1100 & CaOH, CO$_2$, PH$_3$, CH$_4$, MgH, CrH, Na, K, NH$_3$, H$_2$O H$_2$S, CO, PS, SiO, SH \\
\hline
\textit{Eq.} Profiles& 4 & 1300 & CaOH, TiO, H$_2$O CO$_2$, Na, CO, CrH, H$_2$S, MgH, K, PH$_3$, SiO, CH$_4$, NH$_3$, SiS \\
Constant Profiles 10$^{-5}$ & 4 & 1300 & CaOH, TiO, CO$_2$, PH$_3$, MgH, CH$_4$, CrH, NH$_3$, Na, K, H$_2$O, PS, H$_2$S, CO, PO, SiO, SH \\
\hline
\textit{Eq.} Profiles& 4 & 1500 & TiO, H$_2$O, CO$_2$, Na, CO, MgH, CrH, H$_2$S, K, PH$_3$, SiO, CH$_4$, NH$_3$, SiS, SH \\
Constant Profiles 10$^{-5}$ & 4 & 1500 & TiO, CO$_2$, PH$_3$, MgH, CH$_4$, CrH, NH$_3$, Na, K, H$_2$O, PS, H$_2$S, CO, SO, PO, SiO, SH, SiS, AlF\\
\hline
Eq. Profiles & 29 & 700 & CO$_2$, CH$_4$, Na, H$_2$O, PH$_3$, CrH, CO, NaOH, K, H$_2$S, SiO, NH$_3$\\
Constant Profiles 10$^{-4}$ & 29 & 700 & SiH$_4$, PH$_3$, NaOH, CO$_2$, CH$_4$, CrH, K, Na, NH$_3$ \\
\hline
\textit{Eq.} Profiles& 29 & 900 & CaOH, CO$_2$, H$_2$O, Na, CrH, CO, PH$_3$, K, H$_2$S, CH$_4$, NaOH, SiO, NH$_3$ \\
Constant Profiles 10$^{-4}$ & 29 & 900 &  CaOH, SiH$_4$, PH$_3$, CO$_2$, NaOH, CH$_4$, CrH, NH$_3$, Na, K, H$_2$O, H$_2$S, \\
\hline
\textit{Eq.} Profiles& 29 & 1100 & CaOH, CO$_2$, H$_2$O, Na, CrH, CO, H$_2$S, K, PH$_3$, MgH, NaOH, SiO, CH$_4$, NH$_3$, SiS \\
Constant Profiles 10$^{-4}$ & 29 & 1100 & CaOH, CO$_2$, PH$_3$, NaOH, CH$_4$, MgH, CrH, NH$_3$, Na, K, H$_2$O, H$_2$S, PS, CO, PO, SiO \\
\hline
\textit{Eq.} Profiles& 29 & 1300 & CaOH, CO$_2$, TiO, H$_2$O, Na, H$_2$S, CO, CrH, MgH, K, PH$_3$, SiO, NaOH, CH$_4$, NaH, NH$_3$, SiS \\
Constant Profiles 10$^{-4}$ & 29 & 1300 & CaOH, TiO, CO$_2$, PH$_3$, NaOH, CH$_4$, MgH, CrH, NH$_3$, NaH, Na, K, H$_2$O, PS, H$_2$S, SO, CO, PO, SiO \\
\hline
\textit{Eq.} Profiles& 29 & 1500 & CO$_2$, TiO, H$_2$O, Na, CO, H$_2$S, MgH, CrH, K, SiO, PH$_3$, AlO, NaOH, NaH, PS, NH$_3$, CH$_4$, SiS, PO, SH \\
Constant Profiles 10$^{-4}$ & 29 & 1500 & TiO, CO$_2$, PH$_3$, AlO, NaOH, CH$_4$, MgH, CrH, NH$_3$, NaH, SO$_2$, Na, K, H$_2$O, PS, H$_2$S, SO, CO, PO, SiO, SH, HF \\
\hline
\textit{Eq.} Profiles & 67 & 700 & CO$_2$, CH$_4$, Na, H$_2$O, CrH, PH$_3$, NaOH, CO, K, H$_2$S \\
% Constant Profiles 10$^{-3}$ & 67 & 700 & \nodata \\
\hline
\textit{Eq.} Profiles & 67 & 900 & CaOH, CO$_2$, H$_2$O, Na, CrH, CO, PH$_3$, H$_2$S, K, NaOH, CH$_4$, SiO \\
% Constant Profiles 10$^{-3}$ & 67 & 900 & \nodata \\
\hline
\textit{Eq.} Profiles& 67 & 1100 & CaOH, CO$_2$, H$_2$O, Na, CrH, H$_2$S, CO, K, PH$_3$, MgH, NaOH, SiO, CH$_4$, NH$_3$, NaH \\
% Constant Profiles 10$^{-3}$ & 67 & 1100 & \nodata \\
\hline
\textit{Eq.} Profiles& 67 & 1300 & CaOH, CO$_2$, H$_2$O, TiO, Na, H$_2$S, CO, CrH, MgH, K, PH$_3$, NaOH, SiO, NaH, CH$_4$, NH$_3$, SiS \\
% Constant Profiles 10$^{-3}$ & 67 & 1300 & \nodata \\
\hline
\textit{Eq.} Profiles& 67 & 1500 & CO$_2$, TiO, H$_2$O, Na, CO, H$_2$S, MgH, CrH, K, SiO, PH$_3$, NaOH, AlO, PS, NaH, PO, NH$_3$, CH$_4$, SiS, SiO$_2$, SH\\
% Constant Profiles 10$^{-3}$ & 67 & 1500 & \nodata \\
\enddata
\tablecomments{*Volume mixing ratio of the species used in the transit depth simulation. $^\triangle$Volume metallicity in solar abundances.}
\end{deluxetable*}
\end{longrotatetable}

\section{Assessment of retrieval convergence} \label{appendix:MAP}

\begin{figure*}[ht!]
\plottwo{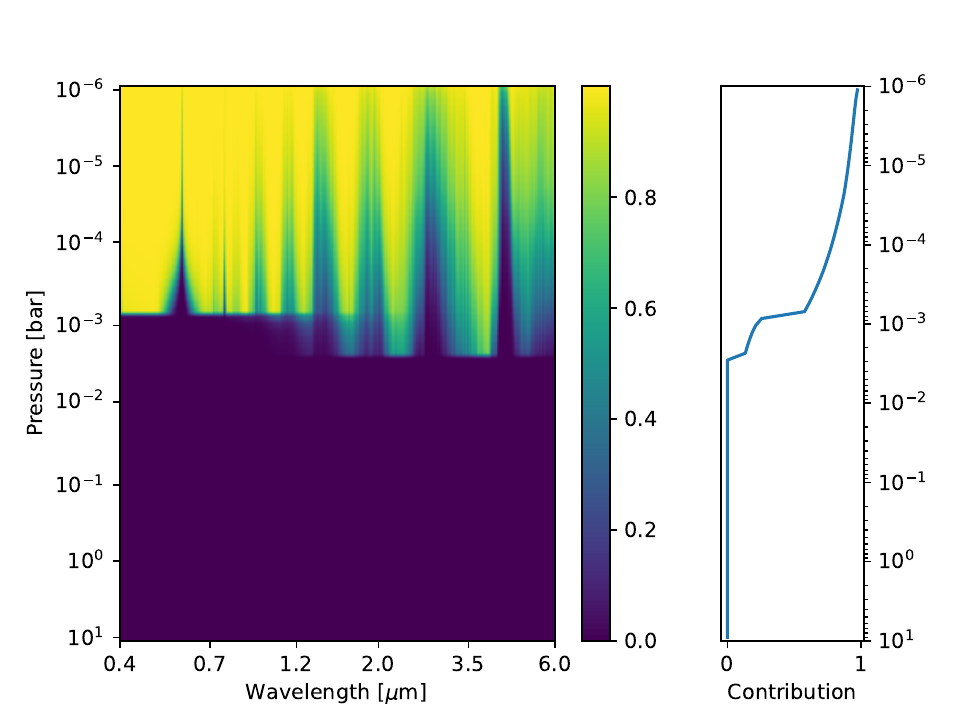}{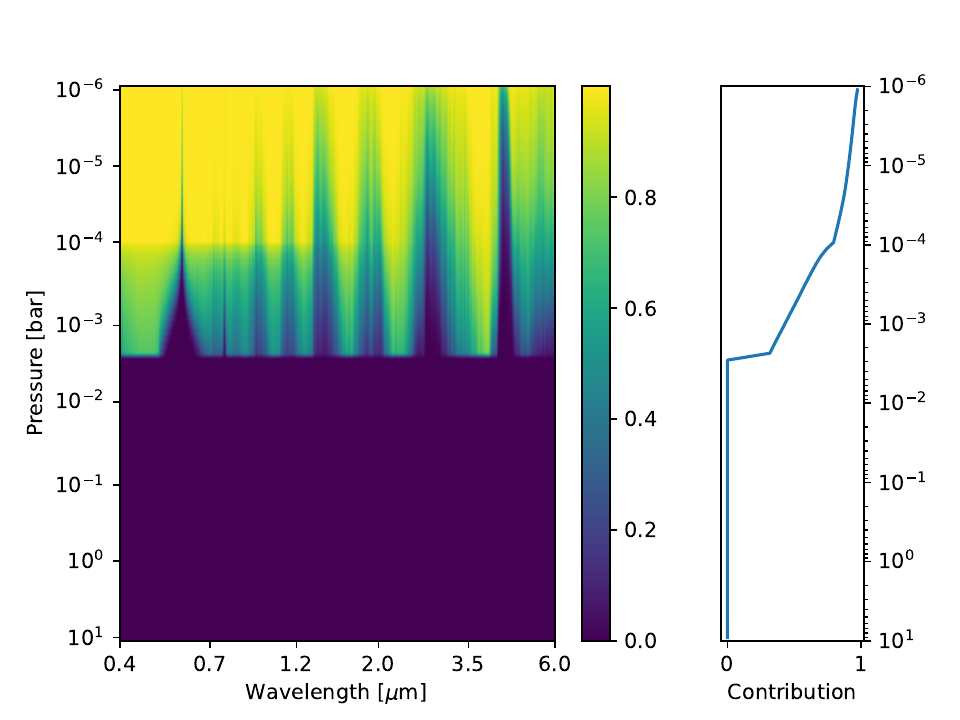}
\caption{Optical depths for the two solutions identified by Step~2 free retrievals in Fig.~\ref{fig:posteriors}. 
\label{fig:retrieved_details}}
\end{figure*}

In both solutions reported in Section~\ref{sec:optimized_retrieval_results}, H$_2$O and CO$_2$ fit the main spectral features, while SiO supports the feature around $4.1 \, \mu\text{m}$. Na and K align with features around 0.6 and 0.8~$\mu\text{m}$ in the optical. Grey clouds provided the continuum from optical to near-infrared. In both cases, the homogeneous particle-size (HPS) clouds contributed to the optical continuum. We show the details of the altitude-dependent optical depths and cloud contribution for the two solutions in Fig.~\ref{fig:retrieved_details}.

The Maximum A Posteriori (MAP) results, representing the statistical local optimisation, are consistent with the medium values shown in Fig.~\ref{fig:posteriors} above each histogram. The MAP values are the sets of parameter values that have the maximum likelihood. In well-constrained retrievals with chi-square posterior distributions, the set of median values is expected to fit the spectrum despite not being from the same retrieval sampling trace, which shows the good convergence of the algorithm. When the parameters follow a non-normal probability distribution, the median values will not necessarily fit the spectrum; however, this should not be the case in our study, given the nature of the parameters we are investigating.
We list the median values of posteriors and the first ten MAP traces in Table~\ref{tab:MAP} of the two retrieval solutions of our reference model in Section~\ref{sec:optimized_retrieval_results}.

\begin{longrotatetable}
\begin{deluxetable*}{lllllllllllllll}
\tablecaption{Reference model retrieved median and MAP values of two solutions. \label{tab:MAP}}
\tablewidth{700pt}
\tabletypesize{\scriptsize}
\tablehead{
\colhead{Value} & \colhead{$R_\mathrm{p}$} & \colhead{$T$} & \colhead{log$\,\chi_\mathrm{H_2O}$} & \colhead{log$\,\chi_\mathrm{CO_2}$} & \colhead{log$\,\chi_\mathrm{SiO}$} & \colhead{log$\,\chi_\mathrm{Na}$} & \colhead{log$\,\chi_\mathrm{K}$} & \colhead{log$\,p_\mathrm{grey}$} & \colhead{log$\,r_\mathrm{HPS}$} & \colhead{log$\,N_\mathrm{HPS}$} & \colhead{log$\,p_\mathrm{HPS,deck}$} & \colhead{log$\,p_\mathrm{HPS,base}$} & \colhead{$\mu$} & \colhead{Weight}\\
\colhead{} & \colhead{[R$_\mathrm{J}$]} & \colhead{[K]} & \colhead{} & \colhead{} & \colhead{} & \colhead{} & \colhead{} & \colhead{[Pa]} & \colhead{[m]} & \colhead{[m$^{-3}$]} & \colhead{[Pa]} & \colhead{[Pa]} & \colhead{[Dalton]} & \colhead{}}
\startdata
 & & & & & & \multicolumn{4}{c}{NIRISS + NIRSpec + NIRCam Solution~1}  & & & & \\
 \hline
Median & 1.209 & 7.782e+02 & -1.629 & -3.118 & -1.818 & -3.024 & -5.434 & 2.406 & -6.755 & 6.906 & 1.921 & 3.142 & 3.377 & \nodata \\
MAP 1& 1.209 & 8.014e+02 & -1.704 & -3.221 & -1.721 & -3.035 & -5.247 & 2.424 & -6.655 & 6.738 & 1.930 & 2.967 & 3.461 & 7.277e-04 \\
MAP 2& 1.209 & 7.898e+02 & -1.657 & -3.195 & -1.762 & -3.055 & -5.332 & 2.399 & -6.697 & 6.851 & 1.956 & 2.181 & 3.426 & 7.201e-04 \\
MAP 3& 1.209 & 7.960e+02 & -1.695 & -3.169 & -1.714 & -3.013 & -5.232 & 2.456 & -6.839 & 7.119 & 1.914 & 2.168 & 3.486 & 6.555e-04 \\
MAP 4& 1.209 & 7.877e+02 & -1.668 & -3.180 & -1.757 & -3.045 & -5.364 & 2.442 & -6.695 & 6.908 & 1.931 & 2.190 & 3.428 & 5.709e-04 \\
MAP 5& 1.208 & 8.032e+02 & -1.714 & -3.259 & -1.728 & -3.070 & -5.499 & 2.456 & -6.835 & 7.103 & 1.955 & 3.099 & 3.439 & 5.512e-04 \\
MAP 6& 1.211 & 7.930e+02 & -1.696 & -3.205 & -1.698 & -3.110 & -5.421 & 2.523 & -6.760 & 7.000 & 2.015 & 2.290 & 3.509 & 4.822e-04 \\
MAP 7& 1.211 & 7.664e+02 & -1.711 & -3.205 & -1.776 & -3.217 & -5.562 & 2.460 & -6.749 & 6.953 & 2.017 & 2.294 & 3.357 & 4.751e-04 \\
MAP 8& 1.210 & 7.952e+02 & -1.662 & -3.105 & -1.716 & -3.039 & -5.674 & 2.439 & -6.682 & 6.825 & 1.919 & 2.182 & 3.510 & 4.524e-04 \\
MAP 9& 1.211 & 7.635e+02 & -1.728 & -3.194 & -1.771 & -3.204 & -5.478 & 2.541 & -6.782 & 6.947 & 2.043 & 3.858 & 3.355 & 4.513e-04 \\
MAP 10& 1.208 & 7.939e+02 & -1.735 & -3.244 & -1.745 & -3.108 & -5.507 & 2.444 & -6.716 & 6.853 & 1.972 & 3.140 & 3.394 & 4.472e-04 \\
\hline
 & & & & & & \multicolumn{4}{c}{NIRISS + NIRSpec + NIRCam Solution~2} & & & & \\
 \hline
Median & 1.208 & 7.640e+02 & -1.687 & -3.157 & -1.857 & -3.239 & -5.942 & 2.403 & -6.290 & 3.817 & 1.156 & 2.792 & 3.264 & \nodata \\
MAP 1& 1.209 & 7.875e+02 & -1.736 & -3.163 & -1.739 & -3.257 & -5.788 & 2.395 & -6.263 & 3.686 & 1.099 & 3.987 & 3.403 & 1.136e-03 \\
MAP 2& 1.208 & 7.769e+02 & -1.749 & -3.201 & -1.783 & -3.322 & -5.740 & 2.443 & -6.192 & 3.499 & 1.006 & 3.514 & 3.318 & 1.134e-03 \\
MAP 3& 1.210 & 7.687e+02 & -1.720 & -3.163 & -1.754 & -3.347 & -6.174 & 2.540 & -6.257 & 3.676 & 1.030 & 3.619 & 3.386 & 9.423e-04 \\
MAP 4& 1.208 & 7.806e+02 & -1.780 & -3.249 & -1.784 & -3.354 & -5.972 & 2.400 & -6.320 & 3.781 & 1.077 & 3.560 & 3.293 & 9.354e-04 \\
MAP 5& 1.208 & 7.759e+02 & -1.740 & -3.235 & -1.789 & -3.352 & -5.993 & 2.449 & -6.372 & 3.888 & 1.037 & 3.369 & 3.311 & 9.112e-04 \\
MAP 6& 1.209 & 7.846e+02 & -1.740 & -3.204 & -1.740 & -3.215 & -5.996 & 2.420 & -6.362 & 3.849 & 1.039 & 3.581 & 3.398 & 8.759e-04 \\
MAP 7& 1.211 & 7.801e+02 & -1.728 & -3.195 & -1.717 & -3.346 & -5.764 & 2.517 & -6.221 & 3.577 & 1.082 & 3.235 & 3.443 & 8.450e-04 \\
MAP 8& 1.207 & 7.845e+02 & -1.671 & -3.167 & -1.819 & -3.242 & -5.786 & 2.361 & -6.382 & 3.990 & 1.010 & 1.770 & 3.321 & 7.704e-04 \\
MAP 9& 1.209 & 7.867e+02 & -1.748 & -3.202 & -1.734 & -3.232 & -5.628 & 2.370 & -6.401 & 3.943 & 1.038 & 3.813 & 3.403 & 7.475e-04 \\
MAP 10& 1.210 & 7.750e+02 & -1.700 & -3.152 & -1.754 & -3.306 & -5.856 & 2.529 & -6.249 & 3.636 & 1.101 & 3.388 & 3.401 & 6.851e-04 \\
\enddata
% \tablecomments{}
\end{deluxetable*}
\end{longrotatetable}

\section{Testing retrieval degeneracy: atmospheric scale height} \label{appendix:mmw}
%% Increased freedom on scale heights
After optimising the retrieval simulations in Step~2 using a minimal set of free parameters (reference model), we introduced higher retrieval freedom by allowing the algorithm to adjust the scale height through the mean molecular weight. By incorporating N$_2$ and Li as representative gases for heavy and light spectrally inactive species, this approach avoids introducing unwanted spectral signatures from the added species. Additionally, we include CO, which was excluded from the reference model in Step~2, to decrease the retrieval dimensionality due to its features largely being masked by H$_2$O and CO$_2$, although it has been suggested in molecular selection alongside other molecules in Step~1.

\begin{figure*}[ht]
\centering
\includegraphics[width=1.\textwidth]{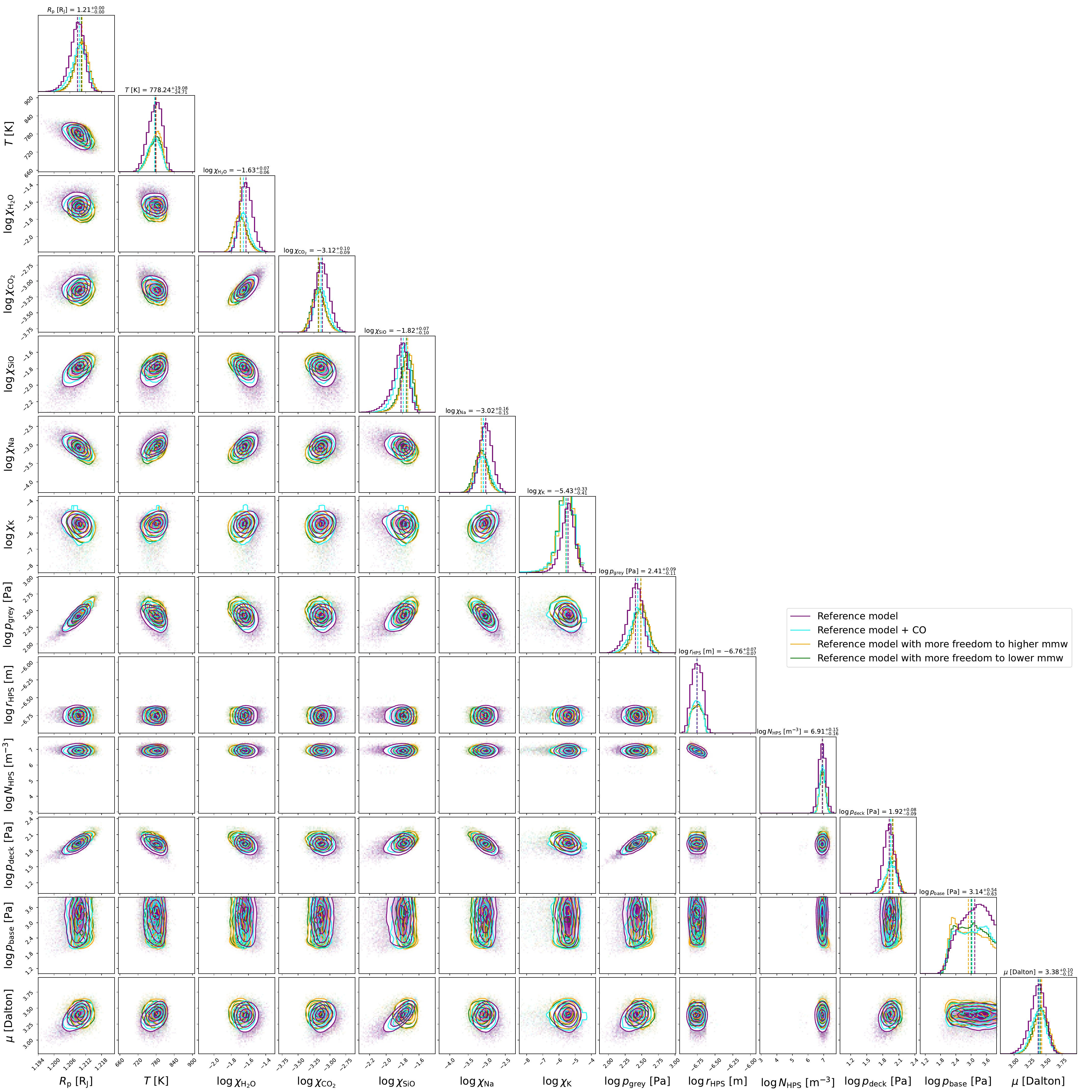}
\includegraphics[width=.20\textwidth]{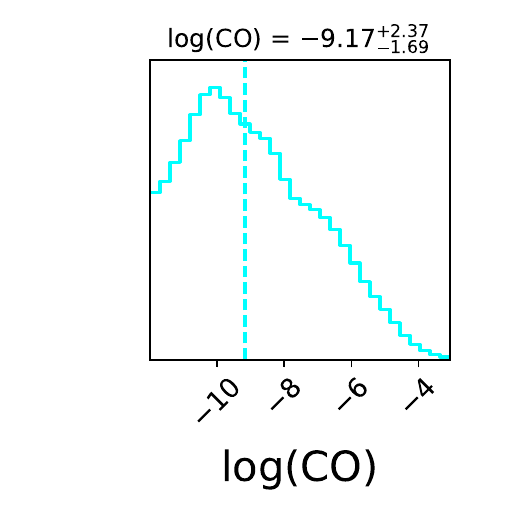}
\includegraphics[width=.20\textwidth]{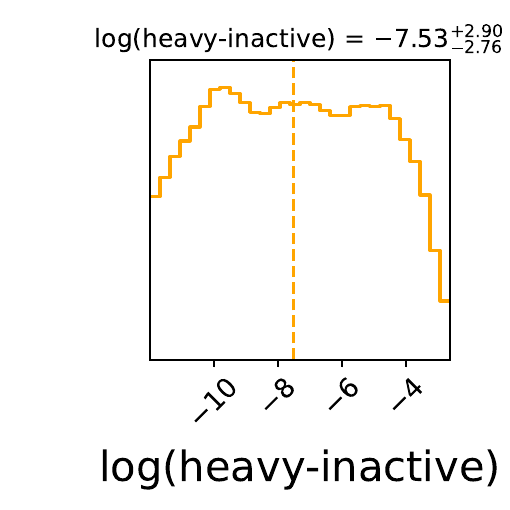}
\includegraphics[width=.20\textwidth]{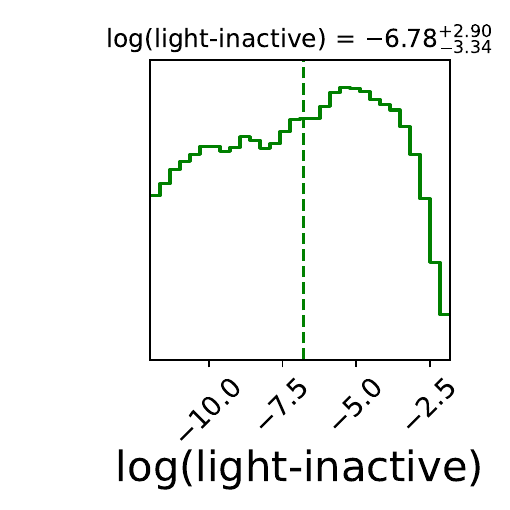}
\caption{Posterior comparison (Solution~1) of retrievals of the reference model in Step~2 (purple), with CO (cyan), with N$_2$ to have more freedom to higher mean molecular weight (yellow) and with Li to have more freedom to lower mean molecular weight (green).} 
\label{fig:posteriors_mmw_sol0}
\end{figure*}

\begin{figure*}[ht]
\centering
\includegraphics[width=1.\textwidth]{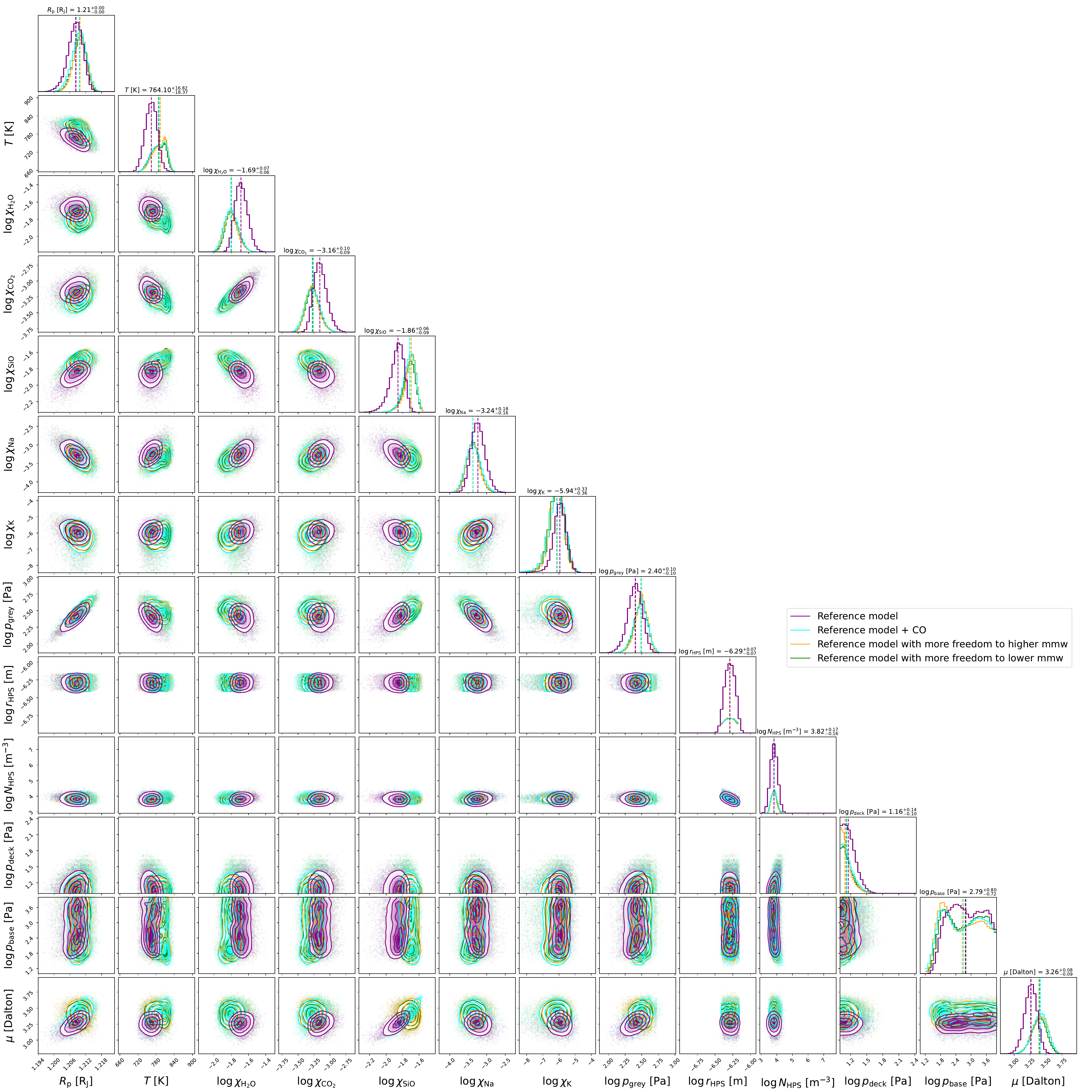}
\includegraphics[width=.20\textwidth]{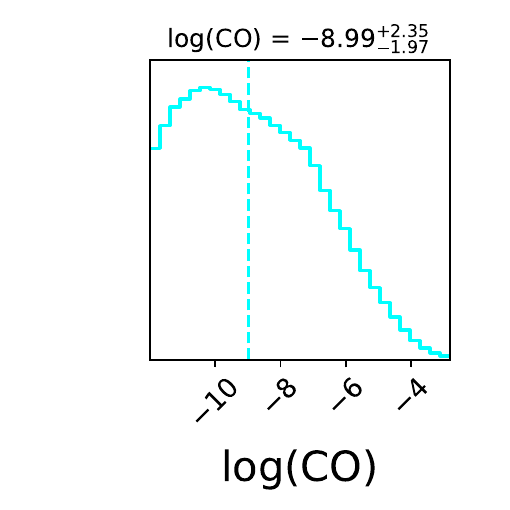}
\includegraphics[width=.20\textwidth]{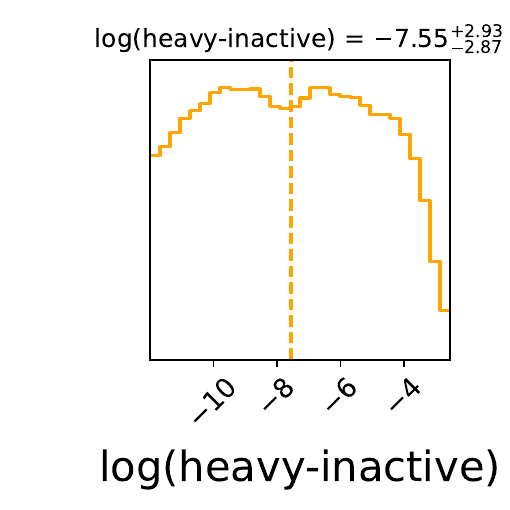}
\includegraphics[width=.20\textwidth]{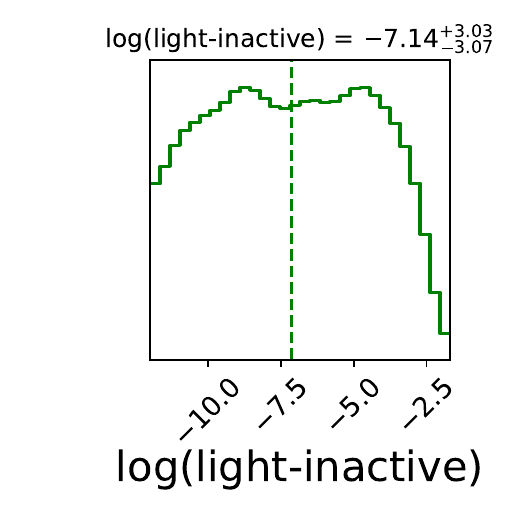}
\caption{Posterior comparison (Solution~2) of the four models in Fig.~\ref{fig:posteriors_mmw_sol0}.}
\label{fig:posteriors_mmw_sol1}
\end{figure*}

We present the posterior distributions in Fig.~\ref{fig:posteriors_mmw_sol0} (Solution~1) and~\ref{fig:posteriors_mmw_sol1} (Solution~2). The results from the other three models show no significant deviation from the reference model in either solution. None of the models constrains the abundances of CO, N$_2$, or Li.

\end{document}